\documentclass[a4paper,11pt]{article}
\usepackage[a4paper,left=2.73cm,right=2.7cm,top=3cm,bottom=3.5cm]{geometry}

\usepackage{amsfonts,amsmath,amssymb,tabstackengine}
\usepackage[usenames, dvipsnames]{color}
\usepackage{array}

\usepackage{color, colortbl,xcolor}
\definecolor{Gray}{gray}{0.95}

\usepackage[colorlinks=true,linktocpage=true,linkcolor=blue,citecolor=blue]{hyperref}
\usepackage{graphicx}

\addtocontents{toc}{\protect\setcounter{tocdepth}{3}}
\numberwithin{equation}{section}

\begin{document}

\begin{titlepage}

\thispagestyle{empty}

\begin{center}

{\LARGE \textbf{Janus and Hades in M-theory}}

\vspace{40pt}
		
{\large \bf Andr\'es Anabal\'on}$^{(1)}$ \,  \large{,} \,  {\large \bf Miguel Chamorro-Burgos}$^{(2)}$  \,  \large{and} \, {\large \bf Adolfo Guarino}$^{(2,3)}$
		
\vspace{25pt}

{\normalsize  
$^{(1)}$ Departamento de Ciencias, Facultad de Artes Liberales, Universidad Adolfo Ib\'{a}\~{n}ez,\\ 
Avda. Padre Hurtado 750, Vi\~{n}a del Mar, Chile.}
\\[7mm]

{\normalsize  
$^{(2)}$ Departamento de F\'isica, Universidad de Oviedo,\\
Avda. Federico Garc\'ia Lorca 18, 33007 Oviedo, Spain.}
\\[7mm]

{\normalsize  
$^{(3)}$ Instituto Universitario de Ciencias y Tecnolog\'ias Espaciales de Asturias (ICTEA) \\
Calle de la Independencia 13, 33004 Oviedo, Spain.}
\\[10mm]

\texttt{andres.anabalon@uai.cl}  \,\, , \,\, \texttt{mig.cha.bur@hotmail.es}  \,\, , \,\, \texttt{adolfo.guarino@uniovi.es}

\vspace{20pt}

\vspace{20pt}
				
\abstract{
Multi-parametric and analytic families of four-dimensional $\,\textrm{AdS}_{3} \times \mathbb{R}\,$ (Janus) and $\,\textrm{AdS}_{3} \,\times\, \mathbb{R}^{+}$ (Hades) solutions are constructed in the SO(8) gauged supergravity that arises from the consistent reduction of eleven-dimensional supergravity on $\,\textrm{S}^7\,$. The solutions are generically non-supersymmetric, involve non-trivial running scalars and preserve a $\,\textrm{U}(1)^4\,$ symmetry. Different patterns of (super) symmetry enhancement occur upon suitable adjustment of the free parameters which further control the boundary conditions of the running scalars. We concentrate on the non-supersymmetric Janus and Hades solutions with $\,\textrm{SU}(3) \times \textrm{U}(1)^2\,$ symmetry and provide their higher-dimensional description in terms of M-theory fluxes and membranes. Special attention is paid to a class of such Hades solutions
dubbed ``ridge flows" which resemble dielectric rotations of Coulomb branch flows previously investigated in the literature.
}

\end{center}

\end{titlepage}

\tableofcontents

\hrulefill
\vspace{10pt}

\section{Introduction: Janus and Hades}

Janus solutions in string/M-theory were originally introduced in the context of type IIB supergravity as a simple deformation of the $\,\textrm{AdS}_{5} \times \textrm{S}^{5}\,$ background involving a non-trivial dilaton profile \cite{Bak:2003jk}. The deformation breaks the $\,\textrm{SO}(2,4)\,$ isometries of $\,\textrm{AdS}_{5}\,$ to the $\,\textrm{SO}(2,3)\,$ isometries of $\,\textrm{AdS}_{4}\,$, but preserves the $\,\textrm{SO}(6)\,$ isometries of the round $\,\textrm{S}^{5}\,$. Soon after, a holographic interpretation of the solutions in \cite{Bak:2003jk} was proposed in terms of a planar $\,(1+2)$-dimensional interface in super Yang--Mills (SYM) separating two half-spaces with different coupling constants \cite{Clark:2004sb}. The supersymmetric Janus was constructed in \cite{Clark:2005te} using a 5D effective SO(6) gauged supergravity approach. Its ten-dimensional incarnation was put forward in \cite{DHoker:2006vfr}, which provided the gravity dual of the $\,\mathcal{N}=1\,$ three-dimensional interface with OSp$(1|4)$ superconformal symmetry first anticipated in \cite{Clark:2004sb} and then constructed in \cite{DHoker:2006qeo}. This supersymmetric Janus turns out to break the symmetry of the original (non-supersymmetric) Janus down to at least $\,\textrm{SU(3)} \subset \textrm{SO}(6)\,$. The supersymmetric Janus solution with $\,\textrm{SO}(4)\,$ symmetry dual to the $\,\mathcal{N}=4\,$ interface with OSp$(4|4)$ superconformal symmetry was constructed in \cite{DHoker:2007zhm}. However it was only recently that the supersymmetric Janus with $\,\textrm{SU}(2) \times  \textrm{U}(1)\,$ symmetry dual to the $\,\mathcal{N}=2\,$ interface with OSp$(2|4)$ superconformal symmetry was constructed in five and ten dimensions \cite{Bobev:2020fon}, thus completing the list of Janus solutions dual to the SYM interfaces scrutinised in \cite{DHoker:2006qeo}.

Janus solutions have been much less investigated in the context of M-theory. The first examples were constructed in \cite{DHoker:2009lky} (and generalised in \cite{Bachas:2013vza}) as half-maximal deformations of the $\,\textrm{AdS}_{4} \times \textrm{S}^{7}\,$ background preserving a subgroup $\,\textrm{SO}(4) \times \textrm{SO}(4) \subset \textrm{SO}(8)\,$ of the isometries of the round $\,\textrm{S}^{7}\,$. This time the deformation breaks the $\,\textrm{SO}(2,3)\,$ isometries of $\,\textrm{AdS}_{4}\,$ to the $\,\textrm{SO}(2,2)\,$ isometries of $\,\textrm{AdS}_{3}\,$,  and the Janus can still be holographically understood as an $\,\mathcal{N}=(4,4)\,$ two-dimensional interface with $\,\textrm{OSp}(4|2) \times \textrm{OSp}(4|2)\,$ superconformal symmetry in ABJM theory \cite{Aharony:2008ug} despite the absence of a dilaton field in the M-theory context \cite{DHoker:2009lky,Bobev:2013yra,Kim:2018qle}. Interestingly, it was shown in \cite{Bobev:2013yra} that the $\,\textrm{SO}(4) \times \textrm{SO}(4)\,$ symmetric Janus can be alternatively found using a 4D effective SO(8) gauged supergravity description. Using this 4D approach, a supersymmetric Janus with $\,\textrm{SU}(3) \times \textrm{U}(1)^2\,$ symmetry dual to an $\mathcal{N}=(0,2)$ interface with $\,\textrm{OSp}(0|2) \times \textrm{OSp}(2|2)\,$ superconformal symmetry was constructed in \cite{Bobev:2013yra} using numerical methods, for which 11D uplift formuli were provided in \cite{Pilch:2015dwa}. More numerical Janus solutions were also presented in \cite{Bobev:2013yra} by studying the $\,\textrm{G}_{2}$-invariant sector of the SO(8) gauged supergravity.\footnote{See \cite{Karndumri:2020bkc} for a numerical study of Janus solutions in the one-parameter family of $\omega$-deformed SO(8) gauged supergravities \cite{Dall'Agata:2012bb}. See also \cite{Suh:2018nmp,Karndumri:2021pva} for a similar study in the context of massive IIA compactified on $\,\textrm{S}^{6}\,$ and its effective description in terms of the ISO(7) gauged supergravity \cite{Guarino:2015jca,Guarino:2015qaa,Guarino:2015vca}.}

Amongst the various interesting questions raised in the discussion section of \cite{DHoker:2009lky} we will provide a positive answer to that of whether exact M-theory Janus solutions exist with no supersymmetry. We will use the four-dimensional SO(8) gauged supergravity that arises upon reduction of eleven-dimensional supergravity on $\,\textrm{S}^{7}\,$ \cite{deWit:1982ig,deWit:1986oxb} and construct non-supersymmetric, yet analytic and regular, AdS$_{3}$-sliced domain-wall solutions of the form
\begin{equation}
\label{metric_ansatz_intro}
ds_{4}^{2} = d\mu^{2} + e^{2 A(\mu)} \, ds_{\textrm{AdS}_{3}}^{2} \ ,
\end{equation}
for which the metric function $\,A(\mu)\,$ depends arbitrarily on three real constants $\,\alpha_{i} \in \mathbb{R}^{+}\,$ with $\,{i=1,2,3}\,$. The geometry is supported by three complex scalar fields $\,z_{i}(\alpha_{i},\beta_{i};\mu)\,$ which depend on three additional compact parameters, or phases $\,\beta_{i} \in [0,2\pi]\,$, and develop non-trivial profiles along the radial coordinate $\,\mu\,$ transverse to the domain-wall. The effective 4D gauge coupling $\,g\,$ -- which relates to the inverse radius of $\,\textrm{S}^{7}\,$ -- and the set of real parameters $\,(\alpha_{i},\beta_{i})\,$ fully determine a particular Janus configuration.

The Janus parameters $\,(\alpha_{i},\beta_{i})\,$ specify the boundary values of the complex scalars at $\,{\mu \rightarrow \pm \infty}\,$. In particular, the parameters $\,\beta_{i}\,$ encode the source/VEV and bosonic/fermionic nature of the dual operators turned on on each side of the interface living at the boundary. A generic choice of Janus parameters breaks all the supersymmetries and the $\,\textrm{S}^{7}\,$ isometry group down to its Cartan subgroup $\,\textrm{U}(1)^4 \subset \textrm{SO}(8)\,$. On the contrary, the very special choice $\,\alpha_{i}=0\,$ $\forall i\,$ trivialises the Janus and the maximally supersymmetric AdS$_{4}$ vacuum of the SO(8) supergravity that uplifts to the $\,\textrm{AdS}_4 \times \textrm{S}^7\,$ Freund--Rubin background of eleven-dimensional supergravity with a round $\,\textrm{S}^{7}\,$ metric is recovered \cite{Freund:1980xh}. Interestingly, (super) symmetry enhancements occur upon suitable identifications between the parameters. For instance, the supersymmetric Janus of \cite{DHoker:2009lky,Bobev:2013yra} with $\,\textrm{SO}(4) \times \textrm{SO}(4)\,$ symmetry is recovered upon setting two of the $\,\alpha_{i}\,$ parameters to zero. In this work we will pay special attention to the Janus with $\,{\alpha_{1}=\alpha_{2}=\alpha_{3}}\,$ and $\,{\beta_{1}=\beta_{2}=\beta_{3}}\,$ which is non-supersymmetric and features an $\,{\text{SU}(3) \times\text{U}(1)^2}\,$ symmetry enhancement. We will present the uplift of this 4D Janus to eleven-dimensions providing, to the best of our knowledge, the first example of an exact M-theory Janus with no supersymmetry.

In addition to the Janus, we will construct another class of analytic solutions -- we refer to them as flows to Hades following standard terminology in the literature -- which are non-supersymmetric and display a singularity at $\,\mu =0\,$ where the $\,e^{2 A(\mu)}\,$ factor in (\ref{metric_ansatz_intro}) shrinks to zero size and the complex scalars run to the boundary of moduli space. Some similar curved-sliced \cite{Bobev:2013yra} and flat-sliced \cite{Cvetic:1999xx,Pope:2003jp,Pilch:2015vha,Pilch:2015dwa} singular flows have been constructed within the $\,\textrm{SO}(8)\,$ gauged supergravity and argued to holographically describe an interface between a superconformal ABJM phase and a non-conformal phase with potentially interesting physics.\footnote{The scalar potential of the maximal $\,\textrm{SO}(8)\,$ gauged supergravity is bounded above by its value at the maximally supersymmetric AdS$_{4}$ vacuum thus satisfying the \textit{good} condition of \cite{Gubser:2000nd}.} In their simplest realisation, these flat-sliced singular flows in M-theory are the analogue of the type IIB flows to the Coulomb branch of $\,\mathcal{N}=4\,$ SYM investigated in \cite{Cvetic:1999xx,Freedman:1999gp,Freedman:1999gk,Gubser:2000nd}.

There are similarities and differences between the Janus and the Hades. As for the Janus, the Hades solutions depend on a set of six parameters $\,(\alpha_{i},\beta_{i})\,$. Unlike for the Janus, no supersymmetric limit can be taken on the Hades parameters, and the very special choice $\,\alpha_{i}=0\,$ $\forall i\,$ does not trivialise the Hades to recover AdS$_{4}$. Instead, a special class of Hades flows -- we will refer to them as \textit{ridge flows} adopting the terminology of \cite{Pilch:2015vha} -- appears in this limit. As before, we will concentrate on the simple case with $\,{\alpha_{1}=\alpha_{2}=\alpha_{3}} \equiv \alpha\,$ and $\,{\beta_{1}=\beta_{2}=\beta_{3}}\equiv \beta\,$ for which the flows to Hades feature an $\,{\text{SU}(3) \times\text{U}(1)^2}\,$ symmetry, and present their uplift to eleven-dimensional supergravity.

Special attention will then be paid to the $\,{\text{SU}(3) \times\text{U}(1)^2}\,$ symmetric ridge flows with $\,\alpha=0\,$ for which there is just one free parameter left, \textit{i.e.} the phase $\,\beta \in [0,2\pi]\,$. This phase specifies the boundary values of the complex scalars at $\,\mu \rightarrow \infty\,$ and, therefore, the source/VEV and bosonic/fermionic nature of the dual operators turned on on the ultraviolet (UV) side of the conformal interface. In the infrared (IR) side $\,\mu \rightarrow 0\,$ of the interface, the four-dimensional solution becomes singular and the dual field theory is expected to enter the non-conformal phase. Interestingly, the parameter $\,\beta\,$ determining the boundary conditions of the complex scalars is associated with a $\,\textrm{U}(1)_{\xi}\,$ duality symmetry of the four-dimensional supergravity Lagrangian. However, as originally noticed in \cite{Pope:2003jp} for a class of conventional flat-sliced RG-flows (see also \cite{Pilch:2015vha,Pilch:2015dwa}), the $\,\textrm{U}(1)_{\xi}\,$ changes the physics of the ridge flows once they are uplifted to eleven dimensions: it takes metric modes into three-form gauge field modes.

We will illustrate this phenomenon by analysing in some detail the simple cases of setting $\,\beta= \frac{\pi}{2}\,$ and $\,\beta=0\,$. The corresponding ridge flows are triggered from the UV solely by bosonic VEV's or fermionic sources, respectively. The resulting M-theory ridge flows will be shown to be drastically different as far as the persistence of the singularity and the presence of magnetic M5-branes sources in the background are concerned. Setting $\,\beta= \frac{\pi}{2}\,$ produces a singular M-theory background without magnetic M5-branes sources akin the (flat-sliced) Coulomb branch flows constructed in \cite{Cvetic:1999xx}. Modifying the phase $\,\beta\,$ by acting with $\,\textrm{U}(1)_{\xi}\,$ turns out to induce a transformation on the eleven-dimensional backgrounds that parallels the dielectric rotation of Coulomb branch flows investigated in \cite{Pope:2003jp,Pilch:2015vha,Pilch:2015dwa}. We will look in detail at the limiting case $\,\beta=0\,$ and conclude that the $\,\textrm{U}(1)_{\xi}\,$ transformation totally polarises M2-branes into M5-branes when flowing from the UV to the IR, leaving no M2-branes. We will provide some evidence for this phenomenon to occur also at generic values of $\,\beta\,$.

The paper is organised in four sections plus appendices. In Section~\ref{sec:Janus} we present our multi-parametric $(\alpha_{i},\beta_{i})$-families of analytic Janus and Hades solutions and discuss the ridge flow limit of the latter. We investigate the various possibilities of (super) symmetry enhancement depending on the choice of $\,\alpha_{i}\,$ parameters, as well as the various possibilities of boundary conditions for the complex scalars (sources/VEV's of dual operators) depending on the choice of $\,\beta_{i}\,$ parameters. In Section~\ref{sec:Uplift_11D} we present the uplift of the Janus and Hades solutions with $\,\textrm{SU}(3) \times \textrm{U}(1)^2\,$ symmetry to eleven-dimensional supergravity. We then focus on the ridge flows and discuss some eleven-dimensional aspects of the solutions, like the presence of singularities or the characterisation of the M2/M5-brane sourcing the backgrounds, as a function of the parameter $\,\beta\,$. We summarise the results and conclude in Section~\ref{sec:conclusions}. Two additional appendices accompany the main text which contain technical results regarding the BPS equations as well as some relevant uplift formuli for the STU model. This is the subsector of the four-dimensional maximal $\,\textrm{SO}(8)\,$ gauged supergravity within which we have constructed all the solutions presented in this work.

\section{Four-dimensional Janus and Hades}
\label{sec:Janus}

\subsection{The model}

Our starting point is the $\mathcal{N}=2$ gauged STU supergravity in four dimensions \cite{Cvetic:1999xp}. This theory has a gauge group $\textrm{U}(1)^4$, the maximal Abelian subgroup of $\textrm{SO}(8)$, and can be embedded into the maximal $\mathcal{N}=8$ $\textrm{SO}(8)$-gauged supergravity \cite{deWit:1982ig} as its $\textrm{U}(1)^4$ invariant sector \cite{Cvetic:1999xp}. The field content consists of the $\mathcal{N}=2$ supergravity multiplet coupled to three vector multiplets. Upon setting vector fields to zero, the bosonic Lagrangian reduces to an Einstein-scalar model given by
\begin{equation}
\label{Lagrangian_model_U1^4_Einstein-scalars}
\begin{array}{lll}
\mathcal{L} & = & \left(  \dfrac{R}{2} - V \right)  * 1 - \dfrac{1}{4} \displaystyle\sum_{i=1}^3 \left[
d\varphi_{i} \wedge* d\varphi_{i} + e^{2 \varphi_{i}} \, d\chi_{i} \wedge* d\chi_{i} \right] \\[4mm]
& = & \left(  \dfrac{R}{2} - V \right)  * 1 - \displaystyle\sum_{i=1}^{3}
\dfrac{1}{\left(  1-|\tilde{z_{i}}|^{2} \right)  ^{2}} \, d\tilde{z}_{i}
\wedge* d\tilde{z}_{i}^{*} \ .
\end{array}
\end{equation}
In passing from the first line to the second one in (\ref{Lagrangian_model_U1^4_Einstein-scalars}) we have changed the parameterisation of the scalar  fields $\,z_{i}\,$ ($i=1,2,3$) in the vector multiplets -- which serve as coordinates in the scalar coset geometry $[\textrm{SL}(2)/\textrm{SO}(2)]^3$ -- from the upper-half plane to the unit-disk parameterisation via the field redefinition
\begin{equation}
\label{ztilde&z}
\tilde{z}_{i}=\frac{z_{i}-i}{z_{i}+i} 
\hspace{10mm} \textrm{ with } \hspace{10mm} 
z_{i} = -\chi_{i} + i \, e^{-\varphi_{i}} \ .
\end{equation}
The non-trivial scalar potential in the Lagrangian (\ref{Lagrangian_model_U1^4_Einstein-scalars}) is given by
\begin{equation}
\label{V_U1^4}
V = - \tfrac{1}{2} \, g^{2} \sum_{i} \left(  2 \, \cosh\varphi_{i} + \chi
_{i}^{2} \, e^{\varphi_{i}} \right)  =  g^{2} \, \left( 3 - \sum_{i} \frac{2}{1-|\tilde{z}_{i}|^{2}}   \right) \ ,
\end{equation}
where $\,g\,$ is the gauge coupling in the gauged four-dimensional supergravity. From (\ref{V_U1^4}) one immediately sees that only $|\tilde{z}_{i}|$ enter the
potential. As a result, the Lagrangian (\ref{Lagrangian_model_U1^4_Einstein-scalars}) is
invariant under the three $\,\textrm{U}(1)_{\xi_{i}}\,$ shifts of $\,\arg\tilde{z}_{i}\,$, namely, $\,\delta_{\xi_{i}}\tilde{z}_{i} = i \, \xi_{i} \, \tilde{z}_{i}\,$, with constant parameters $\,\xi_{i}\,$. However, as we will see shortly, the phases $\,\arg\tilde{z}_{i}\,$ will play a central role when discussing boundary conditions for Janus- and Hades-like solutions in this supergravity model.

In this work we will investigate Janus-like solutions for which the space-time metric takes the form
\begin{equation}
\label{metric_ansatz}
ds_{4}^{2} = d\mu^{2} + e^{2 A(\mu)} \, d\Sigma^{2} \ ,
\end{equation}
with $\,\mu\in (-\infty, \infty)\,$ or $\,\mu\in [0, \infty)\,$ being the coordinate along which space-time is foliated with $\Sigma$ slices, and $A(\mu)$ being a scale function. The line element $\,d\Sigma^{2}\,$ describes a globally AdS$_{3}$ space-time of radius $\,\ell= 1\,$. The second-order Euler-Lagrange equations for the scalar fields that follow from the Lagrangian (\ref{Lagrangian_model_U1^4_Einstein-scalars}) read
\begin{equation}
\label{EOM_scalars}
\begin{array}{rll}
\varphi_{i} '' - e^{2 \varphi_{i}} \,(\chi_{i}')^2 + 3 \, A' \, \varphi_{i}' + g^2 \, \left( 2 \, \sinh\varphi_{i} +  e^{\varphi_{i}} \, \chi_{i}^2\right) & = & 0  \ , \\[2mm]
\chi_{i}'' + \left( 3 \, A' + 2  \, \varphi_{i}' \right) \chi_{i}'  + 2 \, g^2 \, e^{-\varphi_{i}} \, \chi_{i}  & = & 0 \ ,
\end{array}
\end{equation}
with $i=1,2,3$ and where primes denote derivatives with respect to the coordinate $\,\mu\,$. The two equations in (\ref{EOM_scalars}) can be expressed  as
\begin{equation}
\label{EOM_scalars_complex}
\tilde{z}_{i}''+3 \, A' \, \tilde{z}_{i}'+2\, \frac{\tilde{z}_i^{*} \, (\tilde{z}_i')^2}{1-|\tilde{z}_i|^2}+2 \, g^2 \, \tilde{z}_i = 0 \ ,
\end{equation}
in terms of the complex scalars $\,\tilde{z}_{i}\,$ in (\ref{ztilde&z}). The Einstein equations impose two additional independent equations given by
\begin{equation}
\label{EOM_Einstein}
\begin{array}{rll}
1 - e^{2 A} \Big[ A'' + \frac{1}{4} \displaystyle\sum_{i} \Big( \, (\varphi_{i}')^2 + e^{2 \varphi_{i}} \, (\chi_{i}')^2 \Big) \, \Big] &=& 0 \ , \\[2mm]
2 + e^{2 A} \Big[  A'' + 3 \, (A')^2 - \tfrac{1}{2} \, g^{2} \displaystyle\sum_{i}  \left( 2 \, \cosh\varphi_{i} + \chi_{i}^2  \,  e^{\varphi_{i}} \right)   \Big] & = & 0 \ .
\end{array}
\end{equation}
Equivalently,
\begin{equation}
\label{EOM_Einstein_complex}
\begin{array}{rll}
2A''+e^{-2A}+V+3 \, (A')^2 + \displaystyle\sum_{i}   \dfrac{\tilde{z}_i' \, (\tilde{z}_i^{*})'}{\left( 1-|\tilde{z}_i|^2 \right)^2} &=&0 \ , \\[2mm]
3 \, e^{-2A} + V + 3 \, (A')^2 - \displaystyle\sum_{i}   \dfrac{\tilde{z}_i' \, (\tilde{z}_i^{*})'}{\left(  1-|\tilde{z}_i|^2 \right)^2}  &=& 0 \ . 
\end{array}
\end{equation}
We will now present analytic and multi-parametric families of Janus and Hades solutions to this system of second-order differential equations.

\subsection{Multi-parametric Janus solutions}

\begin{figure}[t]
\begin{center}
\includegraphics[width=0.50\textwidth]{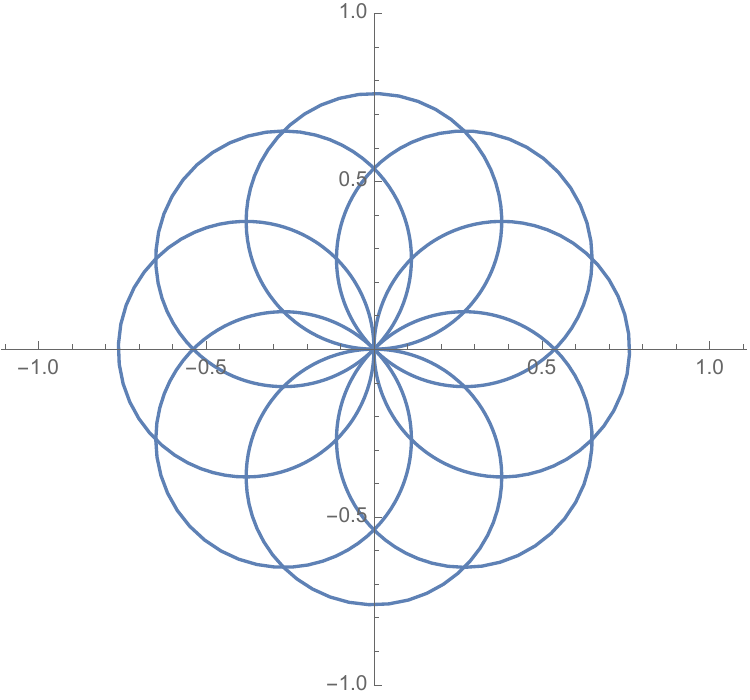} 
\put(0,90){$\textrm{Re}\tilde{z}_{i}$} \put(-105,210){$\textrm{Im}\tilde{z}_{i}$}
\put(-114,101){{\color{red}{$\bullet$}}}
\end{center}
\caption{Parametric plot of $\,\tilde{z}_{i}(\mu)\,$ in (\ref{Janus_solution_U1^4_ztil}) for the Janus solutions with $\,\alpha_{i}=1\,$ and $\,\beta_{i}=\frac{n\pi}{4}\,$ with $\,n=0,\ldots,7\,$. The central red point at $\,\tilde{z}_{i}=0\,$ $\,\forall i\,$ corresponds to the maximally supersymmetric AdS$_{4}$ vacuum and describes the asymptotic values at $\,\mu\rightarrow \pm\infty\,$.}
\label{fig:ztilde_U1^4}
\end{figure}

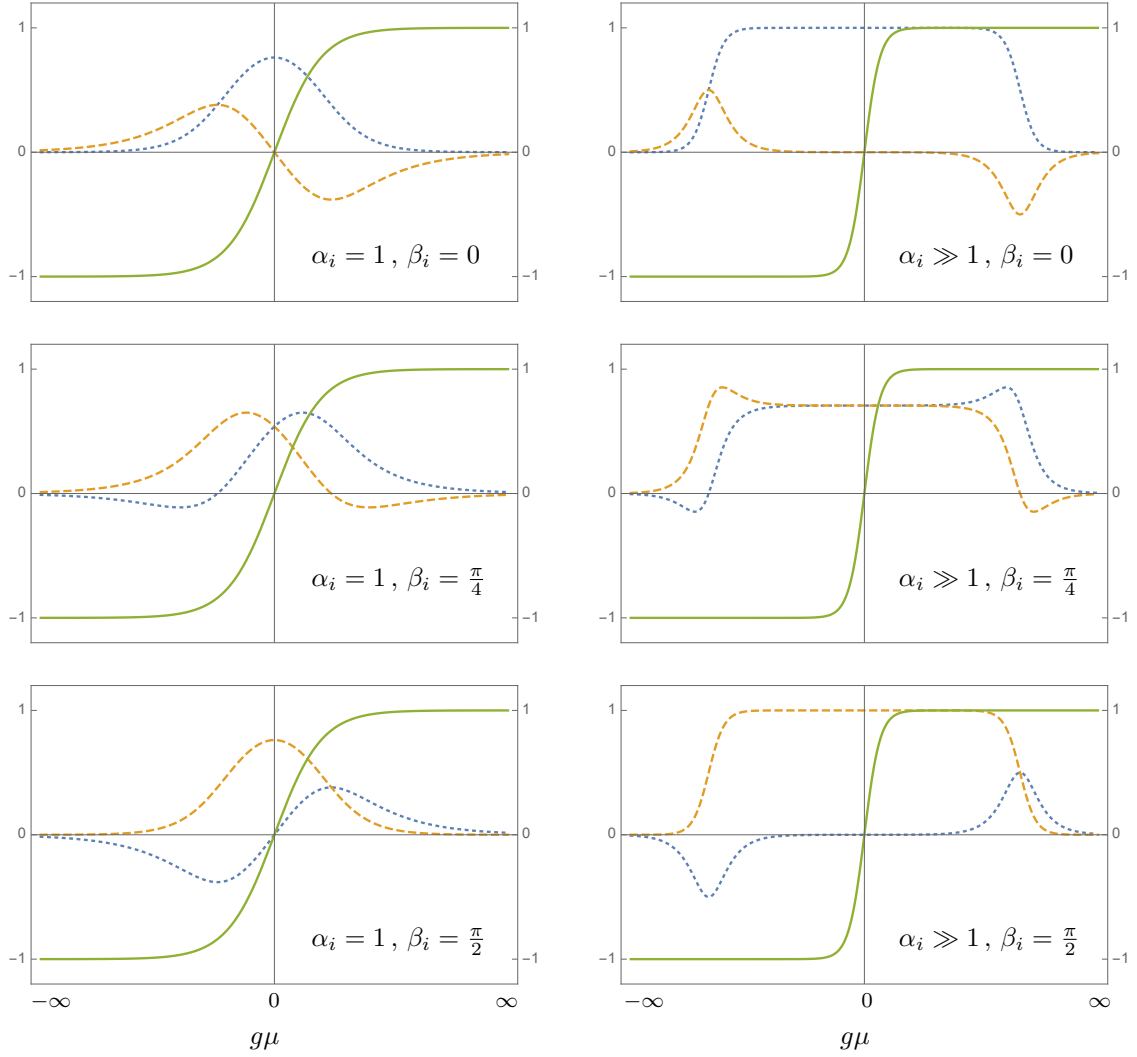
\begin{figure}[t!]
\begin{center}
\includegraphics[width=0.45\textwidth]{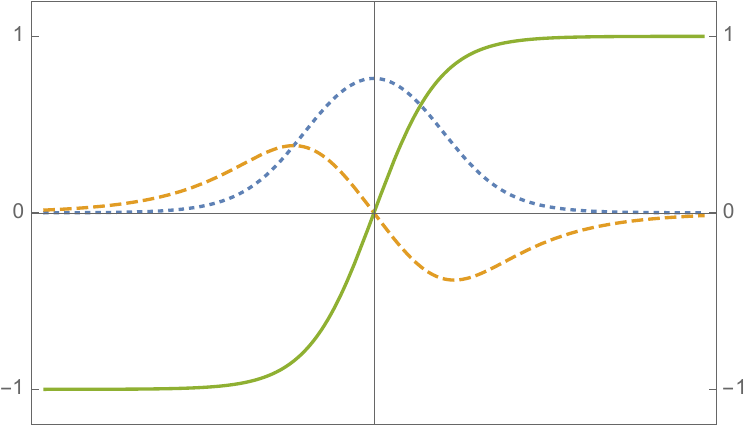} 
\hspace{5mm}
\includegraphics[width=0.45\textwidth]{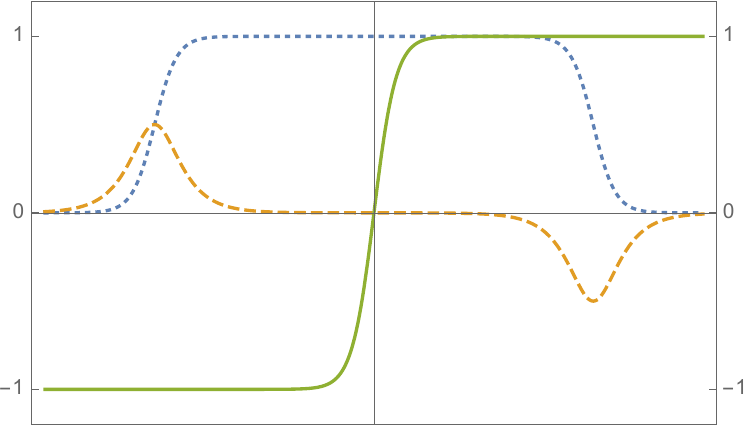}
\put(-307,15){\small{$\alpha_{i}=1 \, , \, \beta_{i}=0$}}\put(-87,15){\small{$\alpha_{i} \gg 1 \, , \, \beta_{i}=0$}}
\\[5mm]
\includegraphics[width=0.45\textwidth]{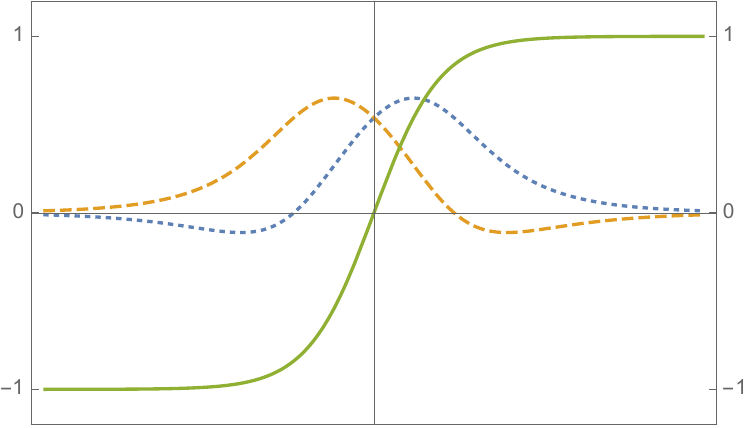} 
\hspace{5mm}
\includegraphics[width=0.45\textwidth]{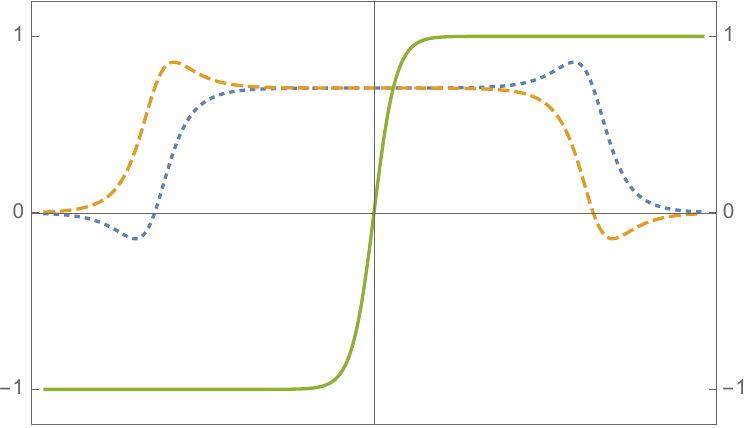} 
\put(-307,22){\small{$\alpha_{i}=1 \, , \, \beta_{i}=\frac{\pi}{4}$}}\put(-87,22){\small{$\alpha_{i} \gg 1 \, , \, \beta_{i}=\frac{\pi}{4}$}}
\\[5mm]
\includegraphics[width=0.45\textwidth]{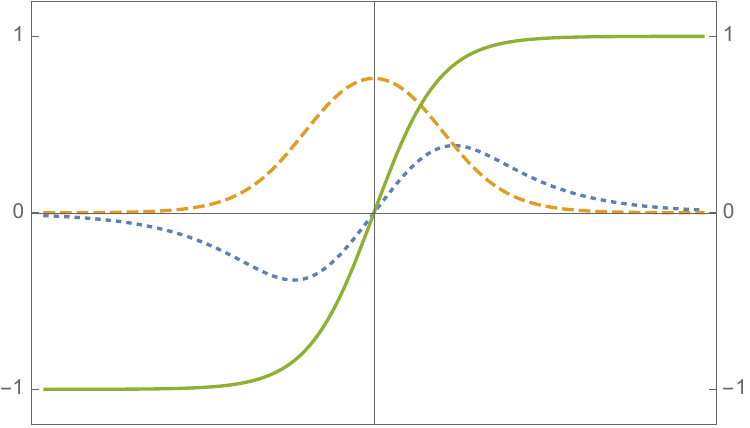} 
\hspace{5mm}
\includegraphics[width=0.45\textwidth]{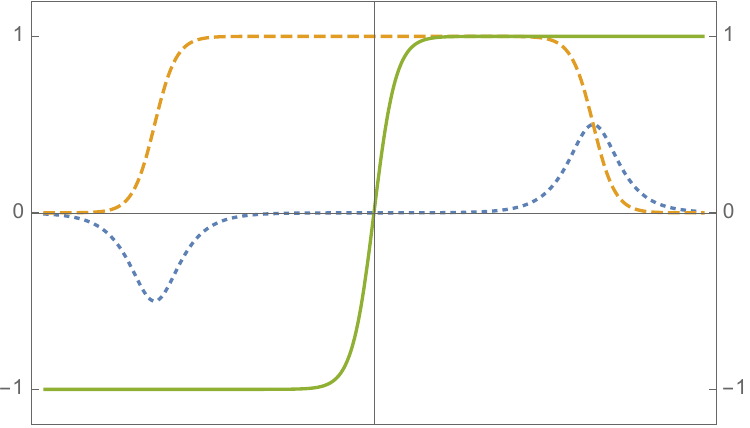} 
\put(-307,15){\small{$\alpha_{i}=1 \, , \, \beta_{i}=\frac{\pi}{2}$}}\put(-87,15){\small{$\alpha_{i} \gg 1 \, , \, \beta_{i}=\frac{\pi}{2}$}}
\put(-412,-8){\scriptsize{$-\infty$}}\put(-323,-8){\scriptsize{$0$}}\put(-238,-8){\scriptsize{$\infty$}}
\put(-330,-22){\small{$g \mu$}}
\put(-190,-8){\scriptsize{$-\infty$}}\put(-101,-8){\scriptsize{$0$}}\put(-16,-8){\scriptsize{$\infty$}}
\put(-109,-22){\small{$g \mu$}}
\end{center}
\caption{Plots of $\,\textrm{Re}\tilde{z}_{i}\,$ (blue dotted line), $\,\textrm{Im}\tilde{z}_{i}\,$ (orange dashed line) and $\,A'(\mu)\,$ (green solid line) as a function of the radial coordinate $\,g \mu \in (-\infty , \infty)\,$ for different values of the Janus parameters $\,(\alpha_{i},\beta_{i})\,$. The limit $\,\alpha_{i} \gg 1\,$ (\textit{i.e.} $\tanh\alpha_{i} \approx 1$) renders the Janus solution singular. In this limit, $\,\tilde{z}_{i}\,$ gets to the boundary of the moduli space which is located at $\,|\tilde{z}_{i}|=1\,$ in the unit-disk parameterisation of (\ref{ztilde&z}).}
\label{fig:Rez&Imz}
\end{figure}

The second-order equations of motion in (\ref{EOM_scalars}) and (\ref{EOM_Einstein}) have a multi-parametric family of analytic Janus solutions. The scale factor in the space-time metric is given by
\begin{equation}
\label{A(mu)_func_U1^4}
e^{2A(\mu)} = {(g k)}^{-2} \cosh^2(g\mu) \ ,
\end{equation}
and
\begin{equation}
\label{k_factor}
k^2= 1 + \sum_{i}\sinh^{2}\alpha_{i}  \, \ge \, 1
\hspace{10mm} \text{ with } \hspace{10mm} \alpha_{i} \in \mathbb{R}^{+} \ .
\end{equation}
Using the unit-disk parameterisation in (\ref{ztilde&z}) to describe the scalar fields in the three vector multiplets, they acquire simple $\mu$-dependent profiles of the form
\begin{equation}
\label{Janus_solution_U1^4_ztil}
\tilde{z}_{i}(\mu) = e^{i \beta_{i}}\, \frac{\sinh\alpha_{i}}{\cosh\alpha_{i} + i \, \sinh(g \mu) }  
\hspace{10mm} \text{ with } \hspace{10mm} \beta_{i} \in [0,2\pi] \ ,
\end{equation}
so that $\,|\tilde{z}_{i}(0)|=\tanh\alpha_{i}\,$. Eqs (\ref{A(mu)_func_U1^4})-(\ref{Janus_solution_U1^4_ztil}) describe a multi-parametric family of Janus solutions parameterised by $3+3$ arbitrary real constants $(\alpha_{i},\beta_{i})$. Importantly, the presence of non-trivial axions $\,\textrm{Im}\tilde{z}_{i}\,$ (spin $0$ pseudo-scalars) turns out to be crucial for the existence of regular Janus solutions, as first noticed in \cite{Bobev:2013yra}. Parametric plots of the complex scalars $\,\tilde{z}_{i}(\mu)\,$ in (\ref{Janus_solution_U1^4_ztil}) are displayed in Figure~\ref{fig:ztilde_U1^4}. The real $\,\textrm{Re}\tilde{z}_{i}\,$ and imaginary $\,\textrm{Im}\tilde{z}_{i}\,$ components of $\,\tilde{z}_{i}\,$ are shown in Figure~\ref{fig:Rez&Imz}. Note the special limiting case of $\,\alpha_{i} \gg 1\,$ (\textit{i.e.} $\tanh\alpha_{i} \approx 1$) for which the flows become singular. In this limit, the complex scalar $\,\tilde{z}_{i}\,$ gets to the boundary of the moduli space, which is located at $\,|\tilde{z}_{i}|=1\,$ in the unit-disk parameterisation of the Lagrangian (\ref{Lagrangian_model_U1^4_Einstein-scalars}), and the scalar potential in (\ref{V_U1^4}) diverges.

On the other hand, the value $\,\alpha_{i}=0\,$ is certainly special. At this value an AdS$_{4}$ maximally supersymmetric solution with radius $\,L_{\text{AdS}_{4}}={g}^{-1}\,$ is recovered with the scalars being fixed at the constant value $\,\tilde{z}_{i}=0\,$. This AdS$_{4}$ vacuum uplifts to the $\,\textrm{AdS}_4 \times \textrm{S}^7\,$ Freund--Rubin background of eleven-dimensional supergravity with a round $\,\textrm{S}^{7}\,$ metric \cite{Freund:1980xh}. Moreover, it describes the near-horizon geometry of a stack of M2-branes and is holographically dual to the three-dimensional ABJM theory \cite{Aharony:2008ug}. When evaluated at this AdS$_{4}$ vacuum, the three $\textrm{U}(1)^{4}$ invariant complex scalars have a normalised mass
\begin{equation}
\label{m^2L^2_AdS4}
m_{i}^2 L^2 = -2 \ ,
\end{equation}
thus lying within the mass range $\, -9/4 < m_{i}^2 \, L^2 <  - 5/4\,$ for wich two possible quantisations of scalar fields in AdS$_{4}$ exist \cite{Klebanov:1999tb}: the mode with conformal dimension $\,\Delta_{i}=\Delta_{-}=1\,$ and the mode with conformal dimension $\,\Delta_{i}=\Delta_{+}=2\,$ (where $\,\Delta_{\pm}\,$ are the two roots of $\,m_{i}^2 \, L^2=(\Delta_{i}-3)\Delta_{i}\,$) can be interpreted as the source and the VEV of the corresponding dual operators (standard quantisation) or \textit{viceversa} (alternative quantisation). However, as shown in \cite{Breitenlohner:1982bm}, proper scalars $\,\textrm{Re}\tilde{z}_{i}\,$ and pseudo-scalars $\,\textrm{Im}\tilde{z}_{i}\,$ must be quantised in exactly opposite ways in order to preserve maximal supersymmetry. And, moreover, only the choice of proper scalars having alternative quantisation yields a perfect matching between the scaling dimensions of the supergravity modes and those of the dual operators in the  M2-brane  theory \cite{Bobev:2011rv} (see footnote~\ref{footnote:operators}).

The class of Janus solutions in (\ref{A(mu)_func_U1^4})-(\ref{Janus_solution_U1^4_ztil}) depends on the set of parameters $\,g\,$ and $\,(\alpha_{i},\beta_{i})\,$. As discussed in \cite{Bobev:2013yra}, the four-dimensional gauge coupling $\,g\,$ sets the scale of the asymptotic AdS$_{4}$ vacuum and, via the AdS/CFT correspondence, the number of M2-branes as well as the rank of the Chern--Simons gauge groups in ABJM theory. The parameters $\,\alpha_{i}\,$ set the height of the bump, \textit{i.e.}  $\,|\tilde{z}_{i}(0)|=\tanh\alpha_{i}\,$, and therefore the strength of the coupling between the (1+1)-dimensional defect and the three-dimensional ambient field theory. The parameters $\,\beta_{i}\,$ set the boundary conditions of the bulk scalars at $\,\mu \rightarrow \pm \infty\,$ and, again via the AdS/CFT correspondence (see footnote~\ref{footnote:operators}), the specific linear combinations of bosonic and fermionic bilinear operators that are activated in the field theory. We will analyse the possible choices of boundary conditions in detail in Section~\ref{sec:boundary conditions}.

Lastly, a study of the supersymmetry preserved by this family of solutions is presented in the Appendix~\ref{app:susy}. The BPS equations (\ref{BPS_A}) and (\ref{BPS_scalars}) are not satisfied by the Janus solution in (\ref{A(mu)_func_U1^4})-(\ref{Janus_solution_U1^4_ztil}) for generic values of $(\alpha_{i} , \beta_{i})$ thus implying that such a solution is generically non-supersymmetric. However, as we will see in a moment, some supersymmetry can be restored upon suitable choice of $(\alpha_{i} , \beta_{i})$, namely, upon suitable adjustment of the Janus boundary conditions.

\subsection{Janus with (super) symmetry enhancements}
\label{sec:Janus_sym_enhancement}

Specific choices of the parameters $\,(\alpha_{i},\beta_{i})\,$ translate into various (super) symmetry enhancements of the general Janus solution in (\ref{A(mu)_func_U1^4}) and (\ref{Janus_solution_U1^4_ztil}).

\subsubsection{\texorpdfstring{$\text{SO}(4) \times\text{SO}(4)$}{SO(4)xSO(4)} symmetry enhancement}
\label{sec:Janus_SO4xSO4}

Setting two vector multiplets to zero, \textit{e.g.} $\,\tilde{z}_{2}=\tilde{z}_{3}=0$, by setting
\begin{equation}
\label{alpha_2_3=0}
\alpha_{2} = \alpha_{3} = 0 \ ,
\end{equation}
and renaming $\tilde{z}_{1} \equiv\tilde{z}\,$, the $\text{SO}(4) \times\text{SO}(4)$ invariant sector of the SO(8) gauged supergravity investigated in Section~$5$ of \cite{Bobev:2013yra} is recovered upon the identification $\tilde{z}=z_{\text{there}}$. The Lagrangian (\ref{Lagrangian_model_U1^4_Einstein-scalars}) reduces to
\begin{equation}
\label{Lagrangian_model_SO4xSO4}
\begin{array}{lll}
\mathcal{L} & = & \left(  \frac{R}{2} - V \right)  * 1 - \frac{1}{4} \left[ (d\varphi)^{2} + e^{2 \varphi} \, (d\chi)^{2} \right] \\[2mm]
& = & \left(  \frac{R}{2} - V \right)  * 1 - \dfrac{1}{\left(  1-|\tilde{z}|^{2} \right)  ^{2}} \, d\tilde{z} \wedge* d\tilde{z}^{*} \ ,
\end{array}
\end{equation}
and the scalar potential in (\ref{V_U1^4}) simplifies to
\begin{equation}
\label{V_SO4xSO4}
V = - \tfrac{1}{2} \, g^{2} \left(  4 + 2 \cosh\varphi+ \chi^{2} \, e^{\varphi} \right)  = -  g^{2} \, \dfrac{3 - |\tilde{z}|^{2}}%
{1-|\tilde{z}|^{2}} \ .
\end{equation}
The Janus solution then reads
\begin{equation}
\label{Janus_solution_SO4xSO4^4_alt}
ds_{4}^{2} = d\mu^{2}+ e^{2 A(\mu)} \, d\Sigma^{2} 
\hspace{10mm} , \hspace{10mm} 
\tilde{z}(\mu) = e^{i \beta} \,
\frac{\sinh\alpha}{\cosh\alpha + i \, \sinh(g \mu) } \ ,
\end{equation}
with
\begin{equation}
\label{A(mu)_func_SO4xSO4}
e^{2A(\mu)} = (g k)^{-2} \cosh^2(g\mu)
\hspace{8mm} \textrm{ and } \hspace{8mm}
k= \cosh \alpha \ge 1\ ,
\end{equation}
where $(\alpha, \beta) = (\alpha_{1} , \beta_{1})$. This solution precisely matches the one presented in Section~$5$ of \cite{Bobev:2013yra} upon the identification $\cosh\alpha=(1-a^{2}_{\text{there}})^{-\frac{1}{2}}$. As noticed therein, the Janus solution is half-PBS and preserves $\,16\,$ real supercharges. From a holographic perspective, the $(1+1)$-dimensional defect dual to the AdS$_{3}$ factor in the geometry features $(4,4)$ supersymmetry and therefore has an $\,\textrm{SO}(4)_{\textrm{R}} \times \textrm{SO}(4)_\textrm{R}\,$ R-symmetry group. We have explicitly verified that the Janus solution satisfies the 1/2-BPS equations (\ref{BPS_A}) and (\ref{BPS_scalars}) for the eight gravitino mass terms (superpotentials) of the maximal theory (see Footnote~\ref{Footnote:axions}). Finally, the original M-theory supersymmetric Janus with $\text{SO}(4) \times\text{SO}(4)$ symmetry was presented in \cite{DHoker:2009lky}.

\subsubsection{\texorpdfstring{$\text{SU}(3) \times\text{U}(1)^2$}{SU(3)xU(1)2} symmetry enhancement}

Identifying the three vector multiplets, namely $\,\tilde{z}_{1}=\tilde{z}_{2}=\tilde{z}_{3} \equiv\tilde{z}\,$, so that
\begin{equation}
\alpha_{1} = \alpha_{2} = \alpha_{3} \equiv\alpha
\hspace{5mm} , \hspace{5mm}
\beta_{1} = \beta_{2} = \beta_{3} \equiv\beta\ ,
\end{equation}
the $\text{SU}(3) \times\text{U}(1)^2$ invariant sector of Section~$6$ of \cite{Bobev:2013yra} (see also \cite{Pilch:2015dwa} for the 11D uplift) is recovered upon the identification $\tilde{z}=z_{\text{there}}$. The
Lagrangian (\ref{Lagrangian_model_U1^4_Einstein-scalars}) simplifies to
\begin{equation}
\label{Lagrangian_model_SU3xU1xU1}
\begin{array}{lll}
\mathcal{L} & = & \left(  \frac{R}{2} - V \right)  * 1 - \frac{3}{4} \left[(d\varphi)^{2} + e^{2 \varphi} \, (d\chi)^{2} \right] \\[2mm]
& = & \left(  \frac{R}{2} - V \right)  * 1 - \dfrac{3}{\left(  1-|\tilde{z}|^{2} \right)  ^{2}} \, d\tilde{z} \wedge* d\tilde{z}^{*} \ ,
\end{array}
\end{equation}
and the scalar potential in (\ref{V_U1^4}) reduces to
\begin{equation}
\label{V_SU3xU1^2}
V = - \tfrac{3}{2}  \, g^{2} \left(  2 \cosh\varphi+ \chi^{2} \, e^{\varphi} \right)  = - 3 \, g^{2} \, \dfrac{1+|\tilde{z}|^{2}}{1-|\tilde{z}|^{2}} \ .
\end{equation}
The Janus solution takes the form
\begin{equation}
\label{Janus_solution_SU3xU1xU1}
ds_{4}^{2} = d\mu^{2}+ e^{2 A(\mu)} \, d\Sigma^{2} 
\hspace{10mm} , \hspace{10mm} 
\tilde{z}(\mu) = e^{i \beta} \, \frac{\sinh\alpha}{\cosh\alpha + i \, \sinh(g \mu) } \ ,
\end{equation}
with
\begin{equation}
\label{A(mu)_func_SU3xU1xU1}
e^{2A(\mu)} = (g k)^{-2} \cosh^2(g\mu) 
\hspace{8mm} \textrm{ and } \hspace{8mm}
k^2= 1 + 3 \sinh^{2}\alpha \ge 1\ .
\end{equation}
This provides an analytic solution in the $\text{SU}(3) \times\text{U}(1)^2 $ invariant sector of the SO(8) maximal supergravity investigated in Section~$6$ of \cite{Bobev:2013yra}. The solution (\ref{Janus_solution_SU3xU1xU1})-(\ref{A(mu)_func_SU3xU1xU1}) satisfies the second-order equations of motion in (\ref{EOM_scalars}) and (\ref{EOM_Einstein}). However, we have verified that the BPS equations (\ref{BPS_A}) and (\ref{BPS_scalars}) are not satisfied for any of the eight gravitino mass terms (superpotentials) in the maximal SO(8) gauged supergravity, so the solution is non-supersymmetric.

\subsection{Janus geometry and boundary conditions}

Let us discuss the geometry of the multi-parametric family of Janus solutions presented in the previous sections. Introducing embedding coordiantes in $\mathbb{R}^{2,3}$, the $k$-family of Janus metrics in (\ref{metric_ansatz}) and (\ref{A(mu)_func_U1^4}) corresponds to
\begin{equation}
\label{embedding_coordinates}
\begin{array}{lll}
X_{0} &=& (g k)^{-1} \, \dfrac{\cos\tau}{\cos\eta} \, \cosh(g \mu) \ , \\[6mm]
X_{4} &=& (g k)^{-1} \, \dfrac{\sin\tau}{\cos\eta} \, \cosh(g \mu)   \ , \\[6mm]
X_{1} &=&  (g k)^{-1} \, \tan\eta \cos\theta \, \cosh(g \mu)  \ , \\[6mm]
X_{2} &=&  (g k)^{-1} \, \tan\eta \sin\theta \, \cosh(g \mu)   \ , \\[6mm]
X_{3} &=&  g^{-1} \,   i \, \textrm{E}(i g \mu \, ; \, k^{-2}) \ ,
\end{array}
\end{equation}
with 
\begin{equation}
k^2 = 1 + \sum_{i}\sinh^{2}\alpha_{i} \, \ge \, 1 \ ,
\end{equation}
and $\textrm{E}(i g \mu \, ; \, k^{-2})$ being the incomplete elliptic integral of the second kind. The solution describes the hyper-surface
\begin{equation}
\label{hypersurface_X}
-X_{0}^2 - X_{4}^2 + X_{1}^2 + X_{2}^2 +(g k)^{-2} \sinh^2(g \mu) = -  (g k)^{-2} \ ,
\end{equation}
where the term $\,(g k)^{-2} \sinh^2(g \mu)\,$ is implicitly given in terms of $X_{3}$ by the last relation in (\ref{embedding_coordinates}). For $\,k=1\,$ one has that $\,i \, \textrm{E}(i g \mu \, ; \,1) = - \sinh(g \mu)\,$ and (\ref{hypersurface_X}) reduces to the hyperboloid describing AdS$_{4}$.

\subsubsection{Global coordinates and boundary structure}

Let us perform a change of coordinates that will help us to understand the Janus geometry in (\ref{metric_ansatz}) and (\ref{A(mu)_func_U1^4})-(\ref{k_factor}), especially its boundary structure. We start by performing a change of radial coordinate to make its range compact
\begin{equation}
\tilde{\mu} = 2 \, k \, \textrm{arctan} \left[ \tanh \left( \frac{g \, \mu}{2}  \right)\right] \ ,
\end{equation}
and then choose global coordinates to describe the AdS$_{3}$ slicing in (\ref{metric_ansatz}). The Janus metric in (\ref{metric_ansatz}) and (\ref{A(mu)_func_U1^4})-(\ref{k_factor}) then becomes (locally) conformal to $\,\mathbb{R} \times \textrm{S}^{3}\,$
\begin{equation}
\label{Janus_metric_original}
ds_{4}^{2} = \frac{(g k)^{-2} }{\cos^2\left( \frac{\tilde{\mu}}{k}\right) \cos^{2}\eta} \,  \left( - d\tau^2  + \cos^{2}\eta \, d\tilde{\mu}^{2} + d\eta^2 + \sin^2\eta \, d\theta^2\right) \ ,
\end{equation}
with
\begin{equation}
\label{global_coords_ranges}
\tau \in (-\infty \, , \infty)
\hspace{5mm} \textrm{ , } \hspace{5mm}
\tilde{\mu} \in [-\frac{\pi k}{2} \, , \frac{\pi k}{2}]
\hspace{5mm} \textrm{ , } \hspace{5mm}
\eta \in [0 \, , \frac{\pi}{2}] 
\hspace{5mm} \textrm{ , } \hspace{5mm}
\theta \in [0 \, , 2 \pi]  \ .
\end{equation}
These are the global coordinates used to describe the original type IIB Janus solution in \cite{Bak:2003jk,Clark:2004sb}. The geometry (\ref{Janus_metric_original}) has a boundary that consists of two hemi-spheres of $\,\textrm{S}^2\,$ at $\,\tilde{\mu} = \pm \tilde{\mu}_{0}\,$, with $\, \tilde{\mu}_{0}=\frac{\pi k}{2} \,$, joined at the $\textrm{S}^1$ equator at $\,\eta = \frac{\pi}{2}\,$. Lastly, using the new radial coordinate $\,\tilde{\mu}\,$, the profiles for the complex scalars in (\ref{Janus_solution_U1^4_ztil}) become
\begin{equation}
\label{Janus_solution_U1^4_new}
\tilde{z}_{i}(\tilde{\mu}) = e^{i \beta_{i}}\, \frac{\sinh\alpha_{i}}{\cosh\alpha_{i} + i \, \tan\left(\frac{\tilde{\mu}}{k}\right) }  \ ,
\end{equation}
so that $\,\tilde{z}_{i}(\tilde{\mu}) \rightarrow 0\,$ when approaching the two hemi-spheres of $\,\textrm{S}^2\,$ at $\,\tilde{\mu} \rightarrow \pm \tilde{\mu}_{0} \,$ in the Janus boundary. Note that $\,\textrm{arg}\left[\tilde{z}_{i}(\tilde{\mu}_{0})\right] - \textrm{arg}\left[\tilde{z}_{i}(-\tilde{\mu}_{0})\right] = \pi$, thus creating an interface discontinuity at the $\,\textrm{S}^{1}\,$ equator where the defect lives.

\subsubsection{\texorpdfstring{AdS$_{3}$}{AdS3} slicing and boundary conditions}
\label{sec:boundary conditions}

In order to investigate the boundary conditions of the family of Janus solution in (\ref{A(mu)_func_U1^4})-(\ref{Janus_solution_U1^4_ztil}) we will perform a regular change of radial coordinate
\begin{equation}
\label{new_coordinate_Janus}
\rho= \sinh({g} \mu) 
\hspace{10mm} , \hspace{10mm}
d\mu = g^{-1} \, \dfrac{d\rho}{\sqrt{\rho^2+1}}  \ ,
\end{equation}
so that the family of Janus solutions in (\ref{A(mu)_func_U1^4})-(\ref{Janus_solution_U1^4_ztil}) becomes\footnote{The Ricci scalar constructed from the metric (\ref{Janus_U1^4_rho_1}) reads
\begin{equation}
\label{Janus_Ricci}
R(\rho) = - 6 \, g^{2} \left( \,  1 +  \frac{\rho^2 + k^2}{\rho^2 + 1} \, \right) \ ,
\end{equation}
%
%
thus ensuring regularity of the Janus geometry within the whole range $\,\rho \in ( -\infty , \infty)\,$.}
\begin{equation}
\label{Janus_U1^4_rho_1}
ds_{4}^{2}=\frac{1}{{g}^{2}} \left(  \frac{d\rho^{2}}{\rho^{2}+1} + 
\frac{\rho^{2}+1  }{ k^2} \, d\Sigma^{2}  \right)
\hspace{5mm} , \hspace{5mm}
d\Sigma^{2} = \frac{1}{\cos^{2}\eta}\left( - d\tau^2 + d\eta^2 + \sin^2\eta \, d\theta^2\right) \ ,
\end{equation}
with
\begin{equation}
\label{global_coords_ranges_2}
\tau \in (-\infty \, , \infty)
\hspace{5mm} \textrm{ , } \hspace{5mm}
\rho  \in (-\infty \, , \infty)
\hspace{5mm} \textrm{ , } \hspace{5mm}
\eta \in [0 \, , \frac{\pi}{2}] 
\hspace{5mm} \textrm{ , } \hspace{5mm}
\theta \in [0 \, , 2 \pi]  \ ,
\end{equation}
and
\begin{equation}
\label{Janus_U1^4_rho_2}
\tilde{z}_{i}(\rho) = e^{i \beta_{i}} \, \frac{\sinh\alpha_{i}}{\cosh\alpha_{i} + i \, \rho} \ .
\end{equation}
The Janus geometry (\ref{Janus_U1^4_rho_1}) has a three-dimensional conformal boundary at $\,\rho \rightarrow \pm \infty\,$ that is conformal to $\,\mathbb{R} \times \textrm{S}^{2}\,$ with a $k$-dependent prefactor $\,(g k)^{-2} \, \rho^2\,$. This is the geometry we will use to analyse the asymptotic behaviour of the $\,\textrm{U}(1)^4\,$ invariant complex scalars (\ref{Janus_U1^4_rho_2}).

When approaching the maximally supersymmetric AdS$_4$ vacuum dual to ABJM theory\footnote{\label{footnote:operators}The 35 pseudo-scalars and 35 proper scalars of the maximal supergravity multiplet are dual to single-trace deformations of ABJM theory \cite{Aharony:2008ug}. More concretely, pseudo-scalars are dual to fermionic bilinears $\,\mathcal{O}_{F}=\textrm{Tr}(\psi^{\dot{A}} \psi^{\dot{B}}) - \frac{1}{8} \delta^{\dot{A}\dot{B}} \textrm{Tr}(\psi^{\dot{C}} \psi^{\dot{C}} )\,$ with $\,\dot{A}=1,\ldots,8\,$ and $\,\textrm{dim}(\mathcal{O}_{F})=2\,$. Proper scalars are dual to bosonic bilinears $\,\mathcal{O}_{B}=\textrm{Tr}(X^{A} X^{B}) - \frac{1}{8} \delta^{AB} \textrm{Tr}(X^{C}X^{C})\,$ with $\,A=1,\ldots,8\,$ and $\,\textrm{dim}(\mathcal{O}_{B})=1\,$.}, the asymptotic behaviour of (\ref{Janus_U1^4_rho_2}) around the endpoints $\,\rho \rightarrow \pm \infty\,$ of the Janus solution reads
\begin{equation}
\label{source&vevs_zt}
\tilde{z}_{i}(\rho) = \dfrac{\tilde{z}_{i,0}}{\rho} +   \dfrac{\tilde{z}_{i,1}}{\rho^2}  + \mathcal{O}\left( \dfrac{1}{\rho^3}\right) 
\hspace{10mm} \textrm{ with } \hspace{10mm} 
i=1,2,3 \ , 
\end{equation}
in terms of normalisable modes $\,\tilde{z}_{i,0}\,$ with $\,\Delta_{i}=1\,$ specified by the parameters $\,(\alpha_{i} , \beta_{i})\,$,
\begin{equation}
\label{zt_0}
\tilde{z}_{i,0} = \sinh\alpha_{i} \, e^{i (\beta_i - \frac{\pi}{2})}  \ ,
\end{equation}
as well as normalisable modes $\,\tilde{z}_{i,1}\,$ with  $\,\Delta_{i}=2\,$. These modes satisfy a set of $\alpha_{i}$-dependent algebraic relations
\begin{equation}
\label{zt_1}
\tilde{z}_{i,1} - i \cosh\alpha_{i}  \, \tilde{z}_{i,0} = 0  \ .
\end{equation}
The on-shell relations (\ref{zt_1}) will help us to characterise the deformations in the field theory dual of the Janus solution upon appropriate manipulation of boundary terms and finite counterterms.

In order to discuss the boundary conditions (\ref{source&vevs_zt})--(\ref{zt_1}) in more detail, we will resort to an expansion of $\,\text{Re}\tilde{z}_{i}\,$ (proper scalars) and $\,\text{Im}\tilde{z}_{i}\,$ (pseudo-scalars) around $\rho\rightarrow\pm\infty$. This yields
\begin{equation}
\label{source&vevs_rho_f}
\begin{array}{llll}
\text{Re}\tilde{z}_{i}(\rho) & = & \dfrac{a^{(v)}_{i,0}}{\rho} +   \dfrac{a^{(s)}_{i,1}}{\rho^2}  + \mathcal{O}\left( \dfrac{1}{\rho^3}\right) & , \\[4mm]
\text{Im}\tilde{z}_{i}(\rho) & = & \dfrac{b^{(s)}_{i,0}}{\rho} + \dfrac{b^{(v)}_{i,1}}{\rho^2}  + \mathcal{O}\left( \dfrac{1}{\rho^3}\right) & ,
\end{array}
\end{equation}
so that
\begin{equation}
\label{zToab}
\tilde{z}_{i,0} = a^{(v)}_{i,0} + i \, b^{(s)}_{i,0}
\hspace{5mm} , \hspace{5mm}
\tilde{z}_{i,1} = a^{(s)}_{i,1} + i \, b^{(v)}_{i,1} \ ,
\end{equation}
with
\begin{equation}
\label{ab_definition}
a^{(v)}_{i,0}=\sinh\alpha_{i} \, \sin\beta_i
\hspace{5mm} , \hspace{5mm}
b^{(s)}_{i,0} = - \sinh\alpha_{i} \, \cos\beta_i
\ .
\end{equation}
The algebraic relations in (\ref{zt_1}) then become
\begin{equation}
\label{ab_algebraic}
a^{(s)}_{i,1}  + \cosh\alpha_{i} \,\,  b^{(s)}_{i,0} = 0
\hspace{5mm} , \hspace{5mm}
b^{(v	)}_{i,1} - \cosh\alpha_{i} \,\, a^{(v)}_{i,0} = 0 \ .
\end{equation}
Note that the independent parameters specifying the boundary conditions in (\ref{ab_definition}) are $\,(\alpha_{i} , \beta_{i})\,$. As a consequence, the coefficients in the expansions (\ref{source&vevs_rho_f}) obey the following two sets of algebraic relations 
\begin{equation}
\dfrac{\left(a^{(s)}_{i,1}\right)^2}{\left(b^{(s)}_{i,0}\right)^2} = 1+ |\tilde{z}_{i,0} |^2
\hspace{8mm} , \hspace{8mm}
\dfrac{\left(b^{(v)}_{i,1}\right)^2}{\left(a^{(v)}_{i,0}\right)^2} = 1+  |\tilde{z}_{i,0} |^2 \  .
\end{equation}
Lastly, following \cite{Bobev:2011rv} (see also \cite{Bobev:2013yra}), we have attached the labels ``source" $\,^{(s)}\,$ and ``VEV" $\,^{(v)}\,$ to the modes in (\ref{source&vevs_rho_f}) to highlight that, in order to preserve maximal supersymmetry, proper scalars should feature the alternative quantisation and pseudo-scalars the standard quantisation. Note that setting $\,\beta_{i}=\pm\frac{\pi}{2}\,$ switches off the sources in (\ref{source&vevs_rho_f}) leaving only the VEV's. This is in agreement with the standard AdS/CFT prescription and renders $\,\tilde{z}_{i,0}\,$ in (\ref{zt_0}) real.

\subsubsection{Janus solutions and boundary conditions}

Let us compute the on-shell variation of the Lagrangian (\ref{Lagrangian_model_U1^4_Einstein-scalars}). A standard computation yields the boundary term
\begin{equation}
\label{deltaS}
\delta S = \displaystyle\sum_{i} \delta S_{i}  =  \displaystyle\sum_{i} \int d^{4}x \,\, \partial_{\mu} \theta^{\mu}_{i} = - \displaystyle\sum_{i}  \int_{\partial M}d^{3}x  \frac{\sqrt{-h}}{\left( 1-\left\vert
\tilde{z}_{i}\right\vert ^{2}\right)^{2}} \, N^{\mu} \, ( \partial_{\mu } \tilde{z}_{i} \, \delta\tilde{z}_{i}^{\ast } + \textrm{c.c.} ) \ ,
\end{equation}
where 
\begin{equation}
\theta^{\mu}_{i}  \equiv  - \frac{\sqrt{-g}}{\left( 1-\left\vert
\tilde{z}_{i}\right\vert^{2}\right)^{2}} \, g^{\mu \nu }  \left( \partial_{\nu} \tilde{z}_{i} \,\,
\delta\tilde{z}_{i}^{\ast} + \textrm{c.c.}  \right) \ ,    
\end{equation}
and c.c stands for complex conjugation. In (\ref{deltaS}) we have introduced the standard foliation $\,g_{\mu \nu }=h_{\mu \nu }+N_{\mu} N_{\nu }\,$ with $\,N_{\mu }=\sqrt{g_{\rho \rho }} \, \delta _{\mu }^{\rho }\,$ being the vector normal to the AdS$_{3}$ leaves. 

Plugging into (\ref{deltaS}) the asymptotic expansion of the scalars in (\ref{source&vevs_zt}) around $\,\rho \rightarrow \pm \infty\,$, and using the asymptotic form of the metric (\ref{Janus_U1^4_rho_1}), we encounter the well known linearly divergent term. In order to regularise the above boundary action and have a well-defined variational principle we introduce, for each complex field $\,\tilde{z}_{i}\,$, the counter-term
\begin{equation}
\label{counterterm}
S_{\textrm{ct},i} = - \, g  \lim_{\rho \rightarrow \pm\infty } \, \int_{\partial M}   
d^{3}x  \sqrt{-h}  \,\,
\tilde{z}_{i} \, \tilde{z}_{i}^{\ast }  \ ,
\end{equation}
so that
\begin{equation}
\label{boundary_contributions}
\delta S_{i} + \delta S_{\textrm{ct},i}  = g^{-2} \, k^{-3} \int_{\partial M}     \left(   \tilde{z}_{i,1} \, \delta \tilde{z}_{i,0}^{\ast }  +  \tilde{z}_{i,1}^{\ast} \, \delta \tilde{z}_{i,0}  \right) \, d\Sigma \ ,
\end{equation}
in terms of the volume element at the boundary $\,d\Sigma=\sqrt{-\gamma} \, d^{3}x \,$ with $\, \sqrt{-\gamma} =\sin\eta \, \cos^{-3}\eta\,$. Substituting the scalar mode parameterisation of (\ref{zToab}) into the boundary contributions in (\ref{boundary_contributions}) one obtains
\begin{equation}
\label{boundary_contributions_ab}
\delta S_i + \delta S_{\textrm{ct},i} =  2 \, g^{-2} \, k^{-3}  \int_{\partial M} \left(   a^{(s)}_{i,1} \, \delta a^{(v)}_{i,0}  +  b^{(v)}_{i,1} \, \delta b^{(s)}_{i,0}  \right) \, d\Sigma \ ,
\end{equation}
with
\begin{equation}
\label{k_factor_alt}
k^2= 1 + \sum_{i}\sinh^{2}\alpha_{i}  = 1 + \sum_{i} |\tilde{z}_{i,0} |^2 \, \ge \, 1 \ .
\end{equation}
In order to remove the $\,k^{-3}\,$ factor in (\ref{boundary_contributions_ab}) we could rescale the radial coordinate as $\,\hat{\rho} = k \, \rho\,$ or, instead, perform the non-linear mode redefinitions
\begin{equation}
\label{rescaled_ab}
\hat{a}^{(v)}_{i,0}  = k^{-1} \, a^{(v)}_{i,0} 
\hspace{5mm} , \hspace{5mm}
\hat{a}^{(s)}_{i,1}  = k^{-2} \, a^{(s)}_{i,1}
\hspace{5mm} , \hspace{5mm}
\hat{b}^{(s)}_{i,0}  = k^{-1} \, b^{(s)}_{i,0} 
\hspace{5mm} , \hspace{5mm}
\hat{b}^{(v)}_{i,1}  = k^{-2} \, b^{(v)}_{i,1}  \ .
\end{equation}
Following the latter prescription, the boundary contribution in (\ref{boundary_contributions_ab}) becomes
\begin{equation}
\label{boundary_contributions_ab_hat}
\delta S_{i} + \delta S_{\textrm{ct},i} \, =  \, 2 \, g^{-2}  \int_{\partial M} \left(   \hat{a}^{(s)}_{i,1} \, \delta \hat{a}^{(v)}_{i,0}  +  \hat{b}^{(v)}_{i,1} \, \delta \hat{b}^{(s)}_{i,0}  \right) \, d\Sigma \ ,
\end{equation}
and, due to the alternative quantisation featured by the proper scalars, we must add an extra boundary term such that
\begin{equation}
\label{boundary_contributions_ab_hat_final}
\delta S_{i} + \delta S_{\textrm{ct},i} - \delta \left( 2 \, g^{-2}  \int_{\partial M}  \hat{a}^{(s)}_{i,1} \, \hat{a}^{(v)}_{i,0}  \right)  \, =  \, 2 \, g^{-2}  \int_{\partial M} \left( \hat{b}^{(v)}_{i,1} \, \delta \hat{b}^{(s)}_{i,0}  -   \hat{a}^{(v)}_{i,0} \, \delta \hat{a}^{(s)}_{i,1} \right) \, d\Sigma \ .
\end{equation}
Therefore, having a well-defined variational principle therefore requires $\,\delta \hat{b}^{(s)}_{i,0} = \delta \hat{a}^{(s)}_{i,1} = 0\,$. Recalling from (\ref{ab_definition})-(\ref{ab_algebraic}) that 
\begin{equation}
b^{(s)}_{i,0} = - \sinh\alpha_{i} \, \cos\beta_i 
\hspace{10mm} \textrm{ and } \hspace{10mm}
a^{(s)}_{i,1} = - \cosh\alpha_{i} \,\,  b^{(s)}_{i,0} \ ,
\end{equation}
we conclude that sources are generically present at the boundary theory of the Janus ($\alpha_{i} \neq 0$) except for the particular choice of boundary conditions $\,\beta_{i}= \pm \frac{\pi}{2}\,$. This implies that every choice of $\,(\alpha_{i}, \beta_{i})\,$ with $\,\beta_{i} \neq \pm \frac{\pi}{2}\,$ corresponds to a different theory with a different value of the sources in the variational principle. On the contrary, when $\,\beta_{i}= \pm \frac{\pi}{2}\,$, the sources are zero on-shell and the boundary theory is unique.

\subsection{Multi-parametric Hades solutions}

\begin{figure}[t]
\begin{center}
\includegraphics[width=0.50\textwidth]{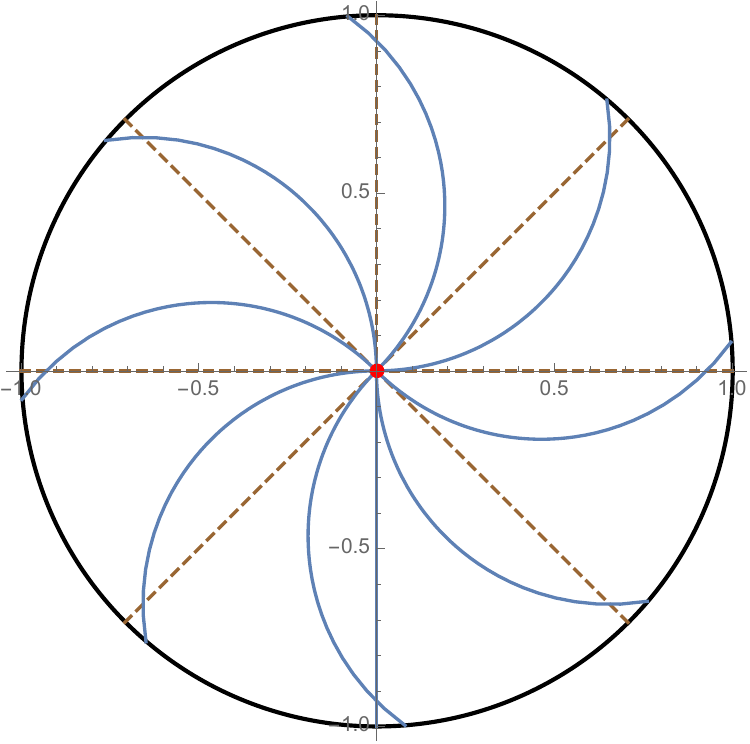} 
\put(10,95){$\textrm{Re}\tilde{z}_{i}$} \put(-105,228){$\textrm{Im}\tilde{z}_{i}$}
\put(-112.5,107.5){{\color{red}{$\bullet$}}}
\end{center}
\caption{Parametric plot of $\,\tilde{z}_{i}(\rho)\,$ in (\ref{Hades_U1^4_rho_2}) for the Hades solutions with $\,\alpha_{i}=1\,$ (blue-solid lines) and the ridge flows with $\,\alpha_{i}=0\,$ (brown-dashed lines) upon setting $\,\beta_{i}=\frac{n\pi}{4}\,$ with $\,n=0,\ldots,7\,$. The central red point at $\,\tilde{z}_{i}=0\,$ $\,\forall i\,$ corresponds to the maximally supersymmetric AdS$_{4}$ vacuum and describes the asymptotic values at $\,\rho\rightarrow \infty\,$. The boundary circle at $\,|\tilde{z}_{i}|=1\,$ corresponds to the singularity at $\,\rho = 1\,$.}
\label{fig:Hades_ztilde_U1^4}
\end{figure}

Starting from the field equations in (\ref{EOM_scalars})-(\ref{EOM_Einstein}) and performing a change of radial coordinate
\begin{equation}
\label{new_coordinate_Hades}
\rho= \cosh({g} \mu) 
\hspace{10mm} , \hspace{10mm}
d\mu = g^{-1} \, \dfrac{d\rho}{\sqrt{\rho^2-1}}  \ ,
\end{equation}
we find a new class of singular solutions of the form
\begin{equation}
\label{Hades_U1^4_rho_1}
ds_{4}^{2}=\frac{1}{{g}^{2}} \left(  \frac{d\rho^{2}}{\rho^{2}-1} + 
\frac{\rho^{2}-1  }{ k^2} \, d\Sigma^{2}   \right)
\hspace{10mm} \textrm{ with } \hspace{10mm}
k^2= - 1 + \sum_{i} \cosh^2\alpha_{i} \ ,
\end{equation}
and
\begin{equation}
\label{Hades_U1^4_rho_2}
\tilde{z}_{i}(\rho) = e^{i \beta_{i}}\, \frac{\cosh\alpha_{i}}{\sinh\alpha_{i} + i \rho} \ .
\end{equation}
These solutions are defined in the domain $\,\rho \in [1,\infty)\,$ and feature a singularity at $\,\rho=1\,$ where the change of radial coordinate in (\ref{new_coordinate_Hades}) is ill-defined, the warping factor in front of the AdS$_{3}$ piece in the geometry collapses to zero size and $\,|\tilde{z}_{i}(1)|=1\,$ (see Figure~\ref{fig:Hades_ztilde_U1^4}). More concretely, the Ricci scalar constructed from the metric (\ref{Hades_U1^4_rho_1}) reads
\begin{equation}
\label{Hades_Ricci}
R(\rho) = - 6 \, g^{2} \left( \,  1 +  \frac{\rho^2 + k^2}{\rho^2 - 1} \, \right) \ ,
\end{equation}
and becomes singular at $\,\rho= 1\,$. An analysis of the BPS equations (\ref{BPS_scalars}) shows that the flows in (\ref{Hades_U1^4_rho_1})-(\ref{Hades_U1^4_rho_2}) turn out to be non-supersymmetric. We will refer to these singular solutions as flows to Hades. This term was coined for singular (flat-sliced) domain-walls dual to conventional RG-flows in \cite{Freedman:1999gp,Gubser:2000nd}.

As previously done for the Janus solution, let us expand $\,\text{Re}\tilde{z}_{i}\,$ (proper scalars) and $\,\text{Im}\tilde{z}_{i}\,$ (pseudo-scalars) around $\,\rho\rightarrow \infty\,$. One finds
\begin{equation}
\label{Hades_source&vevs_rho}
\begin{array}{llll}
\text{Re}\tilde{z}_{i}(\rho) & = & \dfrac{a^{(v)}_{i,0}}{\rho} +   \dfrac{a^{(s)}_{i,1}}{\rho^2}  + \mathcal{O}\left( \dfrac{1}{\rho^3}\right) & , \\[4mm]
\text{Im}\tilde{z}_{i}(\rho) & = & \dfrac{b^{(s)}_{i,0}}{\rho} + \dfrac{b^{(v)}_{i,1}}{\rho^2}  + \mathcal{O}\left( \dfrac{1}{\rho^3}\right) & ,
\end{array}
\end{equation}
with
\begin{equation}
\label{Hades_ab_vevs}
a^{(v)}_{i,0}=\cosh\alpha_{i} \, \sin\beta_i
\hspace{5mm} , \hspace{5mm}
b^{(v	)}_{i,1} = \sinh\alpha_{i} \,\, a^{(v)}_{i,0} \ ,
\end{equation}
and
\begin{equation}
\label{Hades_ab_sources}
b^{(s)}_{i,0} = - \cosh\alpha_{i} \, \cos\beta_i
\hspace{5mm} , \hspace{5mm}
a^{(s)}_{i,1}  = - \sinh\alpha_{i} \,\,  b^{(s)}_{i,0} \ .
\end{equation}
Different choices of the Hades parameters $\,(\alpha_{i},\beta_{i})\,$ translate into different boundary conditions in the expansions (\ref{Hades_source&vevs_rho}). Note that the boundary theory has sources in (\ref{Hades_ab_sources}) generically activated except if setting $\,\beta_{i}=\pm\frac{\pi}{2}\,$.

\subsubsection{Ridge flows with \texorpdfstring{$\,\alpha_{i}=0\,$}{alpha=0}}
\label{sec:ridge_4D}

Unlike for the Janus solutions, setting $\,\alpha_{i}=0\,$ does not recover a regular AdS$_{4}$ vacuum. Instead, the complex scalars in (\ref{Hades_U1^4_rho_2}) reduce to
\begin{equation}
\label{Ridge_U1^4_rho_2}
\tilde{z}_{i}(\rho) = \rho^{-1}  \, e^{i \left( \beta_{i} - \frac{\pi}{2} \right)} \ ,
\end{equation}
and \textit{ridge flows} of the type investigated in \cite{Pilch:2015dwa,Pilch:2015vha} appear with constant $\,\textrm{arg}\tilde{z}_{i} = \beta_{i} - \frac{\pi}{2}\,$ and $\,k^2=2\,$ in the singular geometry (\ref{Hades_U1^4_rho_1}). The $\,\rho \rightarrow \infty\,$ expansion in (\ref{Hades_source&vevs_rho}) and the boundary conditions in (\ref{Hades_ab_vevs})-(\ref{Hades_ab_sources}) also simplify drastically

\begin{equation}
\label{Ridge_source&vevs_rho}
\begin{array}{llll}
\text{Re}\tilde{z}_{i}(\rho) & = & \dfrac{a^{(v)}_{i,0}}{\rho} +   \dfrac{a^{(s)}_{i,1}}{\rho^2}  + \mathcal{O}\left( \dfrac{1}{\rho^3}\right) & , \\[4mm]
\text{Im}\tilde{z}_{i}(\rho) & = & \dfrac{b^{(s)}_{i,0}}{\rho} + \dfrac{b^{(v)}_{i,1}}{\rho^2}  + \mathcal{O}\left( \dfrac{1}{\rho^3}\right) & ,
\end{array}
\end{equation}
with
\begin{equation}
\label{Ridge_ab_definition}
a^{(v)}_{i,0}= \sin\beta_i
\hspace{8mm} , \hspace{8mm}
b^{(s)}_{i,0} = - \cos\beta_i
\hspace{8mm} , \hspace{8mm}
a^{(s)}_{i,1}  = 0
\hspace{8mm} , \hspace{8mm}
b^{(v	)}_{i,1}  = 0 
\ .
\end{equation}
Two special cases are immediately identified. Setting $\,\beta_{i}=0,\pi\,$ renders $\,\tilde{z}_{i}(\rho)\,$ purely imaginary and the ridge flow from the maximally supersymmetric AdS$_{4}$ vacuum at $\,\rho \rightarrow \infty\,$ is triggered by the source modes $\,b^{(s)}_{i,0}\,$ of the pseudo-scalars dual to fermion bilinears. On the contrary, setting $\,\beta_{i}=\pm\frac{\pi}{2}\,$ renders $\,\tilde{z}_{i}(\rho)\,$ purely real and the ridge flow is triggered by the VEV modes $\,a^{(v)}_{i,0}\,$ of the proper scalars dual to boson bilinears. As we will see in Section~\ref{sec:Uplift_11D}, the uplift of these special ridge flows to eleven dimensions will be very different. This is to be contrasted with the situation in four dimensions where the Lagrangian (\ref{Lagrangian_model_U1^4_Einstein-scalars}) is invariant under constant shifts of $\,\beta_{i}\,$. Note also that a shift of the form $\,\beta_{i} \rightarrow \beta_{i} + \pi\,$ amounts to a reflection $\,\rho \rightarrow -\rho\,$ in the respective field $\,\tilde{z}_{i}\,$ in (\ref{Ridge_U1^4_rho_2}) while leaving the Hades metric in (\ref{Hades_U1^4_rho_1}) invariant. Since the domain of the radial coordinate is fixed to $\,\rho \in [1,\infty)\,$, the shift $\,\beta_{i} \rightarrow \beta_{i} + \pi\,$ generically generates a new solution.

A fundamental difference between our ridge flows in (\ref{Hades_U1^4_rho_1}) and (\ref{Ridge_U1^4_rho_2}) and the ones investigated in \cite{Pope:2003jp,Pilch:2015dwa,Pilch:2015vha} is that the ones there have a flat-sliced geometry. Therefore they correspond to conventional holographic RG-flows. Our solutions have an AdS$_{3}$-slicing of the geometry, instead. It was further shown in \cite{Pilch:2015dwa,Pilch:2015vha} that, for the flat-sliced solutions, only a set of discrete values of $\,\textrm{arg}\tilde{z}_{i}\,$ was compatible with supersymmetry. However, if relaxing supersymmetry, any value of $\,\textrm{arg}\tilde{z}_{i}\,$ was permitted. In our non-supersymmetric ridge flows, any possible value of $\,\beta_{i}\,$ is permitted too. Generic flows to Hades with $\,\alpha_{i} \neq 0\,$ and ridge flows with $\,\alpha_{i} = 0\,$ are depicted in Figure~\ref{fig:Hades_ztilde_U1^4}.

\subsubsection{Hades with (super) symmetry enhancements}

As already discussed for the Janus solutions in Section~\ref{sec:Janus_sym_enhancement}, imposing identifications between the complex fields $\,\tilde{z}_{i}(\rho)\,$ translates into different patterns of (super) symmetry enhancements. For example, non-supersymmetric Hades solutions with $\,\textrm{SU}(3) \times \textrm{U}(1)^2\,$ symmetry are obtained upon identifying the three complex scalars, namely, upon setting $\,{\alpha_{1}=\alpha_{2}=\alpha_{3}}\,$ and $\,{\beta_{1}=\beta_{2}=\beta_{3}}\,$ in the general Hades solution (\ref{Hades_U1^4_rho_1})-(\ref{Hades_U1^4_rho_2}).

Supersymmetric Hades solutions with an AdS$_{3}$ slicing have previously been constructed in \cite{Bobev:2013yra} within the $\,\textrm{SO}(4) \times \textrm{SO}(4)\,$ invariant sector of the $\,\textrm{SO}(8)\,$ gauged supergravity. As discussed in Section~\ref{sec:Janus_SO4xSO4}, this sector of the theory is recovered upon setting two of the three complex fields $\,\tilde{z}_{i}\,$ to zero, \textit{i.e.}, $\,\tilde{z}_{2}(\rho)=\tilde{z}_{3}(\rho)=0\,$. However, it is easy to see that this cannot be achieved by tuning the parameters $\,(\alpha_{i},\beta_{i})\,$ in (\ref{Hades_U1^4_rho_2}) to any real value. Instead, one must set two complex fields to zero from the start and search for solutions of the field equations. In this manner, one finds Hades solutions of the form 
\begin{equation}
\label{Hades_no_ridge_SO(4)xSO(4)_rho_1}
ds_{4}^{2}=\frac{1}{{g}^{2}} \left(  \frac{d\rho^{2}}{\rho^{2}-1} + 
\frac{\rho^{2}-1  }{ k^2} \, d\Sigma^{2}   \right) 
\hspace{10mm} \textrm{ with } \hspace{10mm}
k^2= \sinh^2\alpha_{1}
\ ,
\end{equation}
and
\begin{equation}
\label{Hades_no_ridge_SO(4)xSO(4)_rho_2}
\tilde{z}_{1}(\rho) = e^{i \beta_{1}}\, \frac{\cosh\alpha_{1}}{\sinh\alpha_{1} + i \rho} 
\hspace{8mm} , \hspace{8mm} \tilde{z}_{2}(\rho)=\tilde{z}_{3}(\rho)=0 \ ,
\end{equation}
which turn out to solve the BPS equations in (\ref{BPS_A})-(\ref{BPS_scalars}). It is worth emphasising that these supersymmetric Hades with $\,\textrm{SO}(4) \times \textrm{SO}(4)$ symmetry do not belong to the same class of solutions as the non-supersymmetric Hades in (\ref{Hades_U1^4_rho_1})-(\ref{Hades_U1^4_rho_2}). Also, they do not admit a ridge flow limit since setting $\,\alpha_{1}=0\,$ implies having a pathological ($k^2=0$) warping of AdS$_{3}$ in the geometry (\ref{Hades_no_ridge_SO(4)xSO(4)_rho_1}).

\section{Uplift to eleven-dimensional supergravity}
\label{sec:Uplift_11D}

In this section we present the uplift to eleven-dimensional supergravity of the Janus and Hades solutions constructed within the four-dimensional SO(8) gauged supergravity. We use the conventions of \cite{Gauntlett:2002fz} according to which the Lagrangian of eleven-dimensional supergravity \cite{Cremmer:1978km} takes the form
\begin{equation}
\mathcal{L}_{11} = \hat{R} \, \text{vol}_{11} - \tfrac{1}{2} \, \hat{F}_{(4)} \wedge*_{11} \hat{F}_{(4)} - \tfrac{1}{6} \, \hat{A}_{(3)} \wedge\hat{F}_{(4)} \wedge\hat{F}_{(4)} \ .
\end{equation}
A consistent background is then subject to the source-less Bianchi identity
\begin{equation}
\label{BI_F4}
d\hat{F}_{(4)} = 0 \ ,
\end{equation}
as well as the equations of motion
\begin{equation}
\label{EOM_11D}
\begin{array}
[c]{rll}%
d(*_{11} \hat{F}_{(4)}) + \frac{1}{2} \, \hat{F}_{(4)} \wedge\hat{F}_{(4)} & = & 0 \ ,\\[2mm]%
\hat{R}_{MN} - \frac{1}{12} \left(  \hat{F}_{MPQR} \, \hat{F}_{N}{}^{PQR} - \frac{1}{12} \, \hat{F}_{PQRS} \, \hat{F}^{PQRS} \, \hat{G}_{MN} \right)  & = & 0 \ .
\end{array}
\end{equation}
The equation of motion for $\,\hat{F}_{(4)}\,$ in (\ref{EOM_11D}) can be used to introduce the dual flux 
\begin{equation}
\label{F7_definition}
\hat{F}_{(7)} \equiv *_{11} \hat{F}_{(4)} + \tfrac{1}{2} \hat{A}_{(3)} \wedge \hat{F}_{(4)} \ ,
\end{equation}
which therefore obeys the Bianchi identity $\,d\hat{F}_{(7)}=0\,$. The flux in (\ref{F7_definition}) determines the conserved Page charge of M2-branes in the background\footnote{We have set the string length to unity, \textit{i.e.}, $\,\ell_{s} = 1\,$.}
\begin{equation}
\label{M2_brane_charge}
N_{2} = \frac{1}{(2 \pi)^{6}} \int_{M_{7}} \hat{F}_{(7)}  = \frac{1}{(2 \pi)^{6}} \int_{M_{7}}  *_{11} \hat{F}_{(4)} + \tfrac{1}{2} \hat{A}_{(3)} \wedge \hat{F}_{(4)}   \ ,
\end{equation}
where $\,M_{7}\,$ is the internal space. The contribution $\,*_{11} \hat{F}_{(4)}\,$ comes from electric M2-branes and the contribution $\,{\tfrac{1}{2} \hat{A}_{(3)} \wedge \hat{F}_{(4)}}\,$ originates from magnetic M5-branes.

\subsection{\texorpdfstring{$\text{SU}(3) \times\text{U}(1)^2\,$}{SU(3) x U(1)2} invariant sector}
\label{sec:11D_Janus}

The eleven-dimensional uplift of the $\,\text{SU}(3) \times\text{U}(1)^2\,$ invariant sector of the maximal SO(8) supergravity has been worked out in \cite{Pilch:2015dwa,Azizi:2016noi} (see also \cite{Larios:2019kbw}). To describe the internal geometry, we will closely follow the Appendix~B.2 of \cite{Larios:2019kbw} and use intrinsic coordinates on $\,\textrm{S}^{7}\,$ adapted to its seven-dimensional Sasaki--Einstein structure. In these coordinates, the round metric on $\,\textrm{S}^{7}\,$ takes the form
\begin{equation}
\label{metric_round_S7}
ds_{7}^{2} = ds_{\mathbb{CP}_{3}}^{2} + \left( d\psi_{-}+ \sigma_{-} \right)^2 \ ,
\end{equation}
where $\,ds_{\mathbb{CP}_{3}}^{2}\,$ is the Fubini-Study line element (normalised as in \cite{Larios:2019kbw})
\begin{equation}
\label{metric_round_CP3}
ds_{\mathbb{CP}_{3}}^{2} = d\tilde{\alpha}^{2} + \cos^{2} \tilde{\alpha} \, \big( \, ds^{2}_{\mathbb{CP}_{2}} + \sin^{2} \tilde{\alpha} \, (d\tau_{-} + \sigma)^{2} \, \big)
\hspace{6mm} \textrm{ with } \hspace{6mm}
\sigma_{-} = \cos^{2} \tilde{\alpha} \, (d\tau_{-} + \sigma) \ .
\end{equation}
The ranges of the angles in (\ref{metric_round_S7})-(\ref{metric_round_CP3}) are $\,\tilde{\alpha} \in[0,\frac
{\pi}{2}]\,$, $\,\tau_{-} \in[0,2 \pi]\,$ and $\,\psi_{-} \in[0,2 \pi]\,$. Moreover, $\,\sigma\,$ in (\ref{metric_round_CP3}) is the one-form on $\,\mathbb{CP}_{2}\,$ such that $\,d\sigma=2\boldsymbol{J}\,$ with $\,\boldsymbol{J}\,$ being the K\"ahler form on $\,\mathbb{CP}_{2}\,$. The round metric in (\ref{metric_round_S7}) occurs when the scalar field in the four-dimensional Lagrangian (\ref{Lagrangian_model_SU3xU1xU1}) vanishes, \textit{i.e.}, $\,\tilde{z}=0\,$, and the $\,\textrm{AdS}_{4} \times \textrm{S}^{7}\,$ Freund--Rubin vacuum of eleven-dimensional supergravity is recovered \cite{Freund:1980xh}. However, whenever non-vanishing, the scalar $\,\tilde{z}\,$ in (\ref{Janus_solution_SU3xU1xU1}) inflicts a deformation on the Freund--Rubin vacuum so that a new background is generated which displays a smaller $\,\text{SU}(3) \times \text{U}(1)^{2} \subset \textrm{SO}(8)\,$ isometry group.

We are encoding the breaking of isometries caused by $\,\tilde{z}\,$ into a set of metric functions $\,f\,$'s and flux functions $\,h\,$'s. The eleven-dimensional metric takes the form
\begin{equation}
\label{11D_metric}
\begin{array}{lll}
d\hat{s}_{11}^{2} & = & \frac{1}{2} \, f_{1} \, ds_{4}^{2} + 2 \,  g^{-2} \Big[ f_{2} \, d\tilde{\alpha}^{2} + \cos^{2} \tilde{\alpha} \, \big( \, f_{3} \, \, ds^{2}_{\mathbb{CP}_{2}} + \sin^{2} \tilde{\alpha} \, f_{4} \, (d\tau_{-} + \sigma)^{2} \, \big)\\[2mm]
& + & f_{5} \, \big( d\psi_{-} + \cos^{2} \tilde{\alpha} \, f_{6} \, (d\tau_{-} + \sigma) \big)^{2} \Big] \ ,
\end{array}
\end{equation}
with $\,ds_{4}^{2}\,$ given in (\ref{Janus_solution_SU3xU1xU1})-(\ref{A(mu)_func_SU3xU1xU1}). Note that the eleven-dimensional metric  (\ref{11D_metric}) displays an $\,\text{SU}(3) \times \text{U}(1)_{\tau_{-}} \times\text{U}(1)_{\psi_{-}}\,$ symmetry. The $\,\textrm{SU}(3)\,$ factor accounts for the $\,\mathbb{CP}_{2}\,$ isometries and the two $\,\textrm{U}(1)\,$ factors correspond with shifts along the angles $\,\tau_{-}\,$ and $\,\psi_{-}\,$, hence the attached labels. The various metric functions in (\ref{11D_metric}) depend on the complex scalar $\tilde{z}$ in
(\ref{Janus_solution_SU3xU1xU1}) and on the angle $\,\tilde{\alpha} \,$ on S$^{7}$. They are given by
\begin{equation}
\label{f_functions}
\begin{array}{c}
f_{1}^{3} = \dfrac{ (1+\tilde{z}) (1+\tilde{z}^{*})}{(1-|\tilde{z}|^{2})^{3}} \, H^{2}
\hspace{5mm} , \hspace{5mm}
f_{2}^{3/2} = \dfrac{H}{(1+\tilde{z}) (1+\tilde{z}^{*})}
\hspace{5mm} , \hspace{5mm}
f_{3}^{3} = \dfrac{(1+\tilde{z}) (1+\tilde{z}^{*})}{H} \ , \\[6mm]
f_{4}^{3/2} = \dfrac{(1-|\tilde{z}|^{2})^{3}}{ (1+\tilde{z}) (1+\tilde{z}^{*})} \, H \, K^{-\frac{3}{2}} 
\hspace{5mm} , \hspace{5mm}
f_{5}^{3/2} = \dfrac{1}{ (1+\tilde{z}) (1+\tilde{z}^{*})} \, H^{-2} \, K^{\frac{3}{2}} \ , \\[8mm]
f_{6} = \Big[ (1+\tilde{z}) (1+\tilde{z}^{*}) \, H + ( \tilde{z} - \tilde{z}^{*})^{2} \cos(2\tilde{\alpha}) \Big] \, K^{-1} \ ,
\end{array}
\end{equation}
with
\begin{equation}
H = 1+|\tilde{z}|^{2} - ( \tilde{z} +\tilde{z}^{*}) \cos(2\tilde{\alpha}) \hspace{8mm} \text{ and } \hspace{8mm} K = 1+|\tilde{z}|^{4} - 2 \, |\tilde{z}|^{2} \, \cos(4\tilde{\alpha}) \ .
\end{equation}
The round metric on $\,\textrm{S}^{7}\,$ is recovered from (\ref{11D_metric}) upon setting $\,\tilde{z}=0\,$, what implies that all the metric functions $\,H=K=f_{1,\ldots,6}=1\,$. The part of the internal geometry in the upper line of (\ref{11D_metric}) then reconstructs the $\,\mathbb{CP}_{3}\,$ metric in (\ref{metric_round_CP3}).

The eleven-dimensional four-form flux takes a more lengthy expression given in terms of three-, one- and zero-form deformations in four dimensions which we collectively denote $\,h$'s. Adopting the terminology of \cite{Pilch:2015dwa}, the four-form flux naturally splits as
\begin{equation}
\label{11D_F4}
\hat{F}_{(4)} = \hat{F}_{(4)}^{\textrm{st}} + \hat{F}_{(4)}^{\textrm{tr}} \ ,
\end{equation}
with
\begin{equation}
\label{11D_F4_st}
\hat{F}_{(4)}^{\textrm{st}} =
-\frac{1}{2\sqrt{2}} \, g \, h_{1} \, \text{vol}_{4} + \frac{1}{\sqrt{2}} \, g^{-1} \, \sin(2 \tilde{\alpha}) \,\, h^{(3)}_{2} \wedge d\tilde{\alpha} \ ,
\end{equation}
and
\begin{equation}
\label{11D_F4_tr}
\begin{array}{lll}
\hat{F}_{(4)}^{\textrm{tr}} &=& - 2 \sqrt{2} \,  g^{-3} \Big[  \, \sin(2 \tilde{\alpha}) \, h^{(1)}_{3} \wedge d\tilde{\alpha} \wedge d\psi_{-} \wedge(d\tau_{-}+\sigma)\\[2mm]
& + &   \cos^{4} \tilde{\alpha} \,\, h^{(1)}_{4} \wedge (d\tau_{-} + \sigma) \wedge\boldsymbol{J} +   \cos^{2} \tilde{\alpha} \, \cos(2 \tilde{\alpha}) \,\, h_{5}^{(1)} \wedge d\psi_{-} \wedge\boldsymbol{J}\\[2mm]
& + &   \sin(2 \tilde{\alpha}) \, h_{6} \, d\tilde{\alpha} \wedge d\psi_{-} \wedge\boldsymbol{J} +   \cos^{4} \tilde{\alpha} \,\, h_{7} \, \boldsymbol{J} \wedge\boldsymbol{J}\\[2mm]
& + &   \cos^{2}\tilde{\alpha} \, \sin(2 \tilde{\alpha}) \,\, h_{8} \, d\tilde{\alpha} \wedge(d\tau_{-} + \sigma) \wedge\boldsymbol{J} \, \Big] \ .
\end{array}
\end{equation}
For the space-time part in (\ref{11D_F4_st}) we have introduced a zero-form
\begin{equation}
\label{h1_func}
h_{1} = \dfrac{1}{(1-|\tilde{z}|^{2})} \Big( \, 3 \, (1+|\tilde{z}|^{2}) + ( \tilde{z} +\tilde{z}^{*}) \, (1 - 2 \cos(2 \tilde{\alpha}) ) \, \Big) \ ,    
\end{equation}
and a three-form
\begin{equation}
\label{h2_func}
\begin{array}{lll}
h^{(3)}_{2} & = & \dfrac{1}{(1-|\tilde{z}|^{2})^{2}} \Big( (\tilde{z}^{2}-1) *_{4}  d\tilde{z}^{*} + ((\tilde{z}^{*})^{2}-1) *_{4} d\tilde{z} \Big) \ ,
\end{array}
\end{equation}
which has legs along the AdS$_{3}$ factor in the external geometry\footnote{The Hodge dual $\,*_{4}\,$ is defined in four-dimensions using the metric $\,ds^{2}_{4}\,$ in (\ref{Janus_solution_SU3xU1xU1})-(\ref{A(mu)_func_SU3xU1xU1}).}. For the transverse part in (\ref{11D_F4_tr}) we have introduced a set of one-forms
\begin{equation}
\label{11D_F4_one-forms}
\begin{array}{lll}
h^{(1)}_{3} & = & \dfrac{i}{2} \left(  \dfrac{d\tilde{z}^{*} }{(1+\tilde{z}^{*})^{2}} - \dfrac{d\tilde{z} }{(1+\tilde{z})^{2}} \right)  \ ,\\[4mm]
h^{(1)}_{4} & = & h^{(1)}_{5} \,\, = \,\, i \, H^{-2} \, \Big( \left(  1 - 2 \cos(2 \tilde{\alpha}) \, \tilde{z}^{*} + (\tilde{z}^{*})^{2} \right)  \, d\tilde{z} - \left(  1 - 2 \cos(2 \tilde{\alpha}) \, \tilde{z} + \tilde{z}^{2} \right)  \, d\tilde{z}^{*}
\Big) \ ,
\end{array}
\end{equation}
together with zero-forms
\begin{equation}
\begin{array}{lll}
h_{6} & = & i \, 4 \, H^{-2} \, (\tilde{z}^{*}-\tilde{z}) \dfrac{(1+|\tilde{z}|^{2})}{(1+\tilde{z})(1+\tilde{z}^{*})} \Big( 1+ |\tilde{z}|^{2} + (\tilde{z}+\tilde{z}^{*}) \sin^{2}\tilde{\alpha} \Big) \ ,\\[4mm]
h_{7} & = & -i \, 2 \, H^{-1} \, (\tilde{z}^{*}-\tilde{z}) \ ,\\[4mm]
h_{8} & = & i \, 2 \, H^{-2} \, (\tilde{z}^{*}-\tilde{z}) \, \Big( 1+ |\tilde{z}|^{2} + (\tilde{z}+\tilde{z}^{*}) \sin^{2}\tilde{\alpha} \Big) \ .
\end{array}
\end{equation}
The zero-forms $\,h_{6}\,$, $\,h_{7}\,$ and $\,h_{8}\,$ determine the purely internal components in (\ref{11D_F4_tr}) and vanish if $\,\tilde{z}^{*}=\tilde{z}\,$. Also the one-form deformations in (\ref{11D_F4_one-forms}) vanish in this case so that $\,\hat{F}_{(4)}^{\textrm{tr}}=0\,$. Lastly, the entire eleven-dimensional flux in (\ref{11D_F4}) preserves an $\,\text{SU}(3)\times\text{U}(1)_{\tau_{-}} \times\text{U}(1)_{\psi_{-}}\,$ symmetry since there is no explicit dependence on the angle $\,\psi_{-}\,$ and, moreover, the two-form $\,\boldsymbol{J}\,$ on $\,\mathbb{CP}_{2}\,$ is not charged under $\,\text{U}(1)_{\tau_{-}}\,$.

To complete the uplift, the above quantities must be evaluated at the value of the complex scalar $\,\tilde{z} \equiv \tilde{z}_{1}=\tilde{z}_{2}=\tilde{z}_{3}\,$ both for the Janus (\ref{Janus_U1^4_rho_2}) and Hades (\ref{Hades_U1^4_rho_2}) solutions. We have explicitly verified that the resulting eleven-dimensional backgrounds in (\ref{11D_metric}) and (\ref{11D_F4}) satisfy the source-less Bianchi identity and equations of motion in (\ref{BI_F4}) and (\ref{EOM_11D}), respectively.

\subsection{\texorpdfstring{$\text{SU}(3) \times\text{U}(1)^2\,$}{SU(3) x U(1)2} symmetric Janus}

\begin{figure}[t]
\begin{center}
\includegraphics[width=0.25\textwidth]{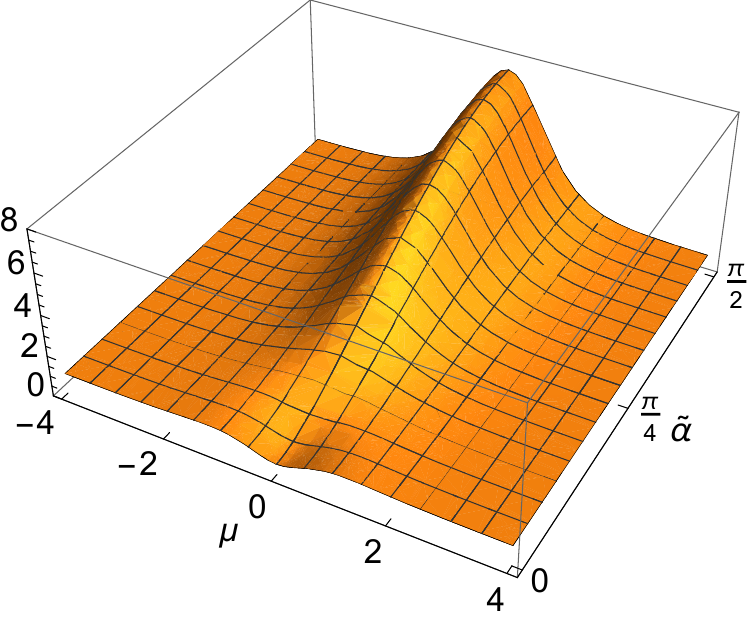} 
\put(-100,100){$f_{1}$} \hspace{10mm}
\includegraphics[width=0.25\textwidth]{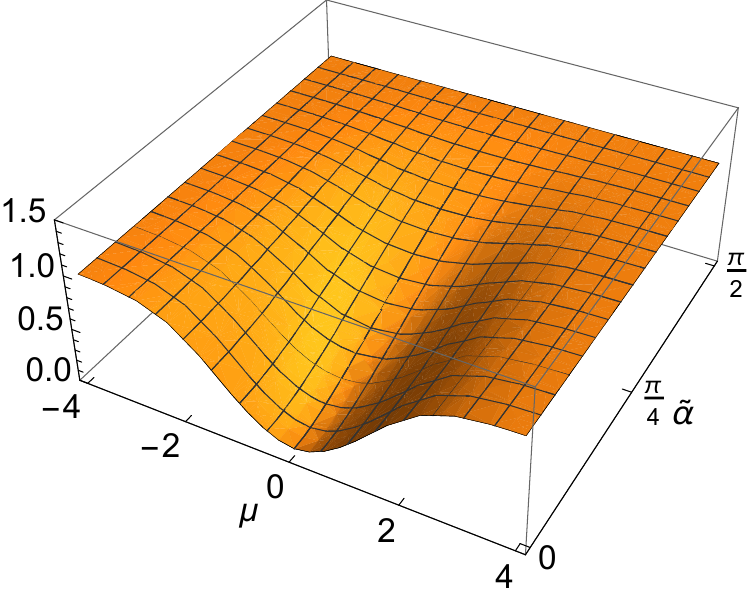} 
\put(-100,100){$f_{2}$} \hspace{10mm}
\includegraphics[width=0.25\textwidth]{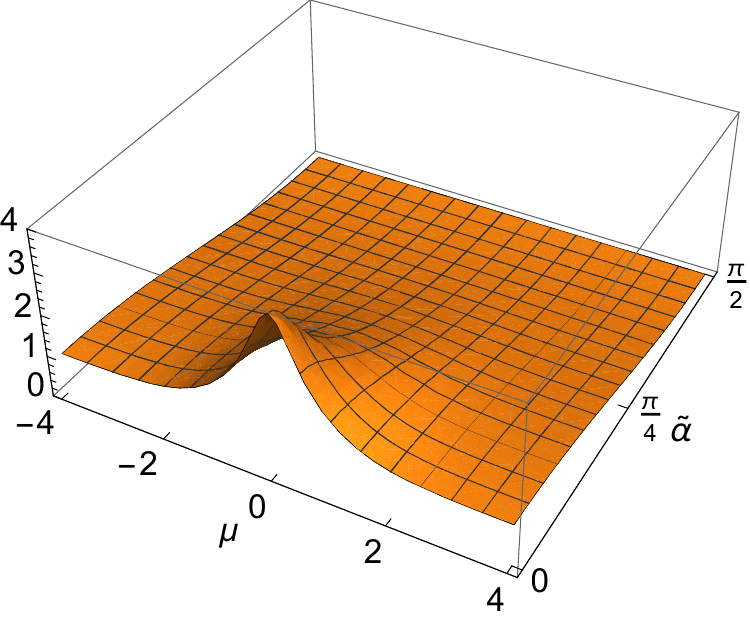} 
\put(-100,100){$\cos^2\tilde{\alpha} \, f_{3}$}
\vspace{5mm}
\includegraphics[width=0.25\textwidth]{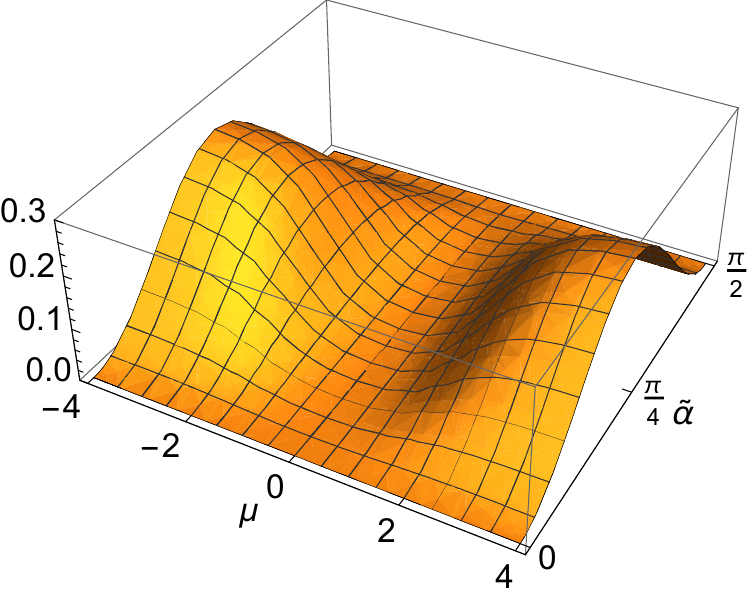} 
\put(-100,100){$\cos^2\tilde{\alpha} \, \sin^2\tilde{\alpha}  \,f_{4}$} \hspace{10mm}
\includegraphics[width=0.25\textwidth]{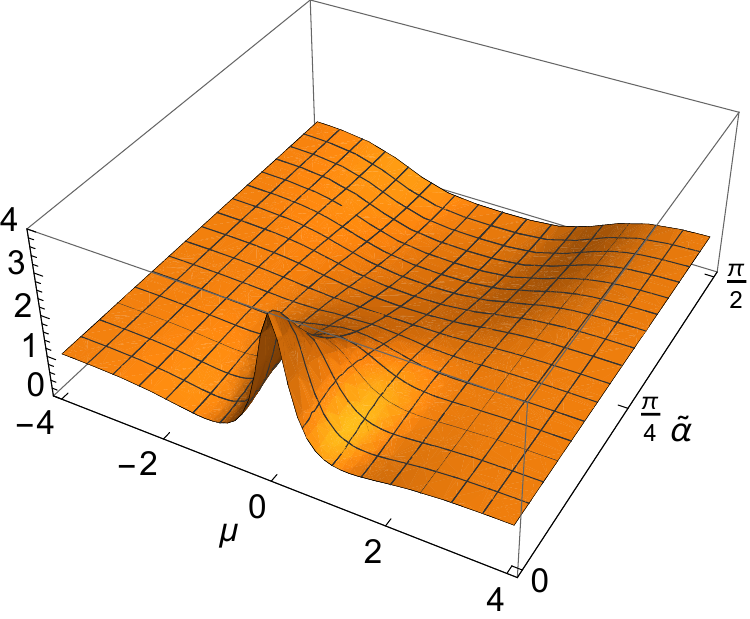} 
\put(-100,100){$f_{5}$} \hspace{10mm}
\includegraphics[width=0.25\textwidth]{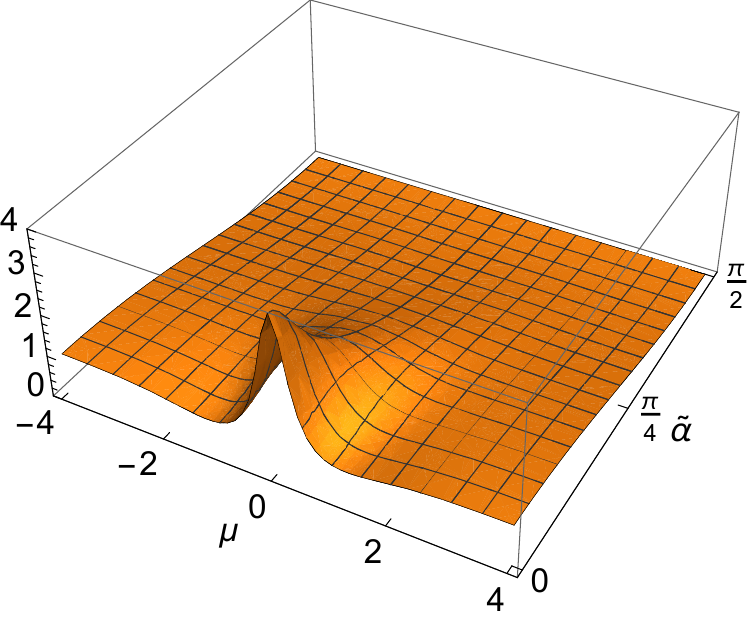} 
\put(-100,100){$\cos^2\tilde{\alpha} \, f_{5}\, f_{6}$}
\end{center}
\caption{Regular metric functions in (\ref{11D_metric}) for the Janus solution with $\alpha=1$ and $\beta=0$.}
\label{fig:f_functions_beta0}
\end{figure}

\begin{figure}[t]
\begin{center}
\includegraphics[width=0.25\textwidth]{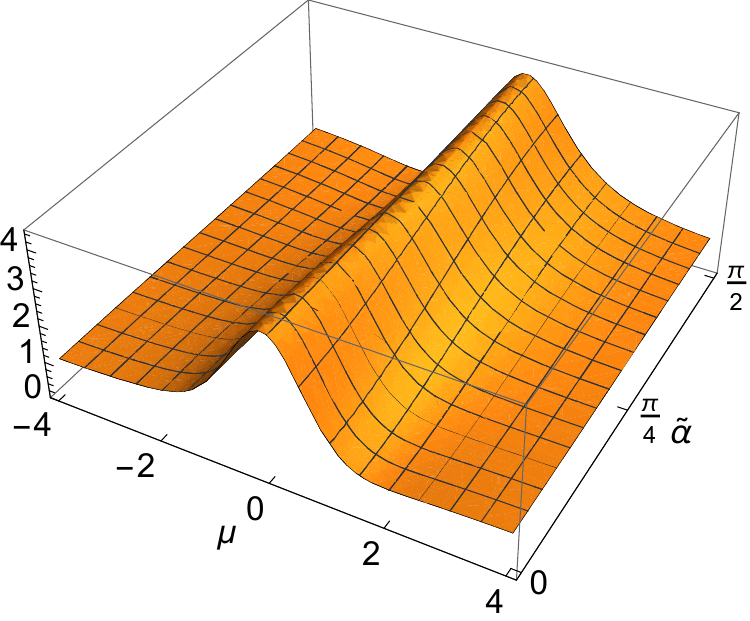} 
\put(-100,100){$f_{1}$} \hspace{10mm}
\includegraphics[width=0.25\textwidth]{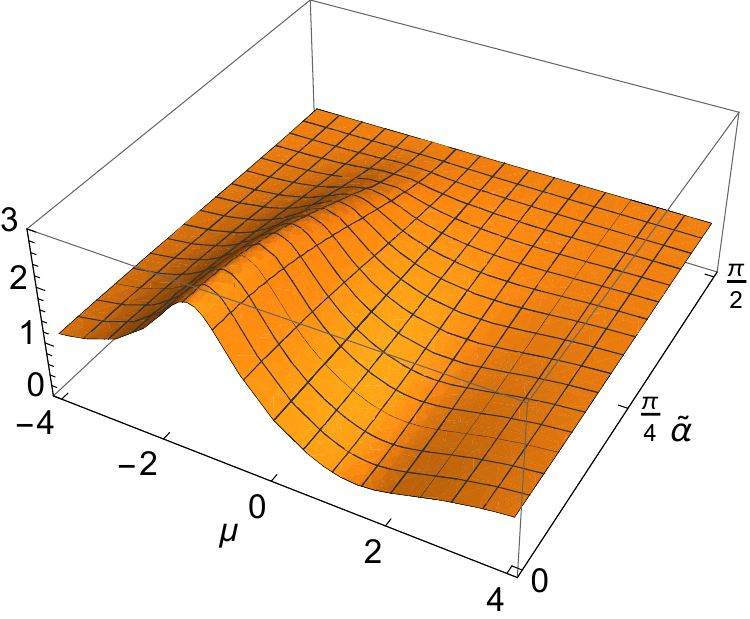} 
\put(-100,100){$f_{2}$} \hspace{10mm}
\includegraphics[width=0.25\textwidth]{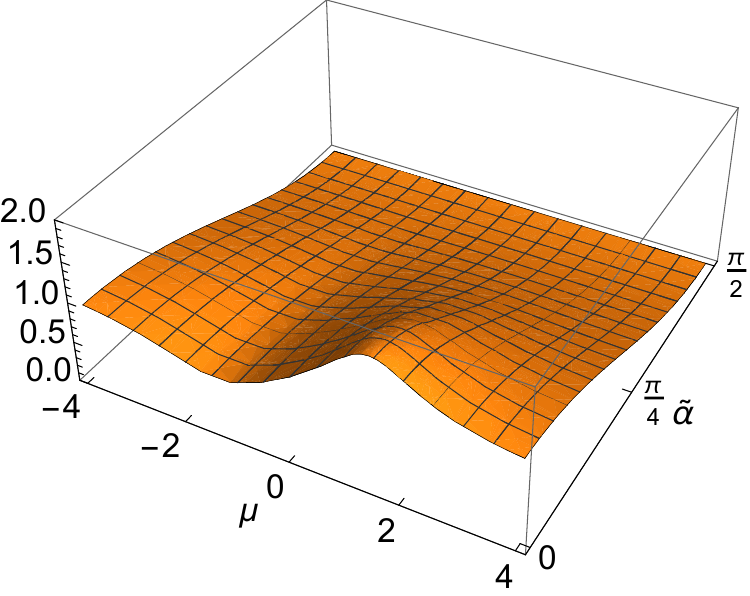} 
\put(-100,100){$\cos^2\tilde{\alpha} \, f_{3}$}
\vspace{5mm}
\includegraphics[width=0.25\textwidth]{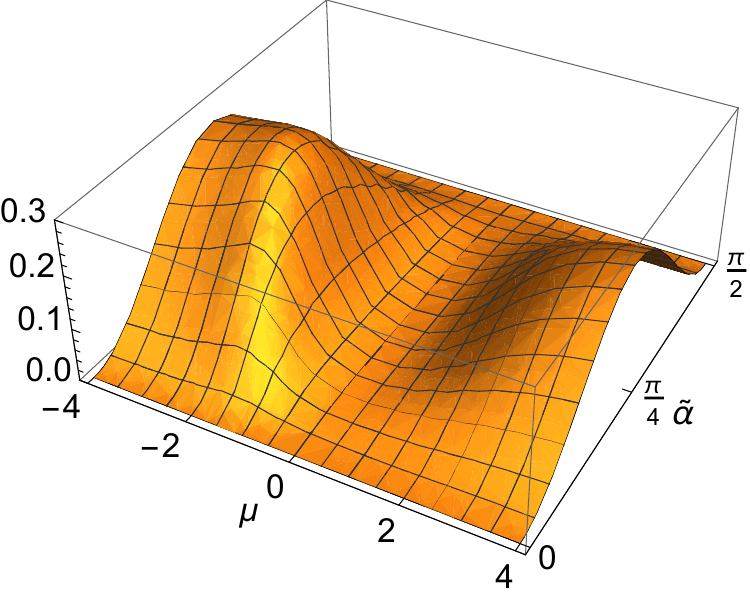} 
\put(-100,100){$\cos^2\tilde{\alpha} \, \sin^2\tilde{\alpha}  \,f_{4}$} \hspace{10mm}
\includegraphics[width=0.25\textwidth]{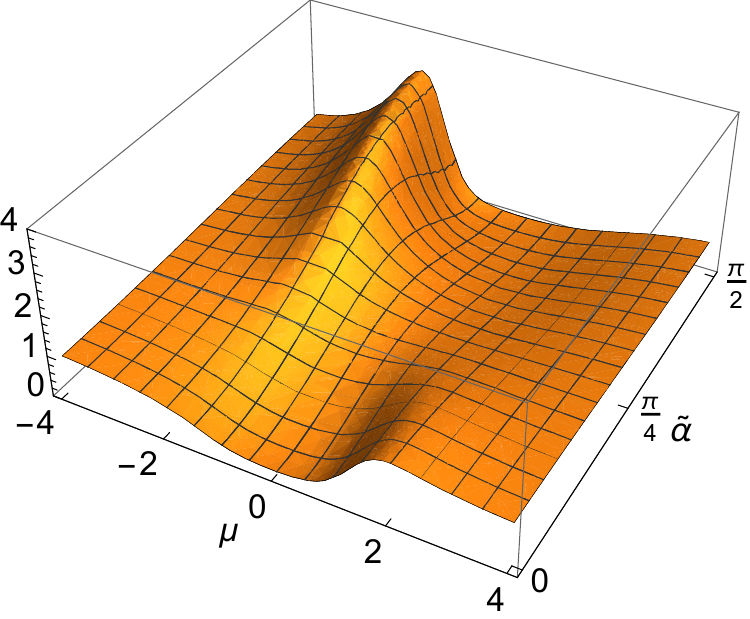} 
\put(-100,100){$f_{5}$} \hspace{10mm}
\includegraphics[width=0.25\textwidth]{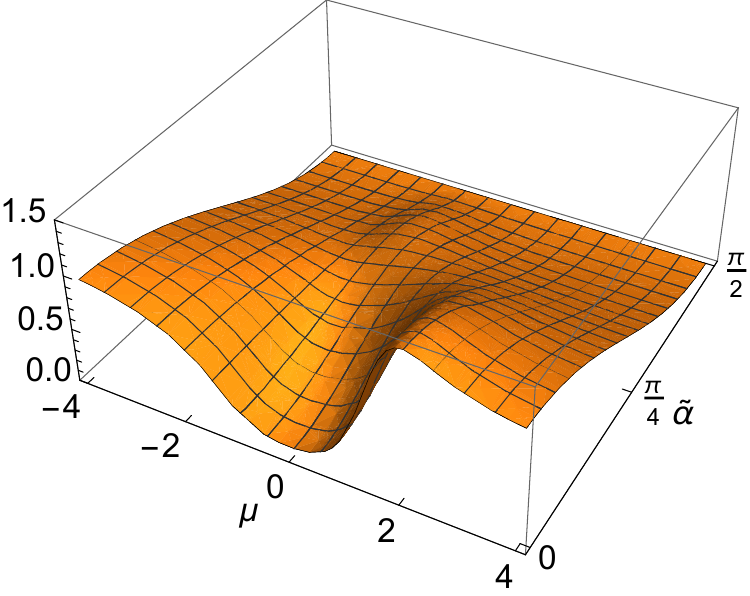} 
\put(-100,100){$\cos^2\tilde{\alpha} \, f_{5}\, f_{6}$}
\end{center}
\caption{Regular metric functions in (\ref{11D_metric}) for the Janus solution with $\alpha=1$ and $\beta=\frac{\pi}{2}$.}
\label{fig:f_functions_betaPi}%
\end{figure}

We have performed the explicit uplift of the analytic and non-supersymmetric Janus solution in (\ref{Janus_solution_SU3xU1xU1})-(\ref{A(mu)_func_SU3xU1xU1}). The resulting eleven-dimensional backgrounds are everywhere regular and depend on the choice of parameters $\,(\alpha, \beta)\,$ specifying the boundary conditions (\ref{ab_definition})-(\ref{ab_algebraic}) of the four-dimensional Janus solution. Plots of the functions entering the metric (\ref{11D_metric}) for $\,\beta=0\,$ and $\,\beta=\frac{\pi}{2}\,$ are depicted in Figure~\ref{fig:f_functions_beta0} and Figure~\ref{fig:f_functions_betaPi}. These two choices respectively activate only sources or VEV's in the Janus boundary conditions (\ref{ab_definition})-(\ref{ab_algebraic}). In addition, the scalar $\,\tilde{z}\,$ in the $\text{SU}(3) \times\text{U}(1)^2\,$ symmetric Janus solution of (\ref{Janus_solution_SU3xU1xU1}) is necessarily complex so that no limit to a real Janus solution exists even in the general case of (\ref{Janus_solution_U1^4_ztil}). This further implies that all the $\,h\,$ functions (and also three- and one-forms) entering $\,\hat{F}_{(4)}^{\textrm{st}}\,$ in (\ref{11D_F4_st}) and $\,\hat{F}_{(4)}^{\textrm{tr}}\,$ in (\ref{11D_F4_tr}) are generically activated.

In order to compute the M$2$-brane charge for the $\,\textrm{SU}(3) \times \textrm{U}(1)^2\,$ symmetric Janus, we first note that the dual seven-form flux can be expressed as
\begin{equation}
\label{F7_general}
\hat{F}_{(7)} = d\hat{\alpha} \wedge h^{(6)}  + \ldots \ , 
\end{equation}
with $\,h^{(6)}=\frac{1}{2}  \boldsymbol{J} \wedge \boldsymbol{J} \wedge d\tau_{-} \wedge d\psi_{-}\,$ being the volume form of $\,M_{6}\,$ spanned by $\,(\mathbb{CP}_{2}, \tau_{-}, \psi_{-})$, and $\,\hat{\alpha}\,$ playing the role of an ``adapted'' angular coordinate threaded by the flux. This adapted coordiante is in general a complicated function
\begin{equation}
\label{hat_alpha}
\hat{\alpha}=\hat{\alpha}(\rho,\tilde{\alpha} \, ; \,\alpha,\beta) \ ,
\end{equation}
that depends on the original coordinates $\,(\rho,\tilde{\alpha})\,$ as well as on the Janus parameters $\,(\alpha,\beta)\,$. Lastly, the ellipsis in (\ref{F7_general}) stand for additional terms with legs on the AdS$_3$ piece of the geometry which do not play a relevant role when computing M$2$-brane charges. Therefore, all the relevant information regarding M$2$-brane charges gets codified into the one-form $\,d\hat{\alpha}\,$ as it defines an adapted angular direction. It is important to highlight that, when taking the limit $\,\rho \rightarrow \pm \infty\,$, one finds that $\,d\hat{\alpha} \propto \sin(2\tilde{\alpha}) \cos^4\tilde{\alpha} \, d\tilde{\alpha}\,$ no longer depends on the Janus parameters $\,(\alpha,\beta)\,$. In this limit, the dual seven-form flux threads the $\,\textrm{S}^7\,$ as required by the asymptotic $\,\textrm{AdS}_{4} \times \textrm{S}^{7}\,$ geometry of the flow.

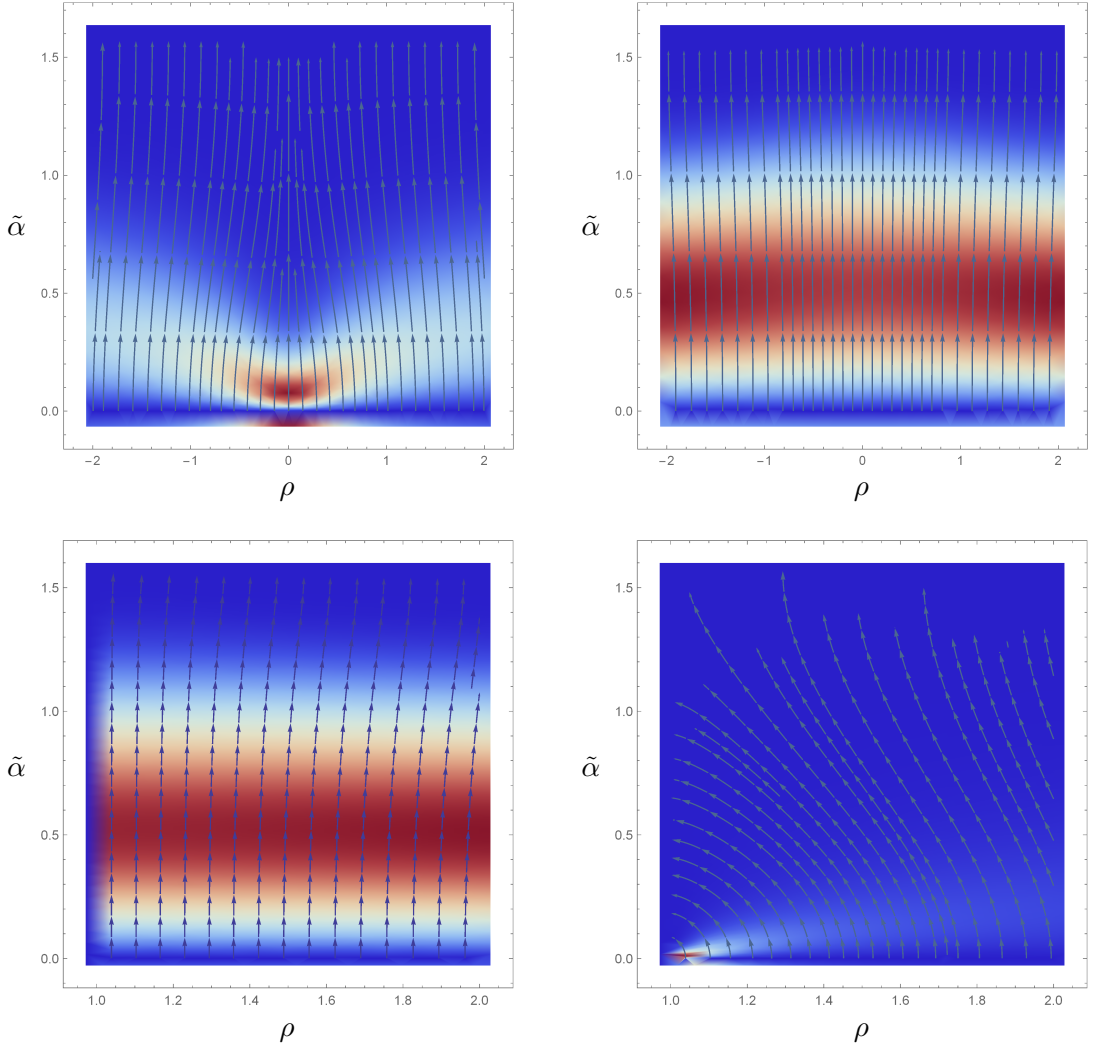
\begin{figure}[ht!]
\begin{center}
\includegraphics[width=0.40\textwidth]{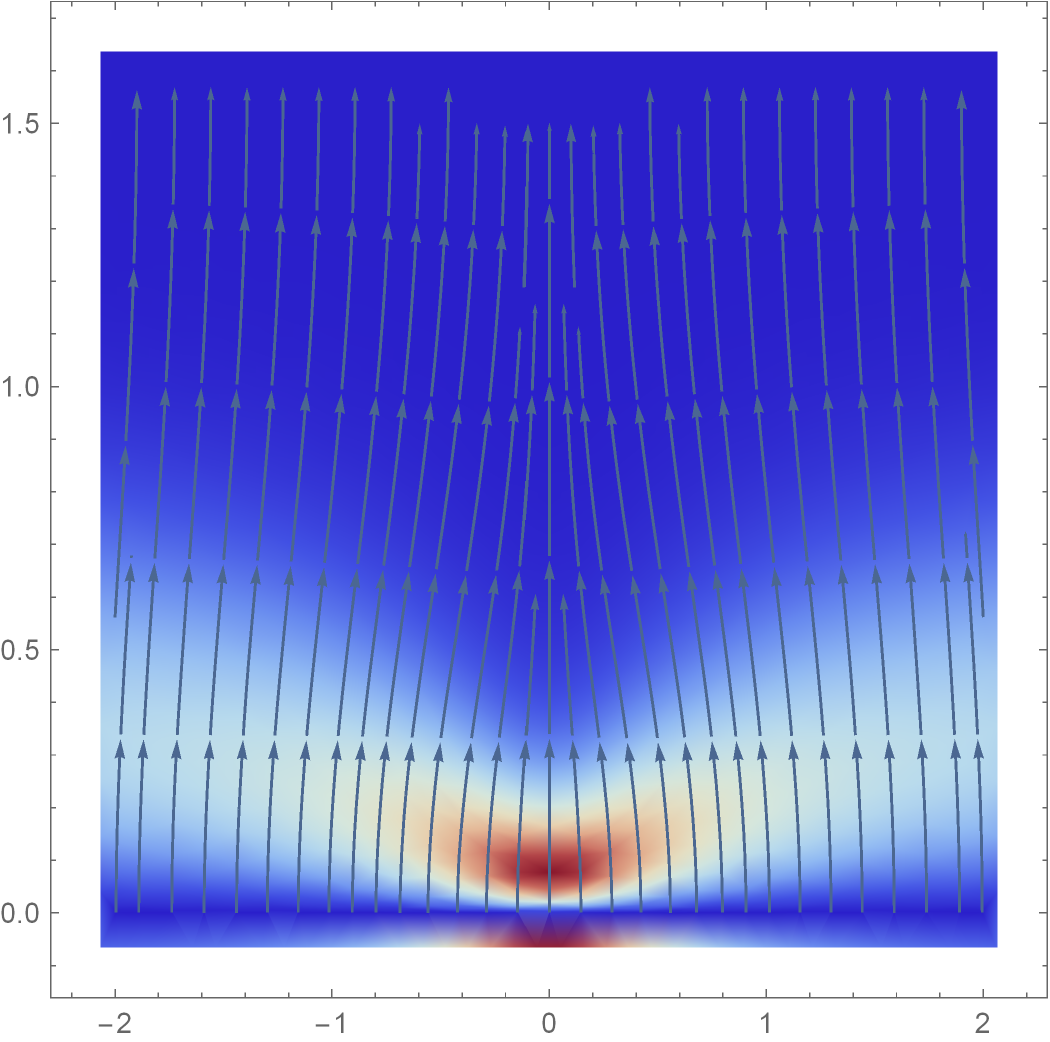} 
\put(-190,88){$\tilde{\alpha}$} 
\put(-88,-10){$\rho$}
\hspace{12mm}
\includegraphics[width=0.40\textwidth]{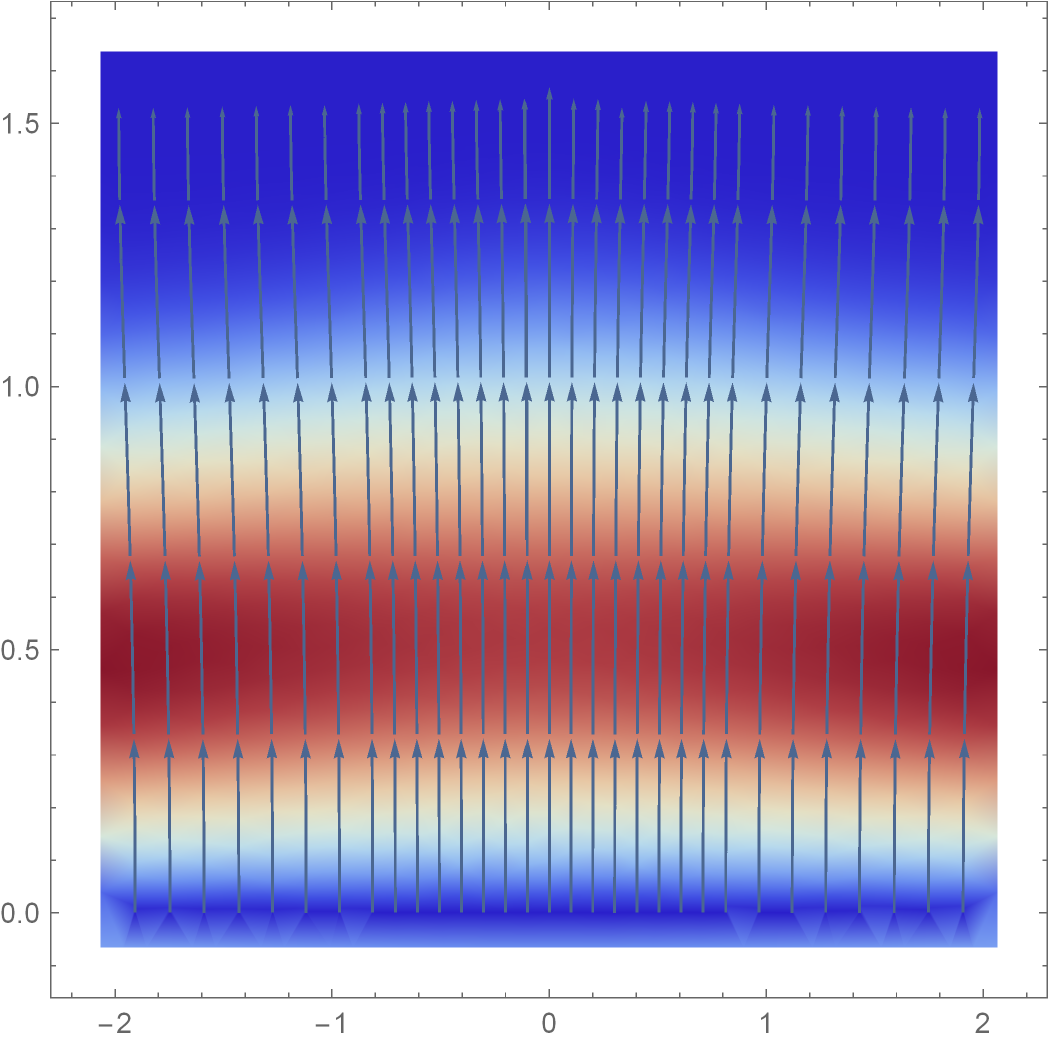} 
\put(-190,88){$\tilde{\alpha}$} 
\put(-88,-10){$\rho$}
\\[5mm]
\includegraphics[width=0.40\textwidth]{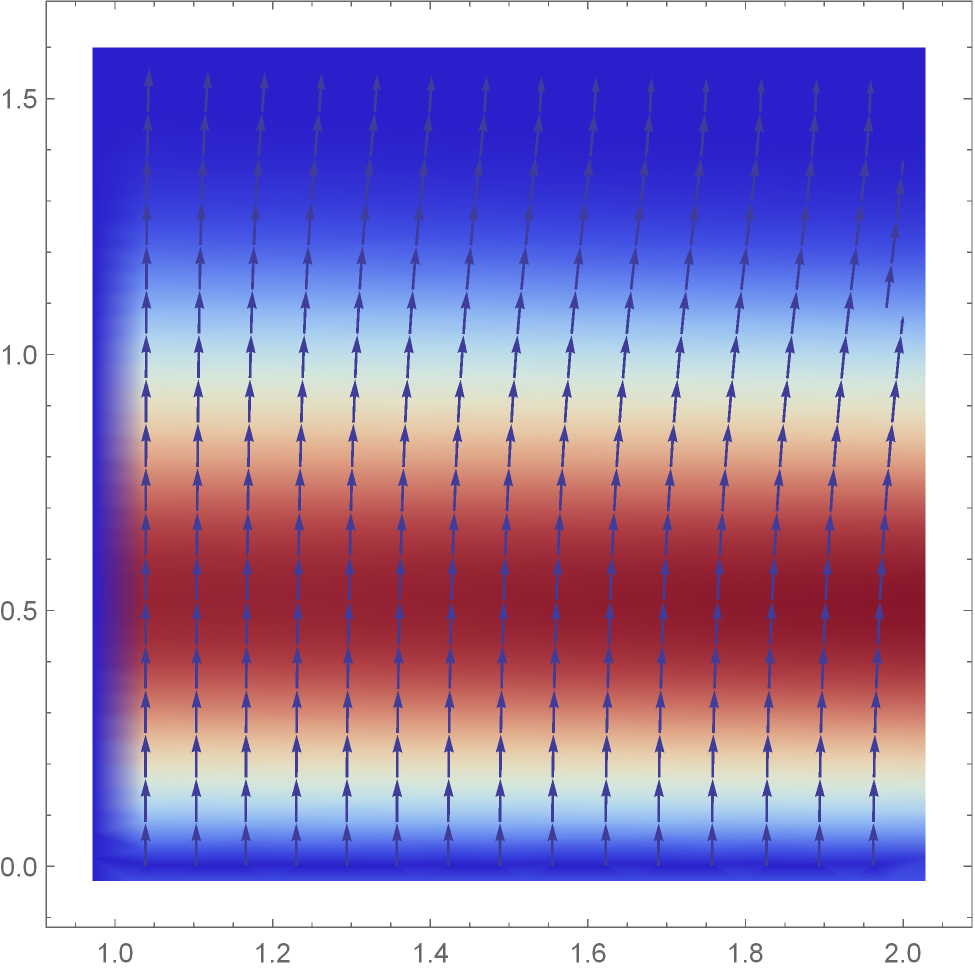} 
\put(-190,88){$\tilde{\alpha}$} 
\put(-88,-10){$\rho$}
\hspace{12mm}
\includegraphics[width=0.40\textwidth]{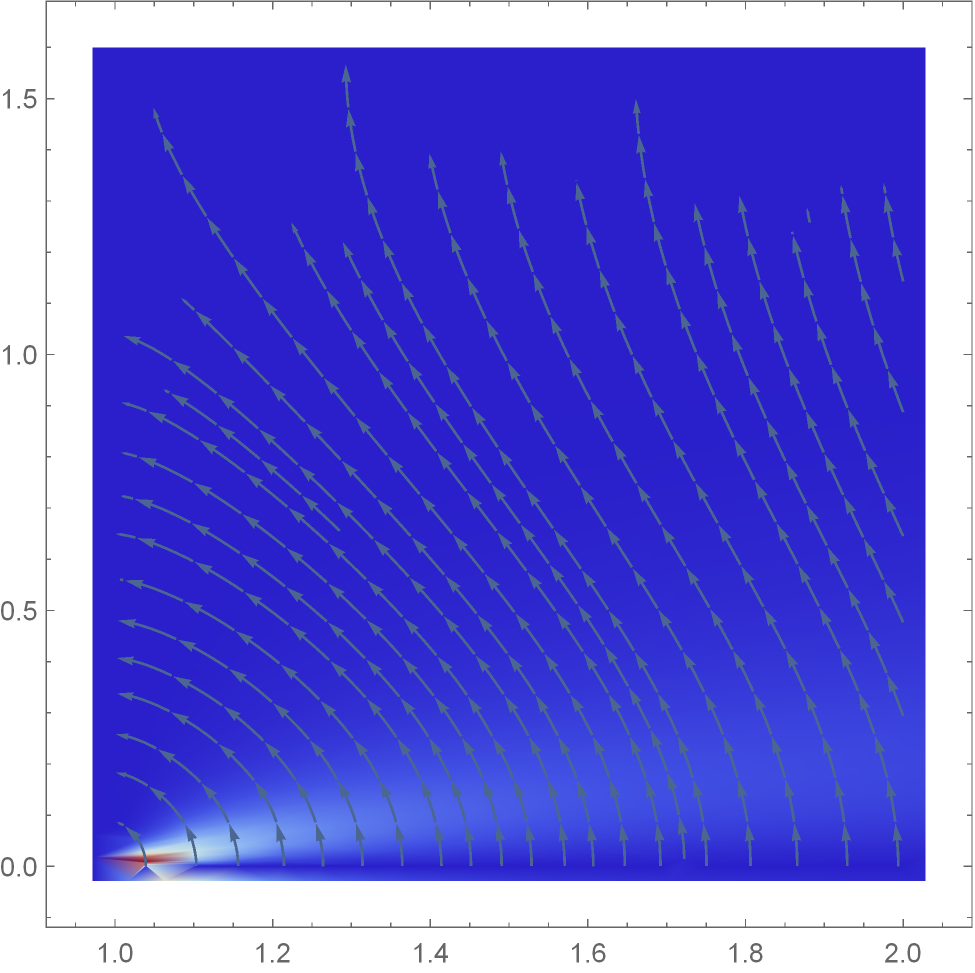} 
\put(-190,88){$\tilde{\alpha}$} 
\put(-88,-10){$\rho$}
\end{center}
\caption{Plots of the vector field $\,\boldsymbol{v} = (\,g \sqrt{\rho^2 \pm 1} \,  \partial_{\rho}\hat{\alpha} \,,\, \partial_{\tilde{\alpha}}\hat{\alpha}\,)\,$ on the strip spanned by $\,(\rho,\tilde{\alpha})\,$. The $\,+\,$ sign must be chosen for the Janus solutions whereas the $\,-\,$ sign corresponds to the Hades solutions as a consequence of the change of radial coordinate $\,d\rho = g \, \sqrt{\rho^2 \pm 1} \, d\mu \,$. Top-Left: Janus flow with $\,\alpha=1\,$ and $\,\beta=0\,$. Top-Right: Janus flow with $\,\alpha=1\,$ and $\,\beta=\pi\,$. Bottom-Left: Ridge flow with $\,\beta = -\frac{\pi}{2}\,$. Bottom-Right: Ridge flow with $\,\beta = \frac{\pi}{2}\,$.}
\label{fig:vec_field}
\end{figure}

The computation of the M$2$-brane charge in the Janus solution gives
\begin{equation}
\label{N2_Janus}
N_2 = \frac{1}{(2\pi)^6}\int_{\Gamma \times \textrm{M}_{6}} \hat{F}_{(7)} = \frac{1}{32 \pi^2}\int_{\partial\Gamma} \hat{\alpha}  = \frac{1}{4\pi^2 g^6} \ ,
\end{equation}
where the relevant curves $\,\Gamma$'s threaded by the purely internal part of the seven-form flux in (\ref{F7_general}) are specified by their tangent vector field $\,\boldsymbol{v} = (\boldsymbol{v}_{\mu},\boldsymbol{v}_{\tilde{\alpha}}) = (g \sqrt{\rho^2+1} \, \partial_{\rho}\hat{\alpha} \,,\, \partial_{\tilde{\alpha}}\hat{\alpha})\,$\footnote{Note that $\,\boldsymbol{v}_{\mu}=\partial_{\mu}\hat{\alpha}=g \sqrt{\rho^2+1} \, \partial_{\rho}\hat{\alpha}\,$ as a consequence of the change of radial coordinate in (\ref{new_coordinate_Janus}).}. For the Janus, all the curves $\,\Gamma\,$ start at $\,\tilde{\alpha}=0\,$ and end at $\,\tilde{\alpha}=\frac{\pi}{2}\,$ pointing at the $\,\tilde{\alpha}\,$ direction on $\,\textrm{S}^{7}\,$ -- see Figure~\ref{fig:vec_field} for an illustration of such curves in various examples --. Since the $\,N_{2}\,$ charge in (\ref{N2_Janus}) is independent of $\,\Gamma\,$ and also of the Janus parameters $\,(\alpha,\beta)$, it matches the one of the $\textrm{AdS}_{4} \times \textrm{S}^7\,$ background controlling the asymptotic behaviour of the (regular) Janus solutions at $\,\rho \rightarrow \pm \infty\,$.

\begin{figure}[t]
\begin{center}
\includegraphics[width=0.45\textwidth]{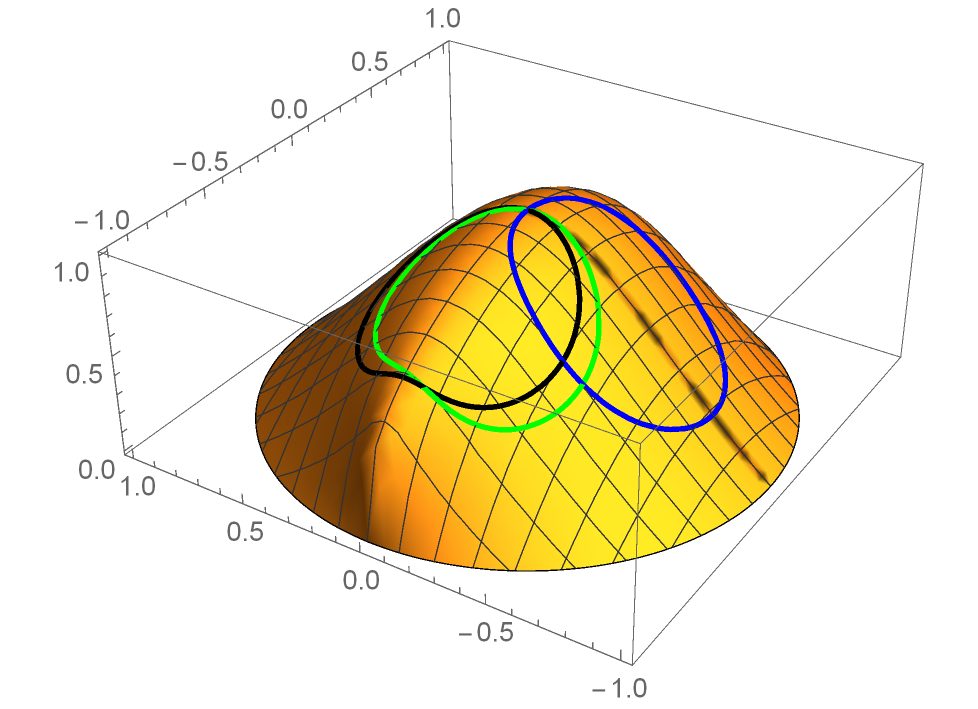} 
\put(-35,30){$\textrm{Re}\tilde{z}$}
\put(-150,10){$\textrm{Im}\tilde{z}$}
\hspace{8mm}
\includegraphics[width=0.42\textwidth]{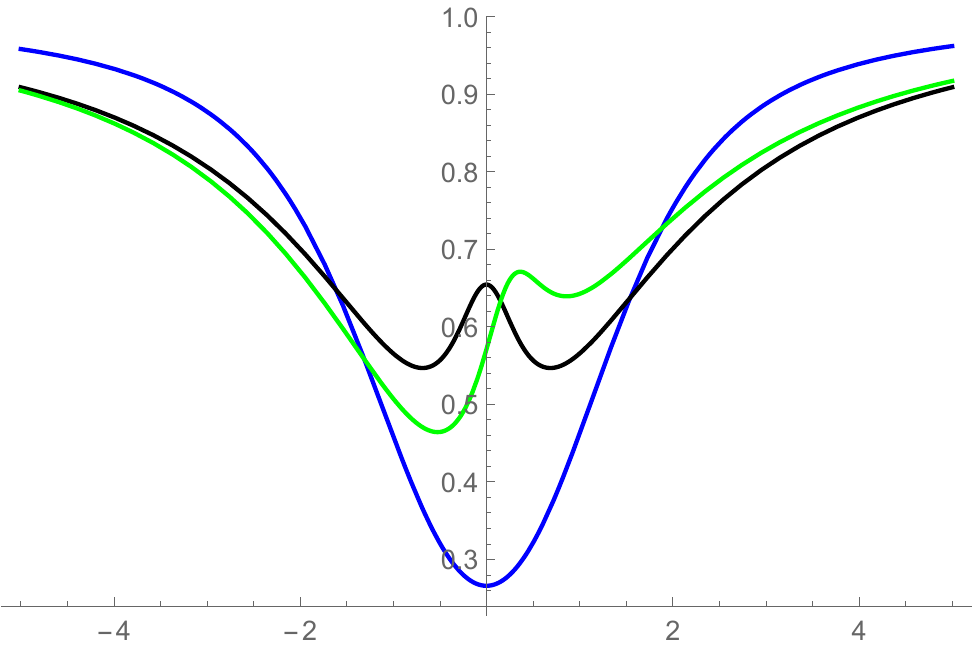} 
\put(0,-5){$\rho$}
\put(-100,135){$\frac{3 \, g^7}{ \, 2^{7/2} \pi^{4}} \, \textrm{vol}_7$}
\end{center}
\caption{Left: Volume of the internal seven-sphere as a function of the complex scalar $\,\tilde{z}\,$ (orange dome). Examples of regular Janus flows (loops) are superimposed. Right: Volume of the internal seven-sphere as a function of the radial coordinate $\,\rho\,$ for the regular Janus solutions. The parameters of the curves are: $\,\alpha=1\,$ and $\,\beta=-\frac{\pi}{2}\,$ (blue line), $\,\beta=\pi\,$ (black line) and $\,\beta=\frac{17}{16}\pi\,$ (green line). Note the presence of two minima (throats) in the black and green lines.}
\label{Fig:Janus_S7}
\end{figure}

Lastly, it is also interesting to compute the volume of the internal manifold $\,\textrm{vol}_7\,$ along the Janus flow as a function of the radial coordinate $\,\rho\,$ and the Janus parameters $\,(\alpha,\beta)\,$. The result is a lengthy expression not very illuminating that we have evaluated and plotted in Figure~\ref{Fig:Janus_S7} for various choices of the Janus parameters. The behaviour is akin a wormhole: the $\,\textrm{S}^7\,$ is a non-contractible seven-manifold  whose volume does not vanish anywhere in the flow along the radial direction $\,\rho\,$. Moreover, for a given value of $\,\alpha\,$, there is a range of the parameter $\,\beta\,$ for which the eleven-dimensional Janus features two throats (see right plot in Figure~\ref{Fig:Janus_S7}).

\subsection{\texorpdfstring{$\text{SU}(3) \times\text{U}(1)^2\,$}{SU3 x U(1)2} symmetric Hades and ridge flows}

Setting $\,\alpha \equiv \alpha_{1}=\alpha_{2}=\alpha_{3}\,$ and $\,\beta \equiv \beta_{1}=\beta_{2}=\beta_{3}\,$ enhances the symmetry of the general Hades solution in (\ref{Hades_U1^4_rho_1})-(\ref{Hades_U1^4_rho_2}) from $\,\textrm{U}(1)^4\,$ to $\,\text{SU}(3) \times\text{U}(1)^2\,$. Setting $\,\alpha \neq 0\,$ renders the running of the scalar field (\ref{Hades_U1^4_rho_2}) along the flow intrinsically complex, as it happened for the Janus case. This again implies that all the $\,h\,$ functions (and also three- and one-forms) entering $\,\hat{F}_{(4)}^{\textrm{st}}\,$ in (\ref{11D_F4_st}) and $\,\hat{F}_{(4)}^{\textrm{tr}}\,$ in (\ref{11D_F4_tr}) are generically activated.

\begin{figure}[t]
\begin{center}
\includegraphics[width=0.50\textwidth]{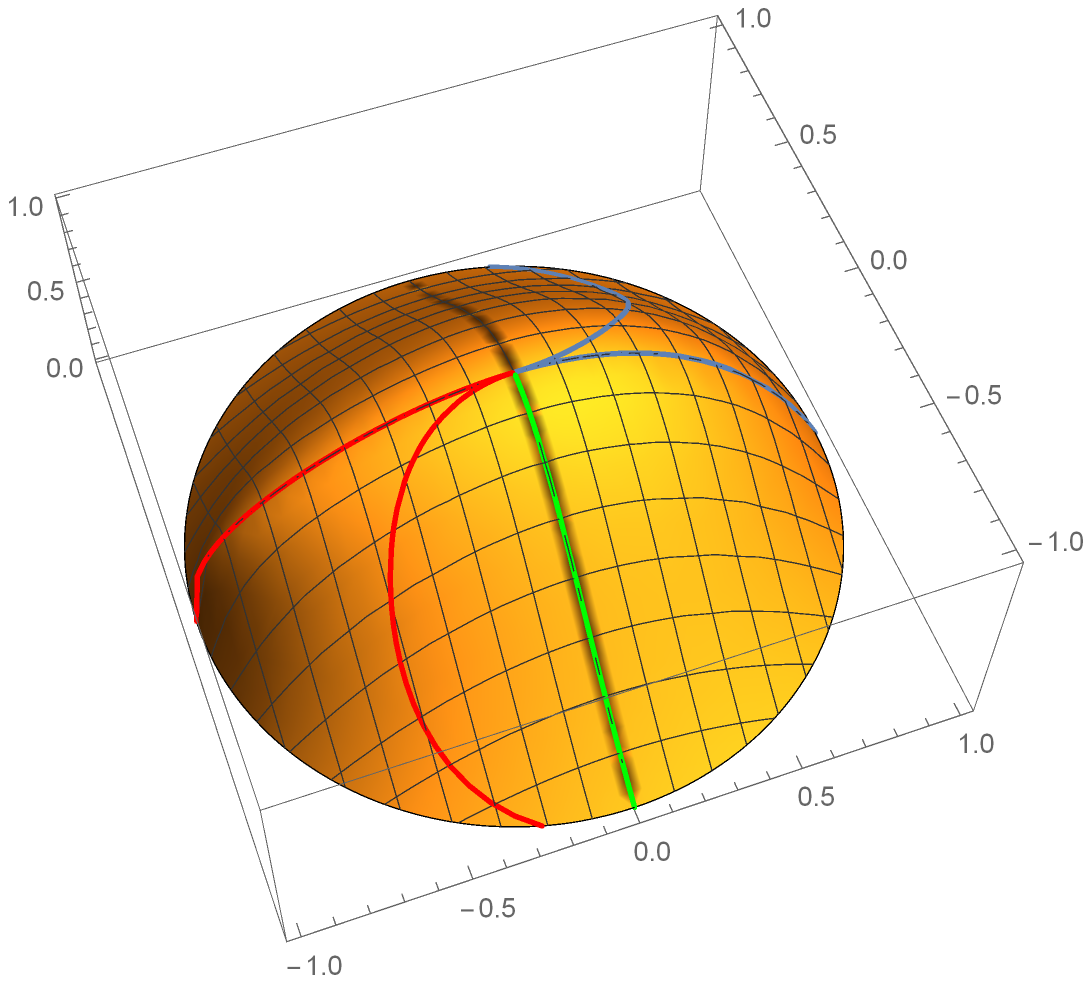} 
\put(-80,20){$\textrm{Re}\tilde{z}$}
\put(-210,60){$\textrm{Im}\tilde{z}$}
\end{center}
\caption{Volume of the internal seven-sphere (orange dome) as a function of the complex scalar $\,\tilde{z}\,$. Examples of singular Hades flows are superimposed with $\,(\alpha,\beta)=(0,0)\,$ (green straight line), $\,{(\alpha,\beta)=(0,\frac{\pi}{2})}\,$ (blue straight line), $\,{(\alpha,\beta)=(0,-\frac{\pi}{2})}\,$ (red straight line), $\,{(\alpha,\lambda)=(2,\frac{\pi}{2}})\,$ (blue curved line) and $\,{(\alpha,\lambda)=(2,-\frac{\pi}{2}})\,$ (red curved line).}
\label{fig:Shell_Hades}
\end{figure}

The decomposition of the seven-form flux $\,\hat{F}_{(7)} \,$ in  (\ref{F7_general}) is still at work for the Hades solutions. The computation of the M$2$-brane charge gives
\begin{equation}
\label{N2_Hades}
N_2 = \frac{1}{(2\pi)^6}\int_{\Gamma \times \textrm{M}_{6}} \hat{F}_{(7)} = \frac{1}{32 \pi^2}\int_{\partial\Gamma} \hat{\alpha}  = \frac{1}{4\pi^2 g^6} \ ,
\end{equation}
so that it matches the one of the $\textrm{AdS}_{4} \times \textrm{S}^7\,$ background controlling the asymptotic behaviour of the Hades solutions at $\,\rho \rightarrow \infty\,$. Some examples of Hades flows on the $\,\tilde{z}\,$ complex plane are displayed in Figure~\ref{fig:Shell_Hades} and superimposed on the volume of the internal seven-sphere.

\subsubsection*{Ridge flows and singularities}

In order to investigate the possible eleven-dimensional resolution of the four-dimensional Hades singularity at $\,\rho=1\,$, we will look at the metric (\ref{11D_metric}) and analyse the relevant function
\begin{equation}
\label{Omega_func}
\Omega \equiv f_{1}^{\frac{1}{2}} \, e^{A} \ ,     
\end{equation}
lying in front of the $\,\textrm{AdS}_{3}\,$ factor of the eleven-dimensional metric describing the world-volume of the (curved) M2-branes in the UV. For simplicity, we will take the limiting case of $\,\alpha=0\,$ and focus on the ridge flows with
\begin{equation}
\label{Ridge_SU3xU1xU1_rho}
ds_{4}^{2}=\frac{1}{{g}^{2}} \left(  \frac{d\rho^{2}}{\rho^{2}-1} + 
\frac{\left(  \rho^{2}-1\right)  }{2} \, d\Sigma^{2}  \right)
\hspace{8mm} \textrm{ and } \hspace{8mm}
\tilde{z}(\rho) =  \rho^{-1} \, e^{i \left( \beta - \frac{\pi}{2} \right)} \ .
\end{equation}
Remarkably, for these flows, the four-dimensional singularity at $\,\rho = 1\,$ gets ameliorated when uplifting the solutions to eleven dimensions provided $\,\beta \neq \pm \frac{\pi}{2}\,$.

The explicit computation of the $\,\Omega\,$ factor in (\ref{Omega_func}) for the ridge flows yields
\begin{equation}
\label{Omega_func_ridge}
\Omega = (2 g)^{-1} \left(  1 + \rho^2 + 2 \, \rho \, \sin\beta \right)^{\frac{1}{6}} \left( 1 + \rho^2 - 2 \, \rho  \, \sin\beta \, \cos(2 \tilde{\alpha})\right)^{\frac{1}{3}} \ .
\end{equation}
Evaluating (\ref{Omega_func_ridge}) at $\,\rho=1\,$ where the four-dimensional singularity is located, one concludes that $\,\Omega\,$ vanishes at $\,(\beta,\tilde{\alpha})=(\frac{\pi}{2},0)\,$ as well as at $\,(\beta,\tilde{\alpha})=(-\frac{\pi}{2},\tilde{\alpha})\,$ $\,\forall \tilde{\alpha}\,$. This hints at a potential pathology at $\,\rho=1\,$ for $\,\beta=\pm \frac{\pi}{2}\,$ which either localises at $\,\tilde{\alpha}=0\,$ or gets delocalised along the interval $\,\tilde{\alpha} \in [0, \frac{\pi}{2}]\,$.\footnote{A similar class of conventional (flat-sliced) RG-flows with $\,{\text{SU}(3) \times\text{U}(1)^2}\,$ symmetry was constructed in \cite{Pilch:2015vha}. For the sake of comparison, there is a redefinition of the relevant parameter given by $\,\zeta_{\tiny{\cite{Pilch:2015vha}}}=\beta-\frac{\pi}{2}\,$. The singularity of the ridge flows we study here would be similar to that of a (yet to be constructed) non-supersymmetric generalisation of the flows in \cite{Pilch:2015vha} with $\,\cos(3\zeta)=+1\,$.} We will look at some limiting examples of ridge flows in order to illustrate their main physical implications.

\subsubsection*{$\circ\,$ Singular $\,\boldsymbol{\beta=\pm\frac{\pi}{2}\,}$ ridge flows:}

The scalar in (\ref{Ridge_SU3xU1xU1_rho}) becomes real when setting $\,\beta=\frac{\pi}{2}\,$. The eleven-dimensional geometry gets simplified to
\begin{equation}
\label{11D_metric_ridge_beta_pi/2}
\begin{array}{lll}
ds_{11}^2 &=& \dfrac{f_{-}^{\frac{2}{3}}}{g^2} \dfrac{(\rho+1)^\frac{2}{3}}{4} \left[ \,  ds_{\textrm{AdS}_{3}}^2 +
\dfrac{2 \, d\rho^2}{(\rho^2-1)^2}  + 8 \dfrac{d\tilde{\alpha}^2}{(\rho+1)^{2}}  \right. \\[6mm]
& + & \left.  \dfrac{8}{f_{-}}   \cos^2\tilde{\alpha} \, \left(  ds^2_{\mathbb{CP}^2} +  \dfrac{(\rho-1)^2}{f_{+}} \,  \sin^{2} \tilde{\alpha} \,  (d\tau_{-} + \sigma)^{2} \right)  \right. \\[6mm]
& + & \left. 
\dfrac{8}{f_{-}}  \,  \left( \dfrac{f_{+}^{\frac{1}{2}}}{\rho+1} d\psi_{-} + \, \dfrac{\rho+1}{f_{+}^{\frac{1}{2}}}  \, \cos^{2} \tilde{\alpha} \, (d\tau_{-} + \sigma) \right)^{2} \, \right]  \ ,
\end{array}
\end{equation}
in terms of the functions
\begin{equation}
\label{f+-_functions}
f_{\pm}= (\rho \pm 1)^2 \mp 4 \, \rho \, \sin^2 \tilde{\alpha} \ .
\end{equation}
Moreover, since the scalar in (\ref{Ridge_SU3xU1xU1_rho}) becomes real, one has that
\begin{equation}
\label{F4tr_ridge_beta_pi/2}
\hat{F}_{(4)}^{\textrm{tr}}=0 \ ,   
\end{equation}
in (\ref{11D_F4_tr}). The non-vanishing contribution to the three-form gauge potential in this case is given by 
\begin{equation}
\label{Ast_ridge_beta_pi/2}
\hat{A}_{(3)}^{\textrm{st}} = \frac{\rho \, (3+\rho+\rho^2) - 2 \, (\rho^2-1) \cos(2\tilde{\alpha})}{8 \, g^3} \, \textrm{vol}_{\textrm{AdS}_{3}} \ ,
\end{equation}
producing a space-time four-form flux in (\ref{11D_F4_st}) of the form
\begin{equation}
\label{F4st_ridge_beta_pi/2}
\hat{F}_{(4)}^{\textrm{st}}=d\hat{A}_{(3)}^{\textrm{st}} = \frac{1}{2 g^3} \left(   
\frac{3 + \rho \, (2 + 3 \, \rho) - 4 \, \rho \,  \cos(2\tilde{\alpha})}{4} \, d\rho   
+ (\rho^2-1) \, \sin(2\tilde{\alpha}) \, d\tilde{\alpha} \right) \wedge  \textrm{vol}_{\textrm{AdS}_{3}}\ . 
\end{equation}
Two facts suggest an interpretation of this flow as a Coulomb branch type flow very much along the line of \cite{Cvetic:1999xx}. Firstly, this singular ridge flow lies in the purely proper scalar sector of maximal supergravity as a consequence of $\,\beta=\frac{\pi}{2}\,$. Namely, it is triggered from the UV solely by the VEV of the proper scalar dual to the boson bilinears. Secondly, the internal flux in (\ref{F4tr_ridge_beta_pi/2}) vanishes all along the flow so there are no magnetic M5-branes sourcing $\,\hat{F}_{(7)}\,$.

Let us now investigate the four-dimensional singularity at $\,\rho=1\,$ from a higher-dimensional perspective. To study the eleven-dimensional geometry around $\,\rho=1\,$ it is convenient to look at the Ricci scalar which, in this case, takes the form
\begin{equation}
\label{Ricci_beta_pi/2}
\hat{R}(\rho) = g^2 \, \frac{ (\rho -1)^2}{3 \, (\rho +1)^{\frac{2}{3}} f_{-}^{\frac{8}{3}}} \,\, r(\rho,\tilde{\alpha}) \ ,
\end{equation}
in terms of the negative-definite function
\begin{equation}
\begin{array}{lll}
r(\rho,\tilde{\alpha}) &=&  
-(9 \rho^4 + 12 \rho^3 + 32 \rho^2 + 16 \rho + 11)
+ 8 \, \rho \, (3 \rho^2 + 2 \rho + 3)   \cos (2 \tilde{\alpha}) \\[2mm]
& & -2 \, (\rho -1) (3 \rho +1) \, \cos(4 \tilde{\alpha}) \ .
\end{array}
\end{equation}
The Ricci scalar in (\ref{Ricci_beta_pi/2}) becomes singular at $\,(\rho,\tilde{\alpha})=(1,0)\,$. On the other hand, the evaluation of the four-form flux in (\ref{F4st_ridge_beta_pi/2}) around the singular value $\,\rho=1\,$ is more subtle. The change of radial coordinate in (\ref{new_coordinate_Hades}) becomes ill-defined and one must resort to the original coordinate $\,\mu\,$ in (\ref{metric_ansatz}) using $\,d\rho = g \, \sqrt{\rho^2-1} \, d\mu \,$. Then, it becomes clear from (\ref{F4st_ridge_beta_pi/2}) that
\begin{equation}
\left. \hat{F}_{(4)}^{\textrm{st}} \right|_{\rho = 1} = 0 \ .
\end{equation}

It is also instructive to look at the flux $\,\hat{F}_{(7)} = d\hat{\alpha} \wedge h^{(6)}  + \ldots\,$ by analysing the expression of the adapted angular variable $\,\hat{\alpha}\,$. In this case it takes the form
\begin{equation}
\hat{\alpha}(\rho,\tilde{\alpha}) = - 8 \, g^{-6} \,  f_{-}^{-1} (\rho-1)^2 \cos^6\tilde{\alpha} \ ,  
\end{equation}
with $\,f_{-}\,$ given in (\ref{f+-_functions}). A plot of the curves $\,\Gamma\,$ is presented in Figure~\ref{fig:vec_field} (bottom-right plot). Note that not all of them start at $\,\tilde{\alpha}=0\,$ and end at $\,\tilde{\alpha}=\frac{\pi}{2}\,$. There are curves that start at $\,\tilde{\alpha}=0\,$ but end at some value $\,0 < \tilde{\alpha} < \frac{\pi}{2}\,$ when reaching the singularity at $\,\rho = 1\,$. These curves display a strong singularity bending: the one-form $\,d\hat{\alpha}\,$ interpolates between being aligned with the $\,\textrm{S}^{7}\,$ direction $\,d\tilde{\alpha}\,$ at $\,\rho \rightarrow \infty\,$ and being aligned with the non-compact direction $\,d\rho\,$ when reaching the singularity at $\,\rho =1\,$.

Finally, recalling the result in Section~\ref{sec:ridge_4D}, setting $\,\beta=-\frac{\pi}{2}\,$ amounts to a reflection of the radial coordinate $\,\rho \rightarrow -\rho\,$ (which implies an exchange $\,f_{+} \leftrightarrow f_{-}\,$) while keeping the domain $\,\rho \in [1,\infty)\,$. This reflection drastically modifies the eleven-dimensional geometry in (\ref{11D_metric_ridge_beta_pi/2}) and (\ref{f+-_functions}) which becomes singular at $\,\rho=1\,$ for any value of the angular coordinate within the interval $\,\tilde{\alpha} \in [0,\frac{\pi}{2}]\,$. This can also be viewed in the eleven-dimensional Ricci scalar which reads
\begin{equation}
\label{Ricci_beta_-pi/2}
\hat{R}(\rho) = g^2 \, \frac{ (\rho+1)^2}{3 \, (\rho -1)^{\frac{2}{3}} f_{+}^{\frac{8}{3}}} \,\, r(-\rho,\tilde{\alpha}) \ .
\end{equation}
Since there is no special value of $\,\tilde{\alpha}\,$ as far as singularities are concerned, the $\,\Gamma\,$ curves constructed from the adapted angular variable 
\begin{equation}
\hat{\alpha}(\rho,\tilde{\alpha}) = - 8 \, g^{-6} \,  f_{+}^{-1} (\rho+1)^2 \cos^6\tilde{\alpha} \ , 
\end{equation}
do not display any bending when approaching $\,\rho=1\,$. These curves are presented in Figure~\ref{fig:vec_field} (bottom-left plot). Lastly, the three-form gauge potential at $\,\beta=-\frac{\pi}{2}\,$ is given by
\begin{equation}
\label{Ast_ridge_beta_minus_pi/2}
\hat{A}_{(3)}^{\textrm{st}} = \frac{\rho \, (3-\rho+\rho^2) + 2 \, (\rho^2-1) \cos(2\tilde{\alpha})}{8 \, g^3} \, \textrm{vol}_{\textrm{AdS}_{3}} \ .
\end{equation}

\subsubsection*{$\circ\,$ Regular $\,\boldsymbol{\beta=0,\pi\,}$ ridge flows:}

The scalar in (\ref{Ridge_SU3xU1xU1_rho}) becomes purely imaginary when setting $\,\beta=0\,$. As a result, this ridge flow is triggered from the UV solely by the source mode of the pseudo-scalar dual to the fermion bilinears.

The eleven-dimensional metric reduces in this case to
\begin{equation}
\label{11D_metric_ridge_beta_0}
\begin{array}{lll}
ds_{11}^2 &=& \dfrac{\rho^2+1}{4 \, g^2} \left[ \,  ds_{\textrm{AdS}_{3}}^2 +
\dfrac{2 \, d\rho^2}{(\rho^2-1)^2}  + 8 \dfrac{d\tilde{\alpha}^2}{\rho^2+1}  \right. \\[6mm]
& + & \left. 8 \, \cos^2\tilde{\alpha} \, \left( \dfrac{1}{\rho^2+1}  ds^2_{\mathbb{CP}^2} + \dfrac{1}{j_{2}} \dfrac{(\rho^2-1)^2}{\rho^2+1} \,  \sin^{2} \tilde{\alpha} \,  (d\tau_{-} + \sigma)^{2} \right)  \right. \\[6mm]
& + & \left. 
\dfrac{8 \, j_{1}}{(\rho^2+1)^3}  \,  \left( \sqrt{\dfrac{j_{2}}{j_{1}}} \, d\psi_{-} + \sqrt{\dfrac{j_{1}}{j_{2}}}  \, \cos^{2} \tilde{\alpha} \, (d\tau_{-} + \sigma) \right)^{2} \, \right]  \ ,
\end{array}
\end{equation}
in terms of the two functions
\begin{equation}
\label{j1_j2_functions}
j_{1} = (\rho^2 + 1)^2 - 4 \, \rho^2 \, \cos(2 \tilde{\alpha}) 
\hspace{8mm} , \hspace{8mm}
j_{2} = (\rho^2 + 1)^2 - 4 \, \rho^2 \, \cos^2(2 \tilde{\alpha}) \ .
\end{equation}
The four-form flux in (\ref{11D_F4}) comes with both space-time and transverse contributions. The former is given by
\begin{equation}
\label{Ast_ridge_beta_0}
\hat{F}_{(4)}^{\textrm{st}}=d\hat{A}_{(3)}^{\textrm{st}}
\hspace{10mm} \textrm{ with } \hspace{10mm}
\hat{A}_{(3)}^{\textrm{st}} = \frac{\rho \, (3+\rho^2)}{8 \, g^3}  \, \textrm{vol}_{\textrm{AdS}_{3}} \ ,
\end{equation}
whereas the latter reads
\begin{equation}
\hat{F}_{(4)}^{\textrm{tr}}=d\hat{A}_{(3)}^{\textrm{tr}} \ ,
\end{equation}
with
\begin{equation}
\label{Atr_ridge_beta_0}
\begin{array}{lll}
\hat{A}_{(3)}^{\textrm{tr}} &=& - \dfrac{4 \sqrt{2}}{g^{3}} \dfrac{\rho}{\rho^2 +1} \Big[ \frac{1}{2} \sin(2\tilde{\alpha}) \, d\tilde{\alpha}  \wedge (d\tau_{-} + \sigma) \wedge d\psi_{-}  \\[2mm]
&& \qquad\qquad\qquad + \cos^4\tilde{\alpha} \, \boldsymbol{J}  \wedge (d\tau_{-} + \sigma) + \cos^2\tilde{\alpha} \, \cos(2\tilde{\alpha}) \, \boldsymbol{J} \wedge d\psi_{-} \Big] \ .
\end{array}
\end{equation}
This signals the presence of both electric M2-branes and magnetic M5-branes at a generic point along the flow.

In order to investigate the four-dimensional singularity at $\,\rho=1\,$ from a higher-dimensional perspective we will look again at the eleven-dimensional Ricci scalar. It reads
\begin{equation}
\label{Ricci_scalar_Hades_beta=0}
\hat{R}(\rho) = g^2 \, \left(1 + \rho^2\right)^{-3} \, \left(1 + 3 \, \rho^2\right) \, \left[ 1 + \rho^2 \left( 8 - \rho^2 \right) \right] \ ,
\end{equation}
and becomes this time independent of the angular variable $\,\tilde{\alpha}\,$. The Ricci scalar in  (\ref{Ricci_scalar_Hades_beta=0}) features no singularity within the domain $\,\rho \in [1 , \infty )\,$. It has a boundary value $\,\hat{R}(\infty) = -3 \, g^2\,$ and changes smoothly until reaching the finite value $\,\hat{R}(1)=4 \, g^2\,$, thus making the eleven-dimensional solution regular. The space-time (\ref{Ast_ridge_beta_0}) and transverse  (\ref{Atr_ridge_beta_0}) components of the three-form gauge potential are both non-zero when approaching the IR region at $\,\rho =1\,$. However, recalling again the change of radial coordinate $\,d\rho = g \, \sqrt{\rho^2-1} \, d\mu \,$, it follows from (\ref{Ast_ridge_beta_0}) that
\begin{equation}
\left. \hat{F}_{(4)}^{\textrm{st}} \right|_{\rho = 1} = 0 \ .
\end{equation}
Therefore, only magnetic M5-branes source the geometry in the deep IR. The same behaviour was observed for the similar, but flat-sliced, $\,\textrm{SU}(3) \times \textrm{U}(1)^2\,$ invariant flows constructed in \cite{Pilch:2015vha}. Such flows were argued to describe how M2-branes in the UV totally dissolve along the flow into magnetic M5-branes, leaving no M2-branes at the core of the regular flows.\footnote{The same type of behaviour was also observed in the flat-sliced dielectric flows with $\,\textrm{SO}(4) \times \textrm{SO}(4)\,$ symmetry of \cite{Pope:2003jp}, although the M2-branes do not totally polarise into M5-branes at the core of these flows.} Moving back to the original radial coordinate
\begin{equation}
\rho = \cosh(g \, \mu) 
\hspace{15mm}  \textrm{ with }  \hspace{15mm} \mu \in [0,\infty) \ ,
\end{equation}
and expanding around $\,\mu = 0\,$ one arrives at
\begin{equation}
\label{11D_metric_ridge_beta_0_IR}
\begin{array}{lll}
\left. ds_{11}^2 \, \right|_{\textrm{IR}} & = & \dfrac{1}{4 \, g^2} \left[ \, \left(\dfrac{4}{(g\mu)^2} + \dfrac{2}{3} + \dfrac{4}{15}(g\mu)^2 + \ldots \right) d(g\mu)^2  +  \left( 2 + (g \mu)^2 + \ldots \right)\, ds_{\textrm{AdS}_{3}}^2 \right. \\[6mm]
& + & \left. 
\left( \dfrac{(g\mu)^4}{2} - \dfrac{(g\mu)^6}{6} + \ldots \right)   \,   \Big( (d\tau_{-} + \sigma) + 2  \cos(2\tilde{\alpha})  \, \left( d\psi_{-} + \frac{1}{2} (d\tau_{-} + \sigma) \right) \Big)^{2} \,  \right. \\[6mm]
& + &
\left.  8 \, \left( d\tilde{\alpha}^2 + \cos^2\tilde{\alpha} \, ds^2_{\mathbb{CP}^2} +  \sin^2(2 \tilde{\alpha}) \, \left(d\psi_{-} + \frac{1}{2}(d\tau_{-} + \sigma) \right)^2 \right)  \right] \ .
\end{array}
\end{equation}
Note that the $\,\mu$-dependent part of the metric only involves the first two lines in (\ref{11D_metric_ridge_beta_0_IR}). This $\,\mu$-dependent part describes a five-dimensional section of the eleven-dimensional geometry that involves the original four coordinates of the ridge flow and an additional $\,\textrm{S}^1\,$ that is non-trivially fibered over a six-dimensional manifold. The latter is described by the last line in (\ref{11D_metric_ridge_beta_0_IR}). Ignoring this fibration, the five-dimensional section of the geometry verifies $\,R^{\textrm{(\tiny{5D})}}_{\mu\nu}= \frac{1}{5} \, R^{\textrm{(\tiny{5D})}} \, g^{\textrm{(\tiny{5D})}}_{\mu\nu}\,$ with $\,R^{\textrm{(\tiny{5D})}} = -20 \, g^2 < 0\,$ at leading order in the radial coordinate $\,\mu\,$. Therefore, up to the non-trivial fibration over the six-dimensional manifold, this ridge flow develops a five-dimensional Einstein geometry in the deep IR.

We will now zoom into the deep IR region ($\mu \rightarrow 0$) by keeping only the leading terms in the parenthesis of the metric (\ref{11D_metric_ridge_beta_0_IR}). Performing a change of coordinate $\,(g \mu)^2 = \sqrt{2}/r\,$ so that $\, r \in [ 0,\infty)$, one finds a deep IR geometry ($r\rightarrow \infty$) of the form
\begin{equation}
\label{11D_metric_ridge_beta_0_deep_IR}
\begin{array}{lll}
\left. ds_{11}^2 \, \right|_{\textrm{IR}} & \approx & \dfrac{1}{4 \, g^2} \left[ \, \dfrac{d\tilde{\tau}_{-}^2+dr^2}{r^2}   + 2 \, ds_{\textrm{AdS}_{3}}^2 \right.  \\[6mm]
& + &
\left.  8 \, \left( d\tilde{\alpha}^2 + \cos^2\tilde{\alpha} \, ds^2_{\mathbb{CP}^2} +  \sin^2(2 \tilde{\alpha}) \, \left(d\psi_{-} + \frac{1}{2}(d\tau_{-} + \sigma) \right)^2 \right)  \right] \ ,
\end{array}
\end{equation}
with
\begin{equation}
d\tilde{\tau}_{-} = (d\tau_{-} + \sigma) + 2  \cos(2\tilde{\alpha})  \, \left( d\psi_{-} + \frac{1}{2} (d\tau_{-} + \sigma) \right) \ .
\end{equation}
As a result, the part of the geometry depending on the coordinate $\,r\,$ in (\ref{11D_metric_ridge_beta_0_deep_IR}) is (locally) the two-dimensional Poincar\'e half-plane (upper half-plane), once again, up to the non-trivial fibration over the six-dimensional manifold and the compactness of $\,\tau_{-} \in[0,2 \pi]\,$. Note that the four-dimensional singularity at $\,r\rightarrow \infty\,$ is an ideal point of the Poincar\'e half-plane and, therefore, is at infinite distance of any other point.

The regularity of the ridge flow at $\,\beta=0\,$ is also reflected in the flux $\,\hat{F}_{(7)} = d\hat{\alpha} \wedge h^{(6)}  + \ldots\,$. The adapted angular variable $\,\hat{\alpha}\,$ simplifies in this case to
\begin{equation}
\hat{\alpha}(\tilde{\alpha}) = - 8 \, g^{-6} \,  \cos^6\tilde{\alpha} \ ,  
\end{equation}
so it is independent of $\,\rho\,$. Therefore, all the $\,\Gamma\,$ curves start at $\,\tilde{\alpha}=0\,$, end at $\,\tilde{\alpha}=\frac{\pi}{2}\,$ and flow parallel to the $\,\textrm{S}^7\,$ angular direction $\,\tilde{\alpha}\,$ without displaying any bending or pathological behaviour.

Finally, as discussed in Section~\ref{sec:ridge_4D}, setting $\,\beta=\pi\,$ amounts to a shift $\,\rho \rightarrow -\rho\,$ in the four-dimensional ridge flow solution while keeping the domain $\,\rho \in [1,\infty)\,$. This reflection of the radial coordinate leaves the eleven-dimensional metric in (\ref{11D_metric_ridge_beta_0}) and (\ref{j1_j2_functions}) invariant.  The three-form gauge potential in (\ref{Ast_ridge_beta_0}) and (\ref{Atr_ridge_beta_0}) simply flips its sign.

\section{Summary and discussion}
\label{sec:conclusions}

In this paper we have presented new analytic families of $\,\textrm{AdS}_{3} \times \mathbb{R}\,$ Janus and $\,\textrm{AdS}_{3} \times \mathbb{R}^{+}\,$ Hades solutions in the $\,\mathcal{N}=2\,$ gauged STU-model in four dimensions \cite{Cvetic:1999xp}. This supergravity model corresponds to the $\textrm{U}(1)^{4}$ invariant sector of the maximal SO(8) gauged supergravity that arises upon reduction of eleven-dimensional supergravity on a seven sphere.

The Janus solutions turn out to be surprisingly simple. Using a radial coordinate $\,\rho  \in (-\infty \, , \infty)\,$, the geometry is given by 
\begin{equation}
\label{Janus_metric_conclus}
g^{2} \, ds_{4}^{2} =  \frac{d\rho^{2}}{\rho^{2}+1} + 
\frac{ \rho^{2} + 1  }{ k^2} \,ds_{\textrm{AdS}_{3}}^{2} \ ,
\end{equation}
in terms of the supergravity gauge coupling $\,g\,$ and three constant parameters $\, \alpha_{i} \in \mathbb{R}^{+} \,$. The latter enter (\ref{Janus_metric_conclus}) through the specific combination
\begin{equation}
\label{k_factor_Janus_conclus}
k^2= 1 + \sum_{i=1}^{3} \sinh^{2}\alpha_{i} \, \ge \, 1 \ .
\end{equation}
The Janus geometry (\ref{Janus_metric_conclus}) is supported by $\rho$-dependent profiles for the three complex scalars in the STU-model. Using the unit-disk parameterisation of the SL(2)/SO(2) scalar coset, they adopt the form
\begin{equation}
\label{Janus_scalar_conclus}
\tilde{z}_{i}(\rho) = e^{i \beta_{i}} \, \frac{\sinh\alpha_{i}}{\cosh\alpha_{i} + i \, \rho}
\hspace{8mm} \textrm{ with } \hspace{8mm} i=1,2,3  \ ,
\end{equation}
and depend on three additional phases $\, \beta_{i} \in [0,2 \pi] \,$. The result is then a six-parameter family $\,(\alpha_{i},\beta_{i})\,$ of Janus solutions in the STU-model which are everywhere regular for arbitrary choices of the parameters $\,(\alpha_{i},\beta_{i})\,$. These are generically non-supersymmetric solutions (they solve second-order equations of motion) but there is a supersymmetry enhancement when two $\,\alpha_{i}\,$ parameters are set to zero. In this limit the supersymmetric Janus with $\,\textrm{SO}(4) \times \textrm{SO}(4)\,$ symmetry of \cite{Bobev:2013yra} is recovered. The very special choice $\,\alpha_{i}=0\,$ $\forall i \,$ sets the three scalars to zero. In this limit the maximally supersymmetric AdS$_{4}$ vacuum of the $\,\textrm{SO}(8)\,$ supergravity is recovered which uplifts to the Freund--Rubin $\,\textrm{AdS}_{4} \times \textrm{S}^{7}\,$ vacuum of eleven-dimensional supergravity \cite{Freund:1980xh}. Note that this vacuum controls the asymptotic behaviour of the Janus solutions at $\,\rho \rightarrow \pm \infty\,$.\footnote{The Janus solutions in (\ref{Janus_metric_conclus})-(\ref{Janus_scalar_conclus}) might resemble the ``boomerang RG flows" studied in \cite{Donos:2017ljs} within the STU-model. These are flows in supergravity both starting and ending at the maximally supersymmetric AdS$_{4}$ vacuum of the SO(8) gauged supergravity, thus being also relevant for ABJM theory. However the Ansatz for the scalar fields in \cite{Donos:2017ljs} explicitly breaks translation invariance in the spatial directions of the dual field theory. This is not the case for the Janus solutions (\ref{Janus_metric_conclus})-(\ref{Janus_scalar_conclus}) which have no dependence on the spatial directions of AdS$_3$.} It is also worth emphasising that the Janus solutions in (\ref{Janus_metric_conclus})-(\ref{Janus_scalar_conclus}) are everywhere regular and genuinely \textit{axionic} in nature: $\,\textrm{Im}\tilde{z}_{i}(\rho) \neq 0\,$ for the solution to exist. This fact makes the study of similar solutions in the Euclidean theory (where pseudo-scalars pick up an extra factor of $\,i\,$ with respect to proper scalars) interesting in the AdS/CFT spirit of \cite{Arkani-Hamed:2007cpn,Bobev:2020pjk}. This could help to understand instanton-like solutions in the context of M-theory, as it has been done for the type IIB non-extremal D-instantons \cite{Bergshoeff:2004fq,Bergshoeff:2004pg,Bergshoeff:2005zf} (see also \cite{Hertog:2017owm}), and perhaps to shed new light on axionic wormholes in M-theory. This issue certainly deserves further investigation.

The Hades solutions are closely related to the Janus solutions and turn out to be very simple too. Using this time a radial coordinate $\,\rho  \in [1 \, , \infty)$, the geometry is given by 
\begin{equation}
\label{Hades_metric_conclus}
g^{2} \, ds_{4}^{2} =  \frac{d\rho^{2}}{\rho^{2} - 1} + 
\frac{  \rho^{2} - 1  }{ k^2} \,ds_{\textrm{AdS}_{3}}^{2} \ ,
\end{equation}
with
\begin{equation}
\label{k_factor_Hades_conclus}
k^2= -1 + \sum_{i=1}^3 \cosh^2\alpha_{i} \ ,
\end{equation}
and the scalar profiles read
\begin{equation}
\label{Hades_scalar_conclus}
\tilde{z}_{i}(\rho) = e^{i \beta_{i}}\, \frac{\cosh\alpha_{i}}{\sinh\alpha_{i} + i \rho} 
\hspace{8mm} \textrm{ with } \hspace{8mm} i=1,2,3  \ .
\end{equation}
Unlike the Janus, the Hades solutions are singular at $\,\rho=1\,$ and do not possess a supersymmetric limit upon tuning of the parameters $\,(\alpha_{i},\beta_{i})\,$. Still the maximally supersymmetric AdS$_{4}$ vacuum controls the asymptotic behaviour of the Hades at $\,\rho \rightarrow \infty\,$. The special limit $\,\alpha_{i}=0\,$ $\forall i \,$ drastically simplifies the Hades solutions giving rise to the so-called ridge flows (see Figure~\ref{fig:Hades_ztilde_U1^4}).

Being obtained within the $\textrm{U}(1)^4$ invariant sector of the massless $\,\mathcal{N}=8\,$ supergravity multiplet in four dimensions, the analytic Janus solutions in (\ref{Janus_metric_conclus})-(\ref{Janus_scalar_conclus}) generalise the supersymmetric ones with $\,\text{SO}(4) \times\text{SO}(4)\,$ symmetry constructed in \cite{Bobev:2013yra}. The non-supersymmetric Hades solutions in (\ref{Hades_metric_conclus})-(\ref{Hades_scalar_conclus}) are genuinely new an cannot be continuously connected with the supersymmetric Hades with $\,\text{SO}(4) \times\text{SO}(4)\,$ symmetry of \cite{Bobev:2013yra} upon tuning of $\,\alpha_{i}\,$. In addition, the Janus and Hades solutions presented in this work can be readily uplifted to eleven-dimensional supergravity using the general results for the oxidation of the STU-model worked out in \cite{Cvetic:1999xp,Azizi:2016noi} and the uplift building blocks collected in the Appendix~\ref{app:general_uplift}. Instead of uplifting the general $\textrm{U}(1)^4$ symmetric Janus and Hades solutions, and for the sake of simplicity, we have restricted to the case 
\begin{equation}
\alpha_{1}=\alpha_{2}=\alpha_{3}=\alpha
\hspace{10mm} \textrm{ and } \hspace{10mm}
\beta_{1}=\beta_{2}=\beta_{3}=\beta
\end{equation}
for which a larger symmetry group $\,\textrm{SU}(3) \times \textrm{U}(1)^2 \subset \textrm{SO}(8)\,$ is preserved by the solutions. The Janus solutions are non-supersymmetric and fully regular, both in four and eleven dimensions, for arbitrary values of the parameters $\,(\alpha,\beta)\,$. The four-dimensional singularity of the Hades may or may not be cured when the solutions are uplifted to eleven-dimensions depending on the choice of parameters $\,(\alpha,\beta)\,$. For example, in the ridge flow limit $\,\alpha=0\,$, the choice $\,\beta=0,\pi\,$ removes the singularity by placing it at infinite distance whereas, if setting $\,\beta=\pm\frac{\pi}{2}\,$, the singularity remains either localised or delocalised in the internal space. It would be interesting to understand the ultimate fate of the singularity in the general Hades solution with $\,\textrm{U}(1)^4\,$ symmetry, as well as to investigate the process of taking the ridge flow limit sequentially on the three scalars $\,\tilde{z}_{i}\,$. Also to further investigate a possible holographic interpretation of these more general flows as interfaces connecting an $\,\mathcal{N}=8\,$ Chern--Simons matter theory to new (non-)conformal phases.

Some open questions and follow-up directions regarding the Janus and Hades presented in this work are immediately envisaged. The first one is the issue of the stability, both perturbative and non-perturbative, of the general class of non-supersymmetric Janus and Hades with $\textrm{U}(1)^4$ symmetry. These solutions can be viewed as AdS$_{3}$ vacua in M-theory, so it would be interesting to investigate their stability in light of the Weak Gravity and Swampland conjectures \cite{ArkaniHamed:2006dz,Ooguri:2016pdq}. In this respect, and unlike for the Hades, the Janus solutions presented here are continuously connected (in parameter space) to the supersymmetric, and thus stable, Janus solutions with $\,{\textrm{SO}(4) \times \textrm{SO}(4)}\,$ symmetry of \cite{Bobev:2013yra}. This could help in improving the stability properties of the generic non-supersymmetric Janus solution at least within some region in the parameter space $\,(\alpha_{i},\beta_{i})\,$.\footnote{See \cite{Eloy:2021fhc} and \cite{Giambrone:2021wsm} for an investigation of this phenomenon in the context of non-supersymmetric  AdS$_{3}$ and AdS$_{4}$ vacua.} Along this line, it would also be interesting to perform a probe brane analysis as a first step towards assessing the non-perturbative stability of the solutions.

The second issue is to understand the higher-dimensional brane picture of the various flows constructed in this work. For a related class of flat-sliced ridge flows, it was shown in \cite{Pilch:2015vha} (motivated by \cite{Pope:2003jp}) that the M2-branes in the UV totally polarise into a $\,(1+3)$-dimensional intersection of M5-branes in the IR generating an AdS$_{5}$ metric at the core of the flow that is non-trivially fibered over a six-dimensional manifold.\footnote{The appearance of a new strongly-coupled IR phase on the M2-brane involving an extra dimension was argued in \cite{Pilch:2015vha} to originate from charged solitons that become massless, very much in the spirit of (massless) type IIA string theory and 11D supergravity.} This phenomenon was signaled by the vanishing of the space-time flux component at the IR endpoint of the flow. In our ridge flows with $\,\textrm{SU}(3) \times \textrm{U}(1)^{2}\,$ symmetry, the expression of the space-time four-form flux (\ref{11D_F4_st}) at generic $\,\beta\,$ is given by
\begin{equation}
\label{F4st_ridge_beta_general}
\hat{F}_{(4)}^{\textrm{st}} = \frac{1}{2 g^3} \left(   
\frac{3 \, (1+\rho^2) + 2 \, \rho \, \sin\beta \, (1- 2 \cos(2\tilde{\alpha}))}{4} \, d\rho   
+ \sin\beta \, (\rho^2-1) \, \sin(2\tilde{\alpha}) \, d\tilde{\alpha} \right) \wedge  \textrm{vol}_{\textrm{AdS}_{3}} \ ,
\end{equation}
so that
\begin{equation}
\left. \hat{F}_{(4)}^{\textrm{st}} \right|_{\rho = 1} = 0 \ ,
\end{equation}
in the deep IR by virtue of the change of radial coordinate $\,d\rho = g \, \sqrt{\rho^2-1} \, d\mu \,$. This suggests a possible interpretation in terms of non-supersymmetric dielectric flows with M2-branes being polarised into intersecting M5-branes. Also, in the case of $\,\beta=0\,$, we have shown the appearence of a five-dimensional geometry in the IR non-trivially fibered over a six-dimensional manifold along the lines of \cite{Pilch:2015vha}. The generalisation to ridge and Hades flows with $\,\textrm{U}(1)^4\,$ symmetry also deserves further investigation.

The third issue has to do with the holographic interpretation of the general Janus and Hades solutions in terms of non-supersymmetric interfaces in the field theory living at the boundary. We have made manifest the strong correlation between the choice of Janus/Hades parameters $\,(\alpha_{i},\beta_{i})\,$ (\textit{i.e.} boundary conditions for the complex scalars $\,\tilde{z}_{i}\,$), the possible emergence of supersymmetry, the source/VEV and bosonic/fermionic nature of the dual operators that are turned on in the interface and the (dis)appearance of gravitational singularities. But much work remains to be done to better understand and characterise the physics of the non-supersymmetric interfaces we have presented.

Finally, it would also be very interesting to construct charged solutions generalising the Janus and Hades constructed in this work, as well as to investigate the effect of including hypermultiplets in the setup thus going beyond the STU-model. We plan to come back to these and related issues in the future.

\section*{Acknowledgements}

We are grateful to Ant\'on Faedo, Carlos Hoyos and Anayeli Ram\'irez for conversations. The research of AA is supported in part by the Fondecyt Grants 1210635, 1221504 and 1181047 and by the FAPESP/ANID project 13231-7. The work of AG and MCh-B is partially supported by the AEI through the Spanish grant PGC2018-096894-B-100 and by FICYT through the Asturian grant SV-PA-21-AYUD/2021/52177.

\appendix

\section{Supersymmetry and BPS equations}
\label{app:susy}

In this appendix we present the BPS equations imposed by supersymmetry. We will use the upper-half plane parameterisation for the complex scalars $z_{i}=-\chi_{i} + i \, e^{-\varphi_{i}}$ as well as the Ansatz for the metric in (\ref{metric_ansatz}), namely, 
\begin{equation}
ds_{4}^{2} = d\mu^{2} + e^{2 A(\mu)} \, d\Sigma^{2} \ .
\end{equation}
The vanishing of the gravitino variation imposes the BPS equation
\begin{equation}
\label{BPS_A}
A^{\prime} = + \sqrt{ |\mathcal{W}|^{2} - \frac{1}{\ell^{2}} \,
e^{-2A} } \ ,
\end{equation}
where we have denoted $A^{\prime}=dA/d\mu$. The real scalar function
\begin{equation}
\label{|W|^2_N2}
|\mathcal{W}|^{2} = \tfrac{1}{2}  \, g^{2} \, e^{K(z,\bar{z})} \, W(z) \, \bar{W}(\bar{z}) \ ,
\end{equation}
with K\"ahler potential 
\begin{equation}
\label{K_pot}
K(z,\bar{z}) = - \sum_{i}\log\left(  - i \, (z_{i} - \bar{z}_{\,\bar{i}}) \right)  \ ,
\end{equation}
and holomorphic superpotential
\begin{equation}
W(z) = \prod_{i} z_{i} - \sum_{i} z_{i} \ ,
\end{equation}
is the gravitino mass square (both $\mathcal{N}=2$ gravitini have the same mass\footnote{\label{Footnote:axions}An explicit computation of the eight gravitino mass terms in the full $\mathcal{N}=8$ theory shows that they come in pairs, namely $8 = 2 + 2_{1} + 2_{2} + 2_{3}$, where the expression of $|\mathcal{W}|^{2}$ for a given pair $2_{i}$ can be obtained from (\ref{|W|^2_N2}) upon the reflection of the axion $\,\chi_{i} \rightarrow- \chi_{i}\,$.}) in terms of which
\begin{equation}
\label{V_from_|W|}
V = \sum_{i} 4 \left[  \left(  \frac{\partial|\mathcal{W}|}{\partial \varphi_{i}} \right)  ^{2} + e^{-2 \varphi_{i}} \left(  \frac{\partial |\mathcal{W}|}{\partial\chi_{i}} \right)  ^{2} \right]  - 3 \, |\mathcal{W}|^{2} \ .
\end{equation}
The vanishing of the remaining fermionic supersymmetry variations imply six additional first-order BPS equations of the form
\begin{equation}
\label{BPS_scalars}
\begin{array}{rll}
\frac{1}{4} \, \varphi_{i}^{\prime} & = & - \dfrac{A^{\prime}}{|\mathcal{W}|} \, \dfrac{\partial|\mathcal{W}|}{\partial\varphi_{i}} + \dfrac{\kappa}{\ell} \, e^{-\varphi_{i}} \, \dfrac{e^{-A}}{|\mathcal{W}|} \, \dfrac{\partial |\mathcal{W}|}{\partial\chi_{i}} \ ,\\[4mm]
\frac{1}{4} \, \chi_{i}^{\prime} & = & - \dfrac{\kappa}{\ell} \, e^{-\varphi_{i}} \, \dfrac{e^{-A}}{|\mathcal{W}|} \, \dfrac{\partial |\mathcal{W}|}{\partial\varphi_{i}} - e^{-2 \varphi_{i}} \, \dfrac{A^{\prime}}{|\mathcal{W}|} \, \dfrac{\partial|\mathcal{W}|}{\partial\chi_{i}} \ ,
\end{array}
\end{equation}
where $i=1,2,3$ and $\kappa= \pm1$ (this sign amounts to change $\mu \rightarrow -\mu$).

\subsubsection*{Supersymmetric Janus}

Using the upper-half plane parameterisation for the complex scalars $z_{i}=-\chi_{i} + i \, e^{-\varphi_{i}}$, the Janus solution in (\ref{Janus_solution_U1^4_ztil}) reads
\begin{equation}
\label{Janus_solution_U1^4}
\begin{array}{rlll}
e^{-\varphi_{i}} & = & \dfrac{-\cosh^{2}({g} \mu)}{1-\sinh^{2}({g} \mu) - 2 \cosh^{2}\alpha_{i} + 2 \, \sinh\alpha_{i} \, \left(  \sinh({g} \mu) \, \sin\beta_{i} + \cosh\alpha_{i} \, \cos\beta_{i} \right)  } & ,\\[6mm]
\chi_{i} & = & \dfrac{2 \, \sinh\alpha_{i} \, \left(   \sinh({g} \mu)  \, \cos\beta_{i} - \cosh\alpha_{i} \, \sin\beta_{i} \right)  }{1-\sinh^{2}({g} \mu) - 2 \cosh^{2}\alpha_{i} + 2 \, \sinh\alpha_{i} \, \left(  \sinh({g} \mu) \, \sin\beta_{i} + \cosh\alpha_{i} \, \cos\beta_{i} \right)  } & .
\end{array}
\end{equation}
Setting $\ell=1$, the BPS equations (\ref{BPS_A}) and (\ref{BPS_scalars}) are not satisfied by the Janus solution in (\ref{A(mu)_func_U1^4})-(\ref{Janus_solution_U1^4}) for generic values of $(\alpha_{i} , \beta_{i})$ thus implying that such a solution is generically non-supersymmetric within this supergravity model. However the identification in (\ref{alpha_2_3=0}) producing a symmetry enhancement to $\,\textrm{SO}(4) \times \textrm{SO}(4)\,$ is compatible with the BPS equations (\ref{BPS_A}) and (\ref{BPS_scalars}) (with $\,\kappa=-1\,$).

\subsubsection*{Supersymmetric Hades}

Using the upper-half plane parameterisation for the complex scalars $\,z_{i}=-\chi_{i} + i \, e^{-\varphi_{i}}\,$ and performing a change of radial coordinate of the form
\begin{equation}
\rho= \cosh({g} \mu) \ ,
\end{equation}
the scalar profiles in the Hades solution (\ref{Hades_U1^4_rho_2}) read
\begin{equation}
\label{Hades_solution_U1^4}
\begin{array}{rlll}
e^{-\varphi_{i}} & = & \dfrac{\sinh^{2}({g} \mu)}{\cosh^{2}({g} \mu) + \cosh(2\alpha_{i}) - 2 \, \cosh\alpha_{i} \, \left(  \cosh({g} \mu) \, \sin\beta_{i} + \sinh\alpha_{i} \, \cos\beta_{i} \right)  } & ,\\[6mm]
\chi_{i} & = & \dfrac{-2 \, \cosh\alpha_{i} \, \left(   \cosh({g} \mu)  \, \cos\beta_{i} - \sinh\alpha_{i} \, \sin\beta_{i} \right)  }{\cosh^{2}({g} \mu) + \cosh(2\alpha_{i}) - 2 \, \cosh\alpha_{i} \, \left(  \cosh({g} \mu) \, \sin\beta_{i} + \sinh\alpha_{i} \, \cos\beta_{i} \right)  } & .
\end{array}
\end{equation}
Setting $\ell=1$, the BPS equations (\ref{BPS_A}) and (\ref{BPS_scalars}) are not satisfied by the Hades solution in (\ref{Hades_U1^4_rho_1})-(\ref{Hades_solution_U1^4}) for any value of $(\alpha_{i} , \beta_{i})$. Therefore, these solutions are non-supersymmetric within this supergravity model.

However, setting from the beginning two out of the three complex fields to the origin of moduli space, \textit{i.e.} $\,z_{2} = z_{3} = i\,$, and changing the geometry to the one in (\ref{Hades_no_ridge_SO(4)xSO(4)_rho_1}), produces a solution that solves the BPS equations (\ref{BPS_A}) and (\ref{BPS_scalars}) (with $\,\kappa=-1\,$) and features an $\,\textrm{SO}(4) \times \textrm{SO}(4)\,$ symmetry.

\section{Uplift of the general \texorpdfstring{$\textrm{U}(1)^4\,$}{U(1)4} symmetric solutions}
\label{app:general_uplift}

The uplift of the $\,\textrm{U}(1)^4\,$ invariant sector of the SO(8) gauged supergravity has explicitly been worked out in Section~$5$ of \cite{Azizi:2016noi}. However, our conventions in this work are slightly different from the ones used in \cite{Azizi:2016noi}. In order to connect both, we start from our Lagrangian in (\ref{Lagrangian_model_U1^4_Einstein-scalars}) and rescale the metric as $\,g_{\mu\nu} = 2 \, \hat{g}_{\mu\nu} \,$ and the gauge coupling as $\,g= \sqrt{2} \,\hat{g}\,$. In this way we obtain the new Lagrangian
\begin{equation}
\label{Lagrangian_Godazgar}
\mathcal{L} = \left(  \hat{R}- \hat{V} \right)  * 1 - \tfrac{1}{2} \displaystyle\sum_{i=1}^3 \left[
d\varphi_{i} \wedge* d\varphi_{i} + e^{2 \varphi_{i}} \, d\chi_{i} \wedge* d\chi_{i} \right] \ ,
\end{equation}
with
\begin{equation}
\label{V_Godazgar}
\hat{V}=-4 \, \hat{g}^2 \sum_{i=1}^3  Y_{i}^2 + \widetilde{Y}_{i}^2 \ ,
\end{equation}
in terms of the quantities
\begin{equation}
\label{Y_U1^4_uplift_quantitites}
Y_{i}^2  =  \frac{1+ |\tilde{z}_{i}|^{2} - (\tilde{z}_{i}+\tilde{z}_{i}^{*}) }{1- |\tilde{z}_{i}|^{2} }
\hspace{6mm} , \hspace{6mm}
\widetilde{Y}_{i}^2  =   \frac{(1+\tilde{z}_{i}) (1+\tilde{z}_{i}^{*}) }{1- |\tilde{z}_{i}|^{2} } \ .
\end{equation}
The Lagrangian in (\ref{Lagrangian_Godazgar})-(\ref{V_Godazgar}) matches precisely the one in eqs (2.1), (2.3), (2.7) and (2.8) of \cite{Azizi:2016noi}. In addition to the quantities in (\ref{Y_U1^4_uplift_quantitites}), three additional quantities
\begin{equation}
\label{b_U1^4_uplift_quantitites}
b_{i} =   - i \frac{\tilde{z}_{i} - \tilde{z}_{i}^{*}}{1- |\tilde{z}_{i}|^{2} } \ ,
\end{equation}
are still needed in order to use the uplift formulae in Section~$5$ of \cite{Azizi:2016noi}. Particularising (\ref{Y_U1^4_uplift_quantitites}) and (\ref{b_U1^4_uplift_quantitites}) either to our general Janus (\ref{Janus_U1^4_rho_1})-(\ref{Janus_U1^4_rho_2}) or Hades (\ref{Hades_U1^4_rho_1})-(\ref{Hades_U1^4_rho_2}) solutions, their eleven-dimensional uplift can be systematically obtained using the results of \cite{Azizi:2016noi}. This goes beyond the scope of this work and we leave it for the future.

\bibliography{references}

\end{document}